\documentclass{ws-ijmpb}
\pdfoutput=1
\usepackage{graphicx}
\usepackage{bm,amssymb}
\usepackage{color}
\begin{document}

\title{A Primer on Surface Plasmon-Polaritons in Graphene}

\author{Yu. V. Bludov$^1$, Aires Ferreira$^2$, N. M. R. Peres$^1$, and M. I. Vasilevskiy$^1$}
\address{$^1$ Physics Department and CFUM, University of Minho, P-4710-057, Braga, Portugal.}

\address{$^2$Graphene Research Centre and Department of Physics, National University
of Singapore, 2 Science Drive 3, Singapore 117542}


\date{\today}
\maketitle

\begin{abstract}
We discuss the properties of surface plasmons-polaritons in graphene and
describe three possible ways
of coupling electromagnetic radiation in the terahertz (THz) spectral range
to this type of surface waves. (i) the attenuated total reflection
 (ATR) method using a prism in the Otto configuration,
(ii) graphene micro-ribbon arrays or monolayers with modulated
 conductivity,  (iii) a metal stripe on top of the graphene layer,
 and (iv) graphene-based gratings. The text provides a number of
original results along with their detailed derivation and discussion.
\end{abstract}



\section{Introduction}
\label{sec_Introduction}

It is since the days of Arnold Sommerfeld, back to 1899, that plasmonic
effects in materials and gratings are investigated both theoretically
and experimentally. Free electron plasma related phenomena in metallic gratings were investigated
as early as 1902 by Wood \cite{Wood}. An attempt to a theoretical
interpretation of Wood's results was first made by Lord Rayleigh \cite{Lord},
followed, three decades latter, by Fano \cite{Fano1,Fano2}. Systematic
studies of plasmonic effects became possible with the work of Kretschmann
\cite{Kretschmann} and Otto \cite{Otto}, who devised two different
methods, using prisms \cite{ExciationIntroduction} on top of or beneath a thin
metallic film, of exciting surface plasmon--polaritons (SPPs). SPPs are evanescent electromagnetic waves coupled to the free electron plasma oscillations, propagating along the surface of a conductor.
Meanwhile, reliable theoretical methods for studying scattering of
electromagnetic radiation (ER) by metallic gratings have been developed
by several authors \cite{Toigo,Chandezon_1982,Chandezon_1999}.

Modern plasmonics gained a renewed interest with the discovery of the anomalously high transmittance through a periodic array of holes with a size
smaller than the diffraction limit \cite{Ebbesen1998}. The effect
was explained invoking surface plasmons (SPs) \cite{Ebbesen2003,Ebbesen2008,theoryarray,theoryarrayII,Stockman2011,PlasmonBook}.
Another impressive SP-related effect is the surface enhanced Raman scattering (SERS) \cite{SERS1,SERS2}.
Beyond the fundamental physics, the plasmonics encompass a wide range of applications
\cite{Zhang}, such as spectroscopy and sensing \cite{SERS1,Langmuir,Amanda2005,Willets2007,Shalabney2011},
photovoltaics \cite{Green2011}, optical tweezers \cite{Reece2008,Juan2011},
nano-photonics \cite{Ebbesen2008,Ozbay2006}, radiation guiding \cite{Han-Bozhevolnyi2013}, transformation and Fourier
optics \cite{Ashkan,Ashkan2,Xu}, etc.
These applications rely on the short wavelength of SPPs (compared propagating photons of the same energy), strong dependence of their dispersion relation on the environment dielectric constant, and high local field intensity associated with localised SPs.

Doped graphene sustains surface plasmons \cite{Shung,SarmaPlasmon,Wunsch,StauberFull,Jablan}
whose frequency is proportional to the 1/4 power of the electronic density,
a result specific to single-layer graphene, and to the 1/2 power of
the wave number, a behavior shared with the 2D electron gas \cite{StauberRS}.
Moreover, the carrier density in graphene can be varied continuously
from nearly zero to $\thickapprox 10^{13}\textrm{cm}^{-2}$ for either type of majority carriers
(electrons or holes)
by the application of an external voltage. This important
ambipolar doping effect has opened the tantalizing prospect of real
time control of SPPs by using a gate. In addition, the transfer of graphene films
to a range of substrates is routine nowadays and has inspired the design of graphene-based
structures with unique plasmonic signatures, such as quantum dots (or anti-dots)
\cite{Mishchenko,Popov,Kinaret,Pedersen,Thongrattanasiri} and nanoribbons
\cite{Nikitin1,Nikitin2,AbajoPlasmons,Raza,Christensen}. In this
respect, the question of whether the classical description of SPs still
holds for finite-size systems, such as quantum dots, is particularly
relevant \cite{Thongrattanasiri}.

The robustness of graphene SPPs with respect to external perturbations,
as well as the interaction of SPPs with individual quantum systems,
are active topics of research: the effect of applied stress in the
graphene plasmon dispersion \cite{Pellegrino,Pellegrino2}, and the
role of SPPs in graphene on the decay rate of nano-emitters \cite{Koppens2011,Nikitin3,Santos,Huidobro,Manjavacas}
are now well established. Plasmons in bilayer graphene have also attracted
attention \cite{Sarma} and the transverse electric (TE) mode spectrum
has recently been obtained \cite{PlasmonsBL}. The plasmon dispersion
relation of a graphene double-layer, two closely separated graphene
sheets, has also been addressed \cite{StauberDL,ProfumoDL}. A natural
extension of the study of propagation of SPPs in monolayer graphene
is the investigation of the same effect in graphene double-layers.
In this case, the two degenerate dispersion relations of each of the
layers hybridize, giving rise to optical and acoustic branches \cite{GanDL,OgnjenDL,WangDL}.

A series of recent experimental works \cite{Schedin2010,LongJuPlasmonics,EchtermeyerPhoto,BasovPlasmons}
triggered a revival of interest to the plasmonic effects in graphene \cite{rmp,rmpPeres}.
In particular, it has been shown that graphene has a strong plasmonic
response in the THz frequency range at room temperature \cite{LongJuPlasmonics}.
THz photonics is emerging as an active field of research \cite{Terahertz}
and graphene may play a key role in THz metamaterials in the near
future. Ju \textit{et al.} have shown \cite{LongJuPlasmonics} that
ER impinging on a grid of graphene microribbons can excite SPPs in
graphene, leading to prominent absorption peaks, whose position can
be tuned by doping. Fei \textit{et al.} mapped the plasmon dispersion
relation of graphene, and demonstrated the plasmonic tunability of
graphene \cite{BasovPlasmons}. In agreement with the theory,\cite{Jablan}
infrared SPPs in graphene have been shown to possess remarkably large
propagation lengths when gauged against more conventional structures
\cite{BasovSPP}. A similar experiment was performed by Chen \textit{et
al.} \cite{KoppensSPP}, where excitation and subsequent detection
of SPPs were achieved. Yan \textit{et al.} \cite{Avouris} have shown
that graphene/insulator stacks can be used as tunable infrared plasmonic
devices, able to work both as a filter and as a polarizer, having
the potential for far-infrared (FIR) and THz photonic devices. As a proof
of principle experiment, it has been shown that this type of devices
can shield 97.5\% of ER at frequencies below 1.2 THz \cite{Avouris}.
Last, the possibility of transforming graphene into a mantle cloak,
working in the THz spectral range, has also been addressed \cite{Chen}.

In general, SPPs cannot be excited by directly shining light on a homogeneous
system due to kinematic reasons: the momentum of a surface polariton is much larger
than that of the incoming light having the same frequency. Therefore,
some type of mechanism is necessary to promote the excitation of SPPs.
The most common mechanisms for SPPs excitation are: (i) attenuated total
reflection (ATR) \cite{YuliyEPL}, (ii) scattering from a topological defect
at the conductor surface \cite{Ebbesen2008,Kuzmenko}, and (iii) Bragg
scattering using diffraction gratings \cite{Ting-Yu_2012} or a periodic
corrugation of the surface of the conductor \cite{Toigo,CorrugatedSLG}.
The method of Ju \textit{et al.} is similar (but not identical)
to the patterning of a metallic grating \cite{EchtermeyerPhoto} on top of
graphene. A theoretical account of the experiment by Ju \textit{et
al.} \cite{LongJuPlasmonics} was given in a recent work \cite{Nikitin1}.
A novel method of SPP excitation consists in using the apex of an illuminated nanoscale
tip \cite{BasovSPP}; this method provides a two-orders of magnitude
enhancement of the in-plane momentum relatively to that of the
impinging ER in free space bypassing the referred kinematic limitation.
It has also been shown that modulation of the optical conductivity
gives rise to efficient ER coupling to SPPs in graphene \cite{nunoSPP}
without the need of a grating. The same principle works with split
\cite{Davoyan2012} or modulated \cite{YuliyPRB} gates. Free space
excitation of SPPs using non-linear optics effects has been proposed
\cite{Shen,Georges} and effectively put in action \cite{Renger}.

The reason why a periodic corrugation allows for the excitation of SPPs
can be understood in analogy with the theory of electrons in a periodic
potential. Here the periodic corrugation plays for SPPs the same role as
the periodic atomic potential plays for electrons, that is, the SPP
momentum is conserved up to a reciprocal lattice vector. The periodic
corrugation provides the missing momentum needed to excite the polariton.
Another way of understanding the effect is to note that the grating gives
rise to a SPP band structure. Then the
folding of the SPP dispersion curve, forming bands in the
first Brillouin zone, makes it possible for an incident electromagnetic wave to
excite a SPP mode associated with the upper bands for the same wavevector. The excitation of SPP modes in the first band still is not possible
in the grating configuration unless the ATR technique is used.

This paper is organized as follows. In Sec.~\ref{sec_maxwell} we write down the macroscopic form of Maxwell's equations, introduce boundary
conditions at an interface containing a graphene sheet, and provide
the definition of TM (transverse magnetic) and TE waves. In Sec.~\ref{sec_Optical_conductivity}
we review the basic properties of the optical conductivity of graphene.
In Sec.~\ref{sec_TM_spectrum} we compute the spectrum of a TM SPP wave providing both numerical and analytical results. In Sec.~\ref{sec_TE_spectrum}
we derive the dispersion relation of a TE SPP
wave. In Sec.~\ref{sec_excitation_by_evanescent} we discuss in detail
the excitation of SPPs on graphene using a prism in the ATR configuration
originally proposed by Otto \cite{Otto}. In Sec.~\ref{sec_rippled}
we discuss scattering of ER by an array of graphene microribbons and other flat graphene structures with periodically modulated conductivity.
Section \ref{sec_metallic contact} is devoted to the SPP excitation by shining light on a metal stripe deposited on graphene.
In Sec.~\ref{sec_rayleigh} we give a formulation of Rayleigh approximation
to scattering of ER by a graphene-based grating due to Toigo \textit{et
al.} \cite{Toigo}, adapted to this particular geometry. There the problems
of both a sine a sawtooth gratings are solved. In Sec.~\ref{sec_metal_grating}
we discuss SPPs on a metal grating coated with graphene. In Sec.~\ref{sec_summary}
we summarize the main results of the paper. In all sections we give
enough details, so that the interest reader may reproduce all the results
by him/herself.


\section{Maxwell's equations and boundary conditions}

\label{sec_maxwell}

\subsection{Planar geometry}

\label{sec_geometry}

The central idea of the emerging area of nanoplasmonics is the use
of SPPs, coupled charge-radiation excitations existing at a dielectric/conductor
interface, for applications ranging from chemical sensors and surface-enhanced
Raman spectroscopy (able to detect a single molecule) to solar cell's
optimization. One of the most attractive features of SPPs is that
they concentrate and guide ER at subwavelength scales. This is appealing
since one can conceive circuitry using metals embedded in dielectrics,
which is able to propagate both electric signals and SPPs \cite{Ozbay2006}.
In graphene, the SPP wavelength can be about 40 times shorter than
the wavelength of the impinging radiation in free space \cite{KoppensSPP}.
\begin{figure}[!ht]
\begin{centering}
\includegraphics[clip,width=9cm]{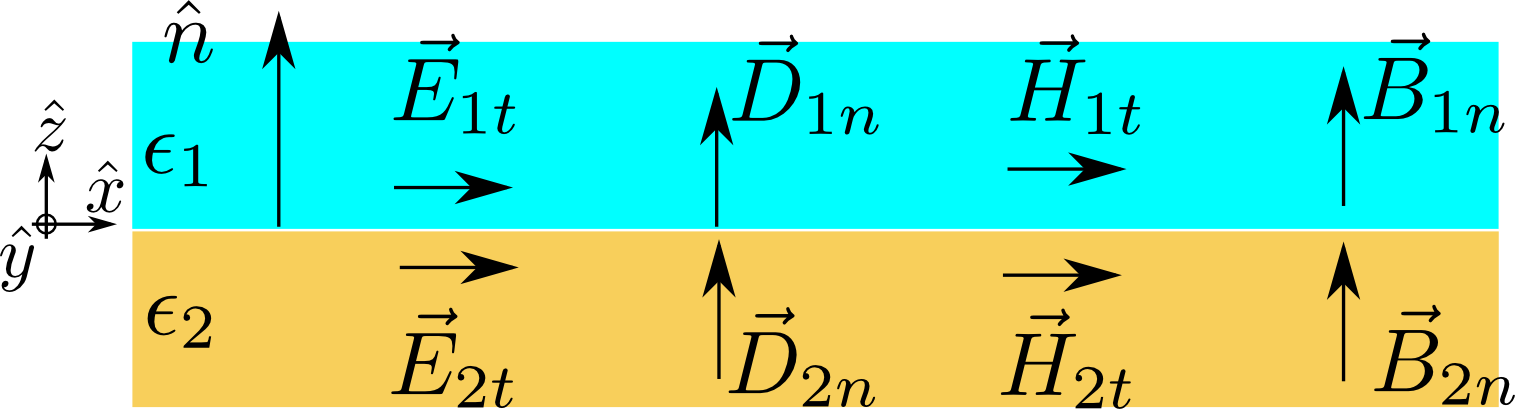}
\par\end{centering}
\caption{Tangential and perpendicular fields at an interface. A graphene sheet
is located between the two dielectrics.}
\label{fig_bc}
\end{figure}

We shall consider a graphene sheet, with conductivity $\sigma$, cladded
by two dielectrics of relative dielectric permittivity $\epsilon_{1}$
and $\epsilon_{2}$. Graphene can be seen as the ultimate conductive
thin film able to support SPPs \cite{semiconductorfilms}. As in any
other conductor, the free charges in graphene couple
to ER. We seek solutions of Maxwell's equations in the form of surface
waves propagating along the graphene sheet. As we will see, graphene supports two
different types of surface waves, the TM (or $p-$polarized) and TE (or
$s-$polarized) SPPs.


\subsection{Maxwell's equations}

\label{sec_Maxwell}

In the MKS system of units, the macroscopic Maxwell's equations read
\begin{eqnarray}
 &  & \vec{\nabla}\cdot\vec{D}=\rho_{f}\,,\label{eq_gauss_law}\\
 &  & \vec{\nabla}\times\vec{E}=-\frac{\partial\vec{B}}{\partial t}\,,\label{eq_faraday_law}\\
 &  & \vec{\nabla}\cdot\vec{B}=0\,,\label{eq_gauss_B}\\
 &  & \vec{\nabla}\times\vec{H}=\vec{J}_{f}+\frac{\partial\vec{D}}{\partial t}\,,\label{eq_ampere_law}
\end{eqnarray}
where $\vec{D}=\epsilon\epsilon_{0}\vec{E}$, $\vec{B}=\mu_{0}\vec{H}$,
$\epsilon$ is the relative dielectric permittivity, $\epsilon_{0}$
and $\mu_{0}$ are the vacuum dielectric and magnetic permeability,
$\rho_{f}$ is the charge density (charge per unit volume), and $\vec{J}_{f}$
is the current density (current per unit area).

In the case of a 2D metal, such as graphene, and assuming that it lies
in the $xy-$plane, we have
\begin{equation}
\vec{J}_{f}=\vec{J}_{s}\delta(z)\,,\qquad\rho_{f}=\rho_{s}\delta(z)\,,
\end{equation}
where $\vec{J}_{s}$ is the surface current density (current
per unit length), $\rho_{s}$ is the surface charge density (charge
per unit area). For a 2D metal with a frequency--dependent (optical) conductivity $\sigma$,
we have $\vec{J}_{s}=\sigma\vec{E}_{t}$, where $\vec{E}_{t}$ lies
in the $xy-$plane as well.

The boundary conditions at the interface between the two media are
(see also Fig.~\ref{fig_bc}):
\begin{eqnarray}
 &  & \vec{E}_{1t}=\vec{E}_{2t}\,,\label{eq:bcond-et}\\
 &  & \vec{H}_{1t}-\vec{H}_{2t}=\vec{J}_{s}\times\hat{n}\,,\label{eq:bcond-ht}\\
 &  & D_{1n}-D_{2n}=\rho_{s}\,,\label{eq:bcond-dn}\\
 &  & B_{1n}=B_{2n}\,.\label{eq_bc_4}
\end{eqnarray}
Since graphene is a two-dimensional system, it enters in the calculation
of the surface wave dispersion relation only through the boundary
conditions. Then, the only quantity we need to known is its optical
conductivity.

Assuming the time dependence of the fields in the form $e^{-i\omega t}$
and considering the absence of free volume currents and charges, Eqs.~(\ref{eq_faraday_law})
and (\ref{eq_ampere_law}) are written as
\begin{eqnarray}
\partial_{y}E_{z}-\partial_{z}E_{y}=i\omega B_{x}\,,\label{eq:Bx}\\
\partial_{z}E_{x}-\partial_{x}E_{z}=i\omega B_{y}\,,\label{eq:By}\\
\partial_{x}E_{y}-\partial_{y}E_{x}=i\omega B_{z}\,,\label{eq:Bz}
\end{eqnarray}
and
\begin{eqnarray}
\partial_{y}B_{z}-\partial_{z}B_{y}=-ic^{-2}\epsilon\omega E_{x}\,,\label{eq:Ex}\\
\partial_{z}B_{x}-\partial_{x}B_{z}=-ic^{-2}\epsilon\omega E_{y}\,,\label{eq:Ey}\\
\partial_{x}B_{y}-\partial_{y}B_{x}=-ic^{-2}\epsilon\omega E_{z}\,,\label{eq:Ez}
\end{eqnarray}
respectively. As mentioned above, Maxwell's equations apply to the dielectrics
with permittivity $\epsilon_{1}$ and $\epsilon_{2}$ surrounding
graphene, and we take into account the relation $\mu_{0}\epsilon_{0}=c^{-2}$.


\subsection{Definition of TE and TM modes and Poynting
vector}

\label{sec_TE_TM}

A conductive surface, depending on certain properties of metal's
optical conductivity, can support transverse electric (TE or $s-$polarized)
or/and transverse magnetic (TM or $p-$polarized) surface waves. These
waves propagate along the metallic interface and decay exponentially
away from it (along the $z-$direction). In the case of graphene,
the free carrier oscillations are confined to the $xy$ plane,
while the electromagnetic field penetrates considerably into the cladding materials.

\begin{figure}[!ht]
\begin{centering}
\includegraphics[clip,width=9cm]{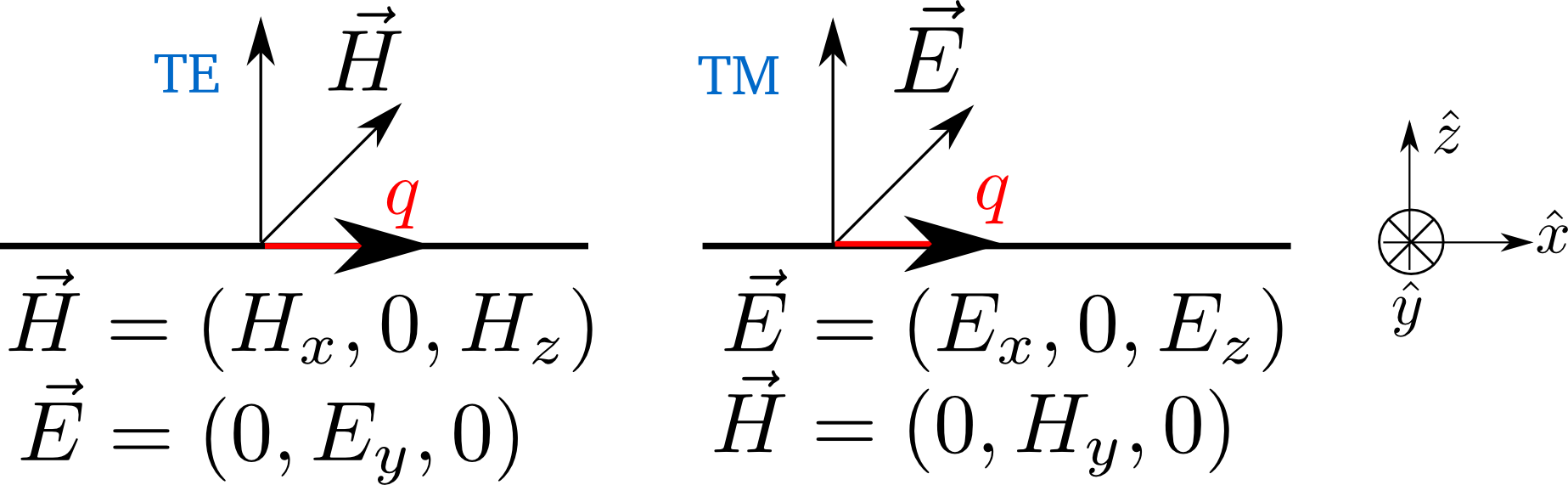} 
\par\end{centering}
\caption{Electric and magnetic fields of transverse electric (TE; left in the
figure) and transverse magnetic (TM; right in the figure) surface
waves. The graphene sheet lies in the $xy-$plane.}
\label{fig_TE_TM}
\end{figure}

In Fig.~\ref{fig_TE_TM} we represent both types of waves propagating
along the $\hat{x}$ direction with momentum $q$. For TE-waves the
electric field is oriented along the $\hat{y}$ direction. Then, we
have
\begin{equation}
\vec{J}_{s}\times\hat{n}=\sigma_{yy}E_{y}\hat{y}\times\hat{n}=\sigma_{yy}E_{y}\hat{x}\,,
\end{equation}
and for TM-waves the magnetic field is oriented
along the $\hat{y}$ direction, so we have
\begin{equation}
\vec{J}_{s}\times\hat{n}=\sigma_{xx}E_{x}\hat{x}\times\hat{n}=-\sigma_{xx}E_{x}\hat{y}\,,
\end{equation}
where we have assumed an anisotropic response. For an isotropic system,
such as unstrained graphene, $\sigma_{xx}=\sigma_{yy}=\sigma$.


For future use, we recall here the definition of the Poynting vector, which is the instantaneous ER energy flux per unit area \cite{OpticsBook},
\begin{equation}
\vec{S}=\vec{E}\times\vec{H}\,,
\end{equation}
where $\vec{E}$ and $\vec{H}$ are assumed to be real. Representing $\vec{E}$ and $\vec{H}$ by harmonic functions we have:
\begin{eqnarray}
\vec{S} & = & \Re(\vec{E}_{0}e^{ikz-i\omega t})\times\Re(\vec{H}_{0}e^{ikz-i\omega t})\nonumber \\
 & = & \frac{1}{2}\Re(\vec{E}_{0}\times\vec{H}_{0}^{\ast})+\frac{1}{2}\Re(\vec{E}_{0}\times\vec{H}_{0}e^{2i(kz-\omega t)})\,.
\end{eqnarray}
The time average of $\vec{S}$, i.e. the directional energy flux density, reads:
\begin{equation}
\langle\vec{S}\rangle=\frac{1}{2}\Re(\vec{E}_{0}\times\vec{H}_{0}^{\ast})\,.
\end{equation}
The calculation of $\langle\vec{S}\rangle$ allows for the determination of the ER reflectance and transmittance at an interface.


\section{Optical conductivity of graphene}

\label{sec_Optical_conductivity}


\subsection{General expression}

\label{sec_sigma}

In this section we provide a brief overview of the optical properties
of graphene. Its optical conductivity is a sum of two
contributions: (i) a term describing interband transitions and (ii)
a Drude contribution, describing intraband processes. At zero temperature
the optical conductivity has a simple analytical expression \cite{nmrPRB06,falkovsky,rmp,rmpPeres,StauberGeim,Polini}.
The inter-band contribution has the form $\sigma_{I}=\sigma_{I}'+i\sigma_{I}''$,
with
\begin{equation}
\sigma_{I}'=\sigma_{0}\left(1+\frac{1}{\pi}\arctan\frac{\hbar\omega-2E_{F}}{\hbar\gamma}-\frac{1}{\pi}\arctan\frac{\hbar\omega+2E_{F}}{\hbar\gamma}\right)\,,
\end{equation}
and
\begin{equation}
\sigma_{I}''=-\sigma_{0}\frac{1}{2\pi}\ln\frac{(2E_{F}+\hbar\omega)^{2}+\hbar^{2}\gamma^{2}}{(2E_{F}-\hbar\omega)^{2}+\hbar^{2}\gamma^{2}}\,,
\end{equation}
where $\sigma_{0}=\pi e^{2}/(2h)$, $e<0$ is the electron charge, $\gamma$ is the relaxation rate,
and $E_{F}>0$ denotes the (local) Fermi level position with respect to
the Dirac point.
The Drude conductivity term is
\begin{equation}
\sigma_{D}=\sigma_{0}\frac{4E_{F}}{\pi}\frac{1}{\hbar\gamma-i\hbar\omega}\, .
\label{eq_sigma_xx_Drude}
\end{equation}
The total conductivity is therefore given by
\begin{equation}
\sigma_{g}=\sigma'+i\sigma''=\sigma_{I}'+i\sigma_{I}''+\sigma_{D}\,.
\end{equation}
\begin{figure}[!ht]
\begin{centering}
\includegraphics[clip,width=10cm]{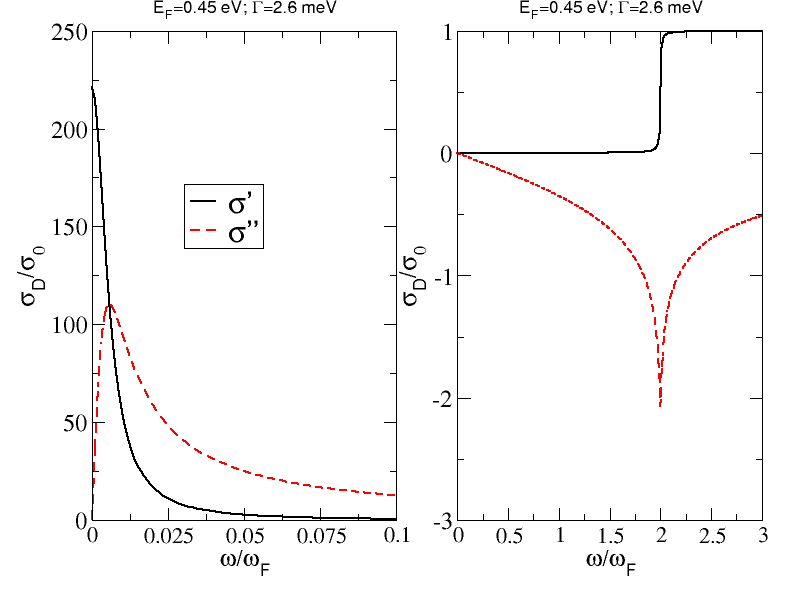}
\par\end{centering}
\caption{Optical conductivity of uniform graphene: Drude (left) and inter-band
(right) contributions. We assume $E_{F}=0.45$~eV and $\Gamma=2.6$~meV.
The solid (dashed) line stands for the real (imaginary) part of the
conductivity. In both panels $\omega_F=E_F/\hbar$.}
\label{fig_conductivity}
\end{figure}
In Fig.~\ref{fig_conductivity} the two contributions, Drude
and interband, are plotted separately for a given value of $E_{F}$ and $\Gamma=\hbar\gamma$.
For heavily doped graphene and for photon energies $\hbar\omega/E_{F}\ll1$,
the optical response is dominated by the Drude term. Therefore, in
what follows we assume
\begin{equation}
\sigma_{g}\approx\sigma_{D}\,,
\end{equation}
since we are interested in this regime of frequencies.
For the frequency range of interest in this work (the THz spectral
range) the above approximation gives accurate results. The only exception
to this approximation will be made when discussing TE-waves in graphene (Sec. \ref{sec_TE_spectrum}).


\subsection{Drude conductivity in a magnetic field}

When a static external magnetic field is considered, the response
of two-dimensional electronic systems to an external time-dependent
field is described by the magneto-optical conductivity tensor $\sigma_{\alpha\beta}$,
where $\alpha,\beta=x,y$ denote the in-plane coordinates. The tensorial
nature of the magneto-optical response is a direct manifestation of
the Lorentz force which, in the presence of a magnetic field, gives
origin to both longitudinal and transverse electronic currents.

We assume that the applied static magnetic field is homogeneous and
transverse to the graphene sheet, $\vec{B}=B\hat{e}_{z}$. In the semiclassical
regime, $\hbar\omega/E_{F}\ll1$, the magneto-optical transport in
graphene can be described in terms of Boltzmann's transport equation. \cite{EOM}
Within this formalism, the electric current is obtained according
to
\begin{equation}
\vec{J}=\frac{e}{\pi^2}\int d^{2}\vec{k}\,\,\delta f_{\vec{k}}\,\vec{v}_{\vec{k}}\,,\label{eq:semi-classical_current}
\end{equation}
where $\delta f_{\vec{k}}$ is the deviation of the carriers' (electrons
or holes) distribution function from the equilibrium Fermi--Dirac distribution,
$f_{0}(\epsilon)$, and
\begin{equation}
\vec{v}_{\vec{k}}=v_{F}(\cos\theta_{\vec{k}},\sin\theta_{\vec{k}})\,,\label{eq:Velocity}
\end{equation}
with $v_{F}\approx10^{6}$~m/s denoting the Fermi velocity in graphene
and $\theta_{\vec{k}}=\arctan(k_{y}/k_{x})$. We remark that both
spin and valley degeneracies have been included in Eq.~(\ref{eq:semi-classical_current}).

In the presence of an electromagnetic field, the distribution function perturbation, $\delta f_{\vec{k}}$,
is the solution of the kinetic equation \cite{Ziman},
\begin{equation}
-e\vec{E}_{\parallel}\cdot\vec{v}_{\vec{k}}\frac{\partial f_{0}}{\partial\epsilon}=\frac{\delta f_{\vec{k}}}{\tau_{\vec{k}}}+\frac{e}{\hbar}\left(\vec{v}_{\vec{k}}\times\vec{B}\right)\cdot\frac{\partial}{\partial\vec{k}}\left[\delta f_{\vec{k}}\right]\,,\label{eq:boltzmann_eq}
\end{equation}
where the standard relaxation approximation,
\begin{equation}
\left.\frac{\partial}{\partial t}\left[\delta f_{\vec{k}}\right]\right|_{\textrm{scatt}}=-\frac{\delta f_{\vec{k}}}{\tau_{\vec{k}}}\:,
\label{eq:relaxation_approx}
\end{equation}
has been assumed. Here $\tau_{\vec{k}}$ denotes the carrier's
relaxation time, and $\vec{E}_{\parallel}=(E_{x},E_{y})$ is the
projection of the electric field onto the graphene plane. The kinetic
equation can be solved exactly by writing $\delta f_{\vec{k}}$
as
\begin{equation}
\delta f_{\vec{k}}=e^{-i\omega t}\vec{k}\cdot\vec{A}_{\vec{k}}\,,\label{eq:g_k}
\end{equation}
and noting that $(\vec{v}_{\vec{k}}\times\vec{B})\cdot\partial_{\vec{k}}\delta f_{\vec{k}}=\vec{v}_{\vec{k}}\cdot(\vec{B}\times\partial_{\vec{k}}\delta f_{\vec{k}})$,
in order to obtain after some algebra
\begin{equation}
\vec{A}_{\vec{k}}=\frac{1}{(1-i\omega\tau_{\vec{k}})^{2}+\omega_{c}^{2}\tau_{\vec{k}}^{2}}\left(\begin{array}{cc}
1-i\omega\tau_{\vec{k}} & -\tau_{\vec{k}}\omega_{c}\\
\tau_{\vec{k}}\omega_{c} & 1-i\omega\tau_{\vec{k}}
\end{array}\right)\vec{\mathcal{E}}_{\vec{k}}\,,\label{eq:vector_pot}
\end{equation}
where $\vec{\mathcal{E}}$ is defined as
\begin{equation}
\vec{\mathcal{E}}_{\vec{k}}=-e\frac{\partial f_{0}}{\partial\epsilon}(E_{x}v_{\vec{k},x},E_{y}v_{\vec{k},y})\,,\label{eq:def_Ex/y}
\end{equation}
and
\begin{equation}
\omega_{c}=ev_{F}^{2}B/|E_{F}|\,\label{eq:cyclotron_freq}
\end{equation}
is the cyclotron frequency in graphene.
Introducing the explicit form of $\delta f_{\vec{k}}$ in Eq.~(\ref{eq:semi-classical_current}),
and assuming $T=0$,
\begin{equation}
-\frac{\partial f_{0}}{\partial\epsilon}=\delta(\epsilon-E_{F})\,,
\label{eq:derivative distribution function}
\end{equation}
one arrives at the semiclassical form of the conductivity tensor:
\begin{eqnarray}
\sigma_{xx}(B,\omega) & = & \frac{e^{2}}{h}\frac{2|E_{F}|}{\hbar}\frac{\gamma-i\omega}{(\gamma-i\omega)^{2}+\left[\omega_{c}(B)\right]^{2}}\,,\label{eq:sigma_xx_semiclass}\\
\sigma_{xy}(B,\omega) & = & -\frac{e^{2}}{h}\frac{2E_{F}}{\hbar}\frac{\omega_{c}(B)}{(\gamma-i\omega)^{2}+\left[\omega_{c}(B)\right]^{2}}\,,\label{eq:sigma_xy_semiclass}
\end{eqnarray}
$\sigma_{yy}(B,\omega)=\sigma_{xx}(B,\omega)$, and $\sigma_{yx}(B,\omega)=-\sigma_{xy}(B,\omega)$.
Note that for simplicity we have expressed the result in terms of
the relaxation's rate $\gamma\equiv\tau_{k_{F}}^{-1}$.

In the presence of a magnetic field, and for pristine graphene ($\hbar\gamma\ll$
energy scales), the Drude peak (i.e.,~the maximum of $\textrm{Re}\,\sigma_{xx}$)
is located at the cyclotronic frequency, reflecting the intuitive
fact that the optical response (absorption) is maximum for impinging
light in resonance with the frequency for cyclotronic motion, that
is,~$\omega=\omega_{c}$. We note that the Drude conductivity for
zero-field {[}Eq.~(\ref{eq_sigma_xx_Drude}){]} is recovered by setting
$B=0$ in Eq.~(\ref{eq:sigma_xx_semiclass}).


\section{Spectrum of TM SPPs in graphene}
\label{sec_TM_spectrum}
\subsection{Dispersion relation}

Let us find the form of a $p$-polarized
surface wave in graphene. We assume a solution of Maxwell's equations
in the form
\begin{eqnarray}
\vec{E} & = & (E_{m,x},0,E_{m,z})e^{iqx}e^{-\kappa_{m}\vert z\vert}\,,\label{eq:specE-field}\\
\vec{B} & = & (0,B_{m,y},0)e^{iqx}e^{-\kappa_{m}\vert z\vert}\,,\label{eq:specH-field}
\end{eqnarray}
where $m=1,2$ refers to the media 1 and 2 (see Fig.~\ref{fig_bc}),
waves are exponentially decaying in both directions away from
the graphene sheet, and we seek the dispersion relation, $\omega=\omega(q)$,
of this type of waves, with $q$ denoting the wave number along
the graphene sheet (see Fig.~\ref{fig_TE_TM}). The wave numbers
$\kappa_{m}$ are yet to be determined. In this case, Maxwell's equations
(\ref{eq:By}), (\ref{eq:Ex}), (\ref{eq:Ez}) yield:
\begin{eqnarray}
(-1)^{m}\kappa_{m}E_{m,x}-iqE_{m,z}=i\omega B_{m,y}\,,\\
(-1)^{m+1}\kappa_{m}B_{m,y}=-i\omega c^{-2}\epsilon_{m}E_{m,x}\,,\\
qB_{m,y}=-\omega c^{-2}\epsilon_{m}E_{m,z}\,,
\end{eqnarray}
which can be solved in terms of the amplitude of the magnetic field
and allows for the determination of $\kappa_{m}$, that is,
\begin{eqnarray}
E_{m,x} & = & i\frac{\kappa_{m}c^{2}}{\omega\epsilon_{m}}B_{m,y}(-1)^{m+1}\,,\label{eq:comp1}\\
E_{m,z} & = & -\frac{qc^{2}}{\omega\epsilon_{m}}B_{m,y}\,,\label{eq:comp2}\\
\kappa_{m}^{2} & = & q^{2}-\omega^{2}\epsilon_{m}/c^{2}\,,\label{eq:disp}
\end{eqnarray}
with $B_{m,y}$ a constant. The SPP spectrum follows
from the boundary conditions (\ref{eq:bcond-et}) and (\ref{eq:bcond-ht}):
\begin{eqnarray}
E_{1,x} & = & E_{2,x}\Leftrightarrow B_{1,y}=-\frac{\kappa_{2}\epsilon_{1}}{\kappa_{1}\epsilon_{2}}B_{2,y}\,,\\
B_{1,y} & = & B_{2_{y}}-\sigma_{xx}E_{1,x}\,,\label{eq_bc_TM_2}
\end{eqnarray}
from which we obtain the dispersion relation (in an implicit form,
since both $\kappa_{m}$ and $\sigma_{xx}$ depend
on frequency)
\begin{eqnarray}
\nonumber
1+\frac{\kappa_{1}\epsilon_{2}}{\epsilon_{1}\kappa_{2}}+i\sigma_{xx}\frac{\kappa_{1}}{\varepsilon_{0}\omega\epsilon_{1}}=0\:;\\
\frac{\epsilon_{1}}{\kappa_{1}}+\frac{\epsilon_{2}}{\kappa_{2}}+i\frac{\sigma_{xx}}{\varepsilon_{0}\omega}=0\,.
\label{eq_W_SPP_2D}
\end{eqnarray}
We note that Eq.~(\ref{eq_W_SPP_2D}) has real solutions only when
the imaginary part of the conductivity is positive. This takes place
when the conductivity is dominated by the Drude contribution. If the
real part of the conductivity is finite (non-zero), the solutions are
necessarily complex. Eq.~(\ref{eq_W_SPP_2D}) gives the spectrum
of the $p-$polarized SPPs in graphene. When the two media are 
the same ($\epsilon_1=\epsilon_2=\epsilon$), we obtain a
simpler relation for the spectrum,
\begin{equation}
1+i\frac{\sigma_{xx}}{2\omega\epsilon_{0}\epsilon}\sqrt{q^{2}-\omega^{2}\epsilon/c^{2}}=0\,.
\end{equation}

Notice that starting from the boundary condition (\ref{eq:bcond-dn})
would lead to the same spectrum. It can be shown as follows. From
the continuity equation
\begin{equation}
\partial_{t}\rho_{s}(x,t)+\vec{\nabla}\cdot\vec{J}_{s}(x,t)=0\,,
\end{equation}
follows $-i\omega\rho_{s}+iq\sigma_{xx}E_{1,x}=0$.
Thus, the boundary condition reduces to $E_{1,z}\epsilon_{1}-E_{2,z}\epsilon_{2}=k_{x}\sigma_{xx}E_{1,x}/(\omega\epsilon_{0})$.
Using the relation (\ref{eq:comp2}), we obtain
\begin{equation}
-B_{1,y}+B_{2,y}=\sigma_{xx}E_{1x}\,,
\label {bc-TM}
\end{equation}
which is the same as Eq. (\ref{eq_bc_TM_2}).


\subsection{Simplified analytical form}

Let us obtain some simple analytical results for the spectrum of the
SPPs in graphene. We assume that the conductivity of graphene is given
by the Drude contribution only. Ignoring absorption ($\gamma=0$),
we have
\begin{equation}
\sigma_{D}\approx i\frac{\nu}{\omega},\qquad\nu=\sigma_{0}\frac{4E_{F}}{\pi\hbar}\,.\label{eq_sigma_approx}
\end{equation}
Furthermore, assuming $\epsilon_{2}=\epsilon_{1}=\epsilon$, the equation
for the SPPs spectrum reads
\begin{equation}
\frac{1}{\kappa}=\frac{\nu}{2\epsilon\epsilon_{0}\omega^{2}}\,.\label{eq_W_SPP_2D_simp}
\end{equation}
Inverting and squaring, we obtain
\begin{equation}
\omega^{4}=\left(\frac{\nu}{2\epsilon\epsilon_{0}}\right)^{2}(q^{2}-\omega^{2}\epsilon/c^{2})\,.\label{eq_W_SPP_2D_simp_f}
\end{equation}
When $\omega\rightarrow0$, we obtain from Eq.~(\ref{eq_W_SPP_2D_simp_f})
$\omega=vq$, where $v=c/\sqrt{\epsilon}$ is the speed of light in
the dielectric. In the electrostatic limit (also dubbed non-retarded
or plasmon approximation) we consider $\omega^{2}/v^{2}\ll q^{2}$,
in which case we obtain
\begin{equation}
\hbar^{2}\omega^{2}=q\frac{\hbar^{2}\nu}{2\epsilon\epsilon_{0}}=\frac{2\alpha E_{F}}{\epsilon}\hbar cq\,,\label{eq_W_Plasmon_graphene}
\end{equation}
where
\begin{equation}
\alpha=\frac{e^{2}}{4\pi\epsilon_{0}\hbar c}\approx\frac{1}{137}
\end{equation}
is the fine-structure constant. We note that Eq.~(\ref{eq_W_Plasmon_graphene})
coincides with the spectrum of plasmons in graphene. Since $E_{F}=v_{F}\hbar k_{F}$,
the spectrum depends on the electronic density as $n_{e}^{1/4}$,
which is specific of graphene and was experimentally confirmed \cite{LongJuPlasmonics}.

We now derive the spectrum of the SPPs taking into account absorption.
In this case we have to use the full form of $\sigma_{D}$. The dispersion relation
now reads,
\begin{equation}
\kappa=\frac{2\epsilon\epsilon_{0}}{\nu}(\omega^{2}+i\gamma\omega)\,.
\end{equation}
In the electrostatic limit we write $\kappa\approx q=q'+iq''$ (assuming $q''\ll q'$), which is now a complex quantity due to a non-zero
$\gamma$. In this case, the SPP spectrum is obtained from
\begin{equation}
q'+iq''=\frac{2\epsilon\epsilon_{0}}{\nu}(\omega^{2}+i\gamma\omega),
\end{equation}
where we have assumed $\gamma\ll\omega$. In terms of $q'$, the spectrum is
\begin{equation}
\omega^{2}=q'\frac{\nu}{2\epsilon\epsilon_{0}}\,,
\end{equation}
which has the same functional form as in Eq.~(\ref{eq_W_Plasmon_graphene}).
The decay of the wave as it propagates in space is characterized by
$q''$,
\begin{equation}
q''=\frac{2\epsilon\epsilon_{0}}{\nu}\gamma\omega\,,
\end{equation}
from which follows
\begin{equation}
\frac{q''}{q'}=\frac{\gamma}{\omega}\,,
\end{equation}
that is, the decay is less pronounced at higher frequencies and smaller $\gamma$.
The confinement of the electromagnetic
field in the $z$ direction (i.e. its penetration depth in the dielectrics
 surrounding the graphene sheet) is given by $\kappa ^{-1}$.
In the electrostatic limit, from Eq.~(\ref{eq_W_Plasmon_graphene}) we have:
\begin{equation}
\kappa \approx q=\omega^{2}\frac{2\epsilon \epsilon_{0}}{\nu}\,.
\end{equation}
In the dielectric the wave number of light is $k=\omega/v$. Comparing
$\kappa$ with $k$ we obtain:
\begin{equation}
\frac{\kappa}{k}\approx\frac{2\epsilon\epsilon_{0}v}{{\nu}}\omega=\frac{\sqrt{{\epsilon}}}{2\alpha}\frac{\hbar\omega}{{E_{F}}}\approx\frac{137}{2}\sqrt{{\epsilon}}\frac{\hbar\omega}{{E_{F}}}\,.
\end{equation}
Thus, the EM penetration depth decreases with $\epsilon$
and with the photon frequency and increases with the Fermi energy.
Considering, as an example, $\omega/(2\pi)=10$ THz, ${E_{F}}=0.2$
eV, and ${\epsilon}=4$, we obtain $\kappa ^{-1}\approx 0.035 k^{-1}$, a fairly high degree of confinement in comparison with the wavelemgth.


\subsection{Numerical results}
\label{sec_TM_numerical}
\begin{figure}[!ht]
\begin{centering}
\includegraphics[clip,width=9cm]{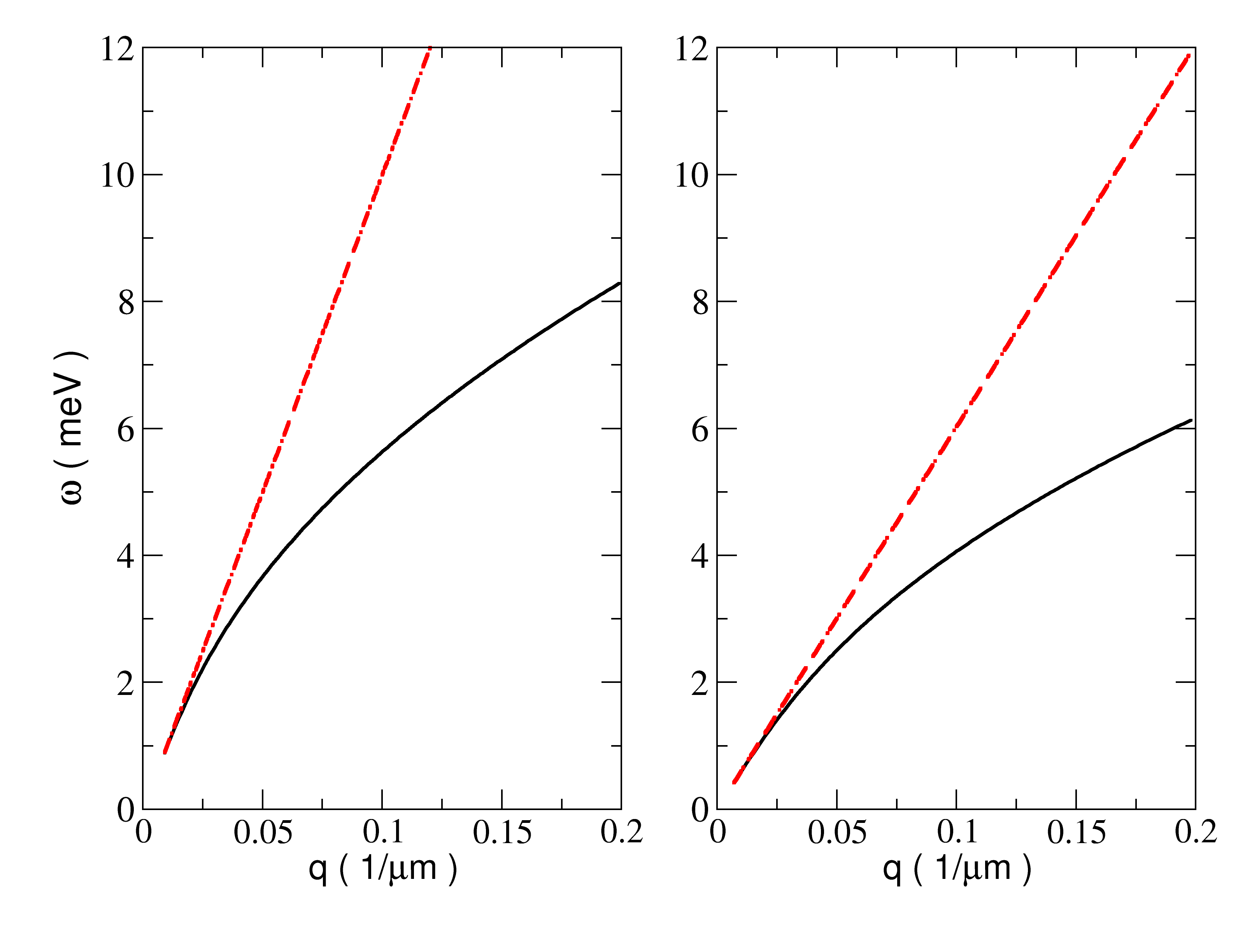}
\par\end{centering}

\caption{SPP dispersion curves calculated taking into account only the imaginary part
of the conductivity (thus $q$ is real). We have chosen $\hbar\gamma=\Gamma=0$
and ${E_{F}}=0.45$ eV. Left: $\epsilon_{1}=3$ and $\epsilon_{2}=4$;
right: $\epsilon_{1}=1$ and $\epsilon_{2}=11$ (silicon). The straight
(dashed) line stands for the light dispersion $\hbar\omega=\hbar cq/\sqrt{\epsilon_{2}}$
in the medium 2 (that with higher dielectric constant).}
\label{fig_spp_spectrum}
\end{figure}

In the general case of different dielectrics, neglecting absorption,
Eq.~(\ref{eq_W_SPP_2D}) can be written as
\begin{equation}
\frac{\epsilon_{1}}{\sqrt{(\hbar cq)^{2}-\epsilon_{1}(\hbar\omega)^{2}}}+\frac{\epsilon_{2}}{\sqrt{(\hbar cq)^{2}-\epsilon_{2}(\hbar\omega)^{2}}}=\frac{4\alpha E_{F}}{(\hbar \omega )^{2}}\,,
\label{eq_W_SPP_2D_numerical}
\end{equation}
which is convenient for numerical purposes. {[}In the numerical calculations
we take $\hbar c=0.2$ eV$\cdot\mu$m; it is also useful to recall
that $\omega/(2\pi)=$1 THz corresponds to an energy of 4.1 meV and
to a wavelength of 300 $\mu$m.{]}

In the non-retarded approximation it is possible to derive from Eq.~(\ref{eq_W_SPP_2D_numerical})
analytical results which coincide with those of the previous section
upon the replacement $\epsilon\rightarrow(\epsilon_{1}+\epsilon_{2})/2\equiv\bar{\epsilon}$,
that is,
\begin{equation}
\hbar\Omega_{p}\approx\sqrt{2\alpha E_{F}\hbar cq/{\bar{\epsilon}}}\,.\label{eq_PP_spectrum_Otto}
\end{equation}
Note that we have used the symbol $\Omega_{p}\equiv\Omega_{p}(q)$
to denote the explicit solution of the dispersion relations; we shall
keep this notation throughout the paper. In general, Eq.~(\ref{eq_W_SPP_2D_numerical})
has no analytical solution. In Fig.~\ref{fig_spp_spectrum} we give
the numerical solution to Eq.~(\ref{eq_W_SPP_2D_numerical}) in two
different cases.

\begin{figure}[!ht]
\begin{centering}
\includegraphics[clip,width=7cm]{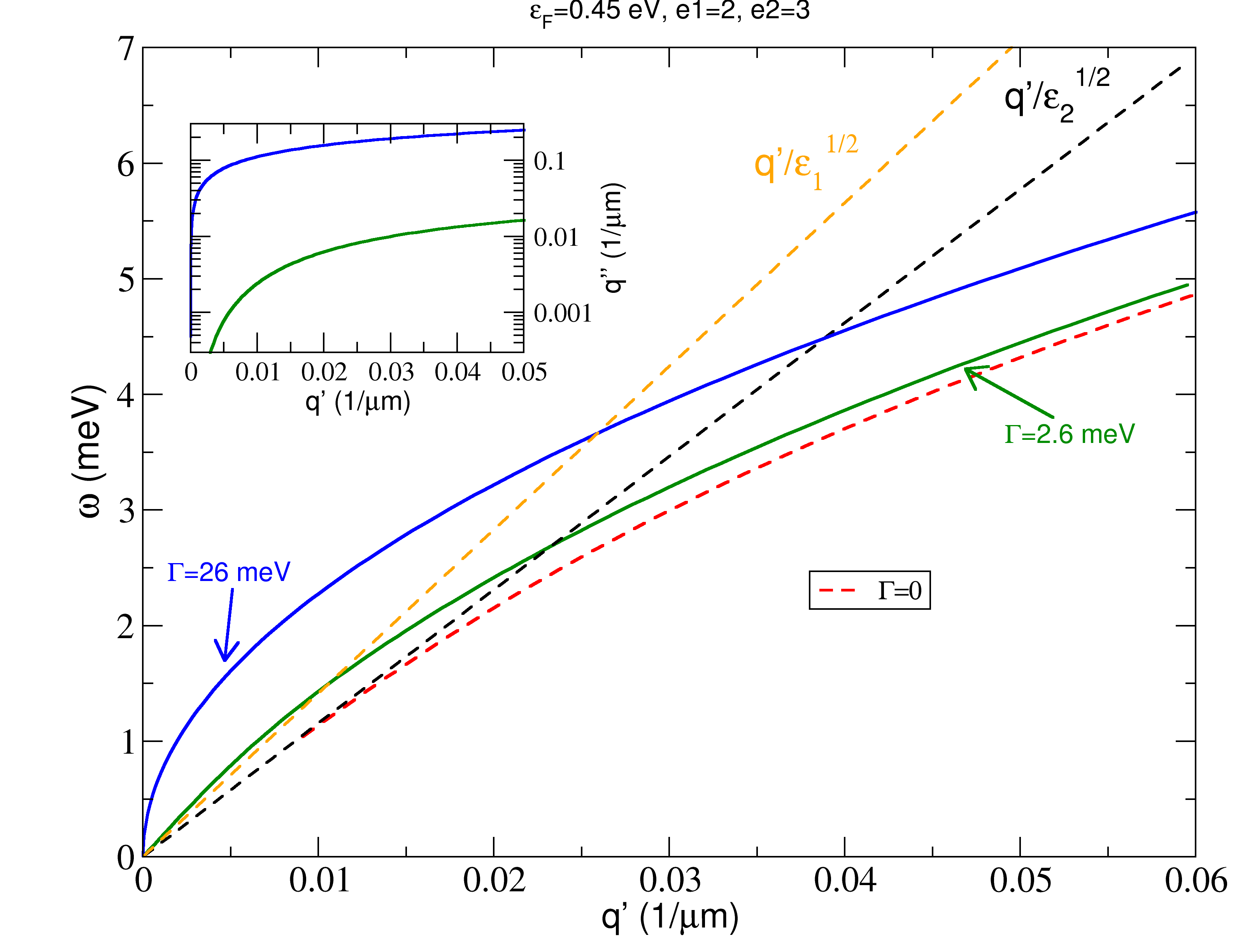}
\par\end{centering}

\caption{Plasmon-polariton dispersion curves for $\epsilon_{1}=3$,
$\epsilon_{2}=2$,
considering the effect of broadening (for different values of $\Gamma$).
Full expression for the conductivity with $E_{F}=0.45\,$eV was included
in the calculation. The red dashed line is for zero broadening. The inset
shows the dependence of the imaginary part of $q$ ($q''$) on the
real part of $q$ ($q'$) (notice the logarithmic scale of the vertical axis).
}
\label{fig_broadening}
\end{figure}

\begin{figure}[!ht]
\begin{centering}
\includegraphics[clip,width=7cm]{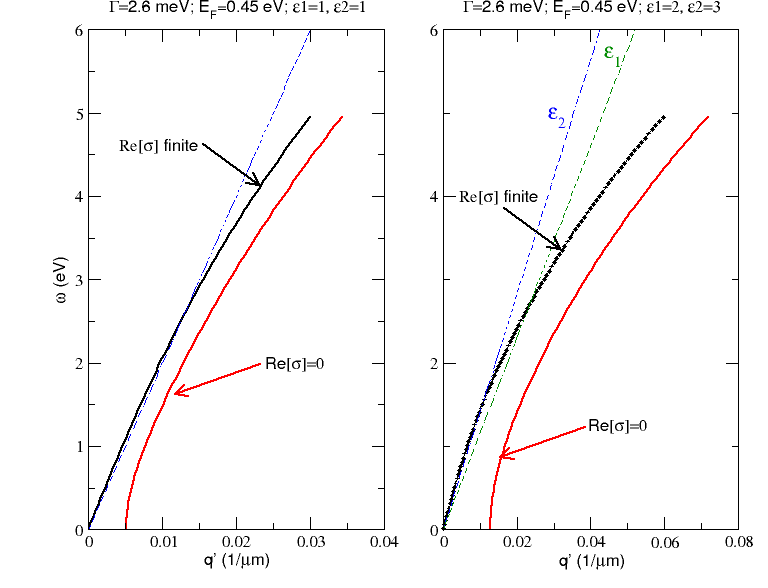}
\par\end{centering}

\caption{SPP dispersion curves calculated considering the real part of the conductivity
either finite or zero (but keeping $\gamma$ finite in the imaginary
part) for cases of equal {(left: $\epsilon_{1}=1$, $\epsilon_{2}=1$)}
and different {(right: $\epsilon_{1}=3$, $\epsilon_{2}=2$)} dielectrics
surrounding  graphene {with chemical potential $E_{F}=0.45\,$eV and
damping $\hbar\gamma=2.6\,$meV.} The straight lines marked $\epsilon_{1}$
and $\epsilon_{2}$ refer to the light dispersion $\hbar\omega=\hbar cq/\sqrt{\epsilon_{1}}$
and $\hbar\omega=\hbar cq/\sqrt{\epsilon_{2}}$ in the dielectrics 1 and 2,
respectively.}

\label{fig_real}
\end{figure}

In Figs.~\ref{fig_broadening} and \ref{fig_real} we present the
low wavenumber part of the SPP spectrum, comparing the numerical
solutions of Eq. (\ref{eq_W_SPP_2D}) obtained with different levels of approximation
to the optical conductivity and considering the role of the damping (or homogeneous broadening) parameter,
$\Gamma=\hbar\gamma$.
In Fig.~\ref{fig_broadening} the effect of the damping is demonstrated,
taking into account the full conductivity $\sigma_{D}$. In this
case the wavenumber $q$ is complex, $q=q'+iq''$, where
$q''$ describes the decay of the SPP as it propagates in space along
the graphene sheet. From this figure we see that the effect of the
increase of the damping is two-fold: it shifts the SPP dispersion
relation toward higher energies, for the same $q'$, and enhances the value of $q''$ (as
expected). If one wants to have long propagation lengths for the SPPs,
then $\Gamma$ must be as small as possible. In the same figure we
also represent the light-lines $\hbar cq/\sqrt{\epsilon_{1}}$ and
$\hbar cq/\sqrt{\epsilon_{2}}$, which correspond to the dispersion relations
for photons propagating in the media 1 and 2, respectively.
It may seem that one should be able
to excite SPPs in graphene by directly
 shining ER on it if $\Gamma $ is sufficiently
large because the dispersion curves in Fig.~\ref{fig_broadening}
intersect the light lines, as most clearly seen for the curve
 corresponding to $\Gamma =26$~meV. This is not true. For such a
 high value of $\Gamma $, plasmon-polaritons are overdamped
(note that $q'<q''$), the "dispersion curve" $\omega $ {\it versus}
$q'=\Re(q)$ is not quite meaningful, and the intersections do not
have their usual physical meaning.

In Fig.~\ref{fig_real} we present the effect of neglecting the real
part of the optical conductivity,
\begin{equation}
{\sigma_{D}\approx i\sigma_{0}\frac{4E_{F}}{\pi\hbar}\frac{\omega}{\gamma^{2}+\omega^{2}}\,.}\label{eq_sigma_xx_rp_neglect}
\end{equation}
The central feature is the vanishing of the dispersion curve for a
finite value of $q'$, which can be expressed as
\begin{equation}
{q'=\frac{\epsilon_{1}+\epsilon_{2}}{4\alpha E_{F}}\frac{\hbar\gamma^{2}}{c}.}
\end{equation}
This is a spurious result. When we use the full Drude conductivity expression,
the dispersion relation vanishes at zero $q'$. Changing the dielectric constants
of the surrounding media changes the value of $q'$ for the same frequency,
as can be seen comparing the left and right panels of Fig. \ref{fig_real}.


\subsection{Structure with two graphene layers }

\label{subsec_transfer_double_layer}
\begin{figure}[!htb]
\begin{centering}
\includegraphics[clip,width=9cm]{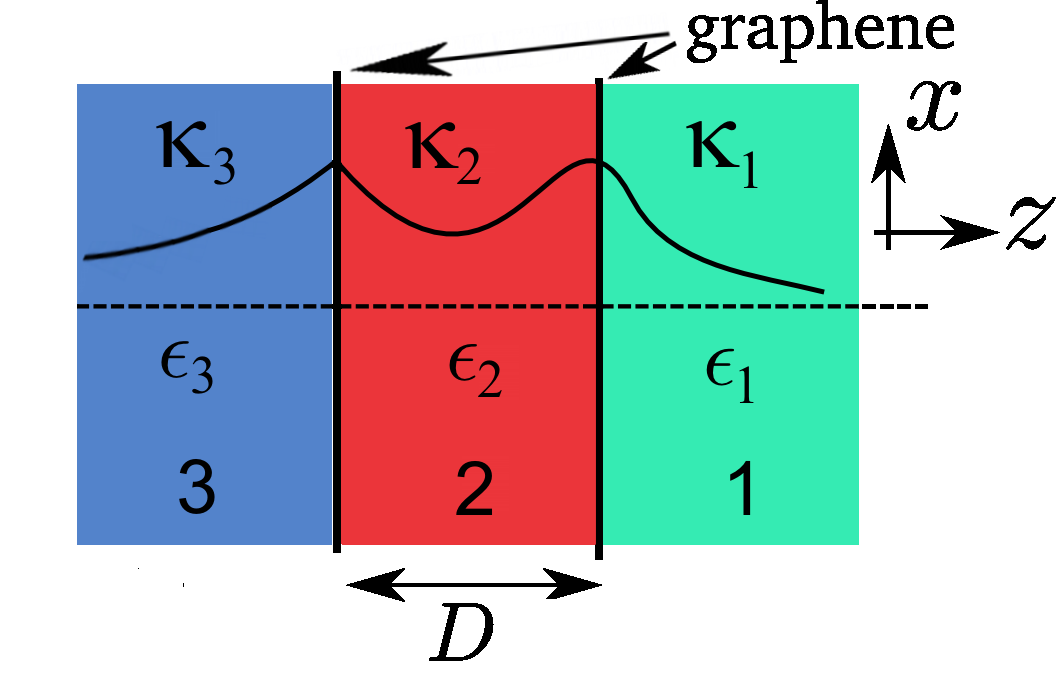} 

\par\end{centering}

\caption{Structure with two graphene layers separated by a dielectric with $\epsilon _2$.  Field profile for the "acoustic" SPP mode is shown qualitatively.}

\label{fig_double_layer_spec}
\end{figure}

In the previous sections we have
studied the SPP spectrum of single layer graphene.
We have shown that the typical value of SPP's energy for wavevector
$q\sim$0.1~$\mu$m$^{-1}$ is $\hbar\omega\approx$4~meV, corresponding
to a frequency of about 1~THz. We would like to investigate the possibility
of shifting the resonance frequency towards higher values. One way
of achieving this is using a double-layer structure as shown in Fig.~\ref{fig_double_layer_spec}.
The two degenerate SPP branches, associated with each
of the layers, hybridize giving rise to a two branch spectrum, with
one branch having higher energy than the bare spectrum for each layer.

We first compute the spectrum of the double layer system. We assume
a configuration as represented in Fig.~\ref{fig_double_layer_spec}
with semi-infinite media 1 and 3 and a dielectric layer 2 of thickness
$D$. Applying the boundary conditions at two interfaces
and requiring exponential decay in the infinity yields:
\begin{equation}
\left[\begin{array}{cccc}
-1 & 1 & 1 & 0\\
\frac{\epsilon_{3}}{\kappa_{3}}+i\frac{\sigma}{\omega\epsilon_{0}} & -\frac{\epsilon_{2}}{\kappa_{2}} & \frac{\epsilon_{2}}{\kappa_{2}} & 0\\
0 & \frac{\epsilon_{2}e^{\kappa_{2}D}}{\kappa_{2}} & -\frac{\epsilon_{2}e^{-\kappa_{2}D}}{\kappa_{2}} & \left(\frac{\epsilon_{1}}{\kappa_{1}}+i\frac{\sigma}{\omega\epsilon_{0}}\right)e^{-\kappa_{1}D}\\
0 & e^{\kappa_{2}D} & e^{-\kappa_{2}D} & -e^{-\kappa_{1}D}
\end{array}\right]\left[\begin{array}{c}
E_{3,x}\\
E_{2+,x}\\
E_{2-,x}\\
E_{1,x}
\end{array}\right]=0\,.
\end{equation}
The spectrum is given by the vanishing of the determinant of the above
matrix, that is,
\begin{eqnarray}
e^{\kappa_{2}D}\left(\frac{\epsilon_{1}}{\kappa_{1}}+i\frac{\sigma}{\omega\epsilon_{0}}+\frac{\epsilon_{2}}{\kappa_{2}}\right)\left(\frac{\epsilon_{3}}{\kappa_{3}}+i\frac{\sigma}{\omega\epsilon_{0}}+\frac{\epsilon_{2}}{\kappa_{2}}\right)=\nonumber \\
e^{-\kappa_{2}D}\left(\frac{\epsilon_{1}}{\kappa_{1}}+i\frac{\sigma}{\omega\epsilon_{0}}-\frac{\epsilon_{2}}{\kappa_{2}}\right)\left(\frac{\epsilon_{3}}{\kappa_{3}}+i\frac{\sigma}{\omega\epsilon_{0}}-\frac{\epsilon_{2}}{\kappa_{2}}\right)\,.\label{eq_eigen_DLayer}
\end{eqnarray}

\begin{figure}[!htb]
\begin{centering}
\includegraphics[clip,width=8cm]{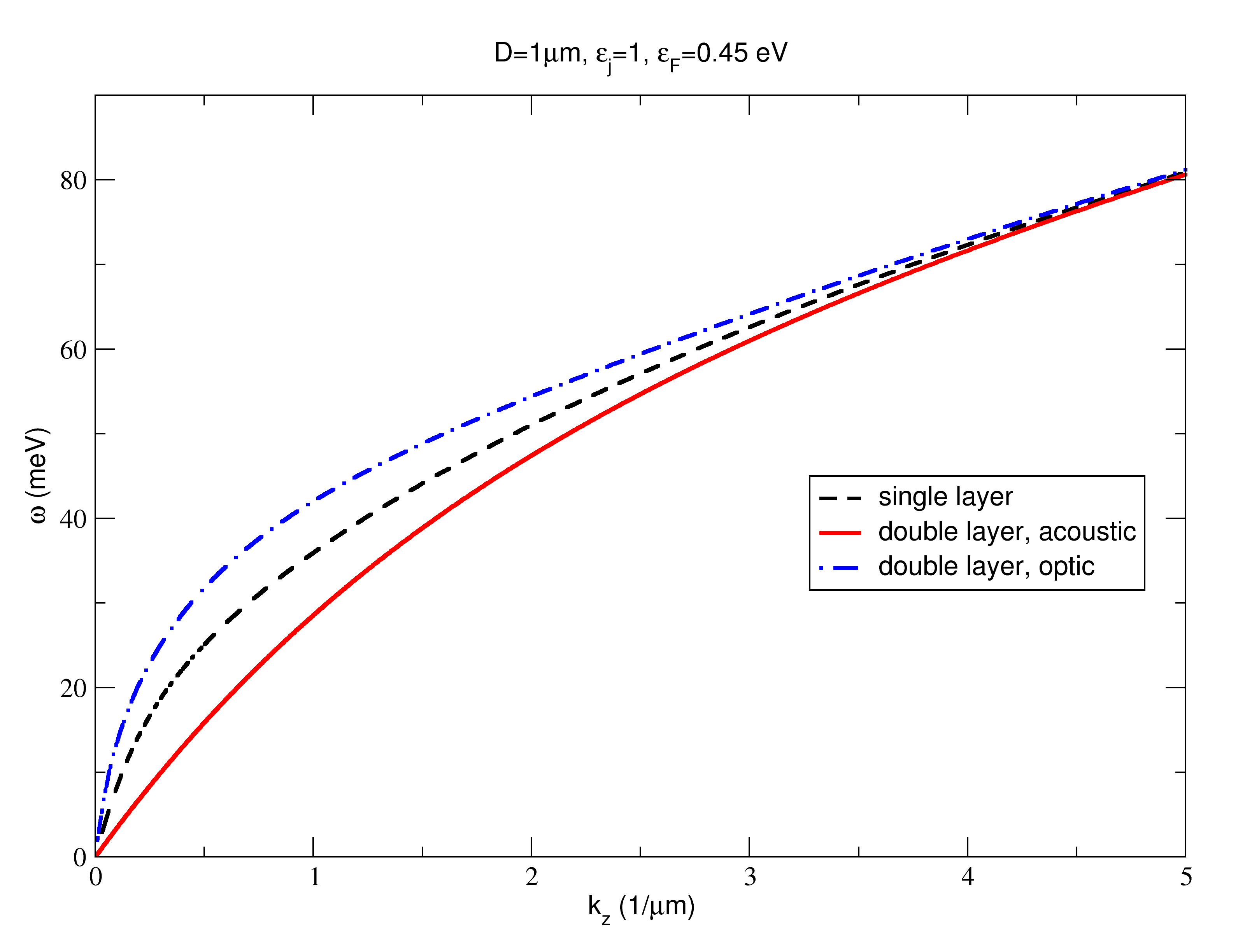} 

\par\end{centering}

\caption{Spectrum of the structure of Fig.~\ref{fig_double_layer_spec}, with its lower (acoustic) and upper
(optical) branches. The dashed curve is the spectrum of a single
layer. The parameters are $\epsilon_{m}=1$ (with $m=1,2,3$), $D=1$~$\mu$m,
and $E_{F}=0.45$~eV. }

\label{fig_Dlayer_spectrum}
\end{figure}

As expected, in the limit $q_{2}d\rightarrow\infty$, the two interfaces decouple
and Eq.~(\ref{eq_eigen_DLayer}) reads as
\begin{equation}
\left(\frac{\epsilon_{1}}{\kappa_{1}}+i\frac{\sigma}{\omega\epsilon_{0}}+\frac{\epsilon_{2}}{\kappa_{2}}\right)\left(\frac{\epsilon_{3}}{\kappa_{3}}+i\frac{\sigma}{\omega\epsilon_{0}}+\frac{\epsilon_{2}}{\kappa_{2}}\right)=0\,.
\end{equation}
Each term in brackets represents the plasmons-polaritons spectrum of one graphene layer {[}c.f.~Eq.~(\ref{eq_W_SPP_2D}){]}, degenerate
if $\epsilon_{1}=\epsilon_{3}$.

In the "symmetric" case of $\epsilon_{1}=\epsilon_{3}$, the dispersion relation (\ref{eq_eigen_DLayer}) can be factorized,
\begin{eqnarray}
\left[\left(\frac{\epsilon_{1}}{\kappa_{1}}+i\frac{\sigma}{\omega\epsilon_{0}}\right)\cosh(\kappa_{2}D/2)+\frac{\epsilon_{2}}{\kappa_{2}}\sinh(\kappa_{2}D/2)\right]\times\nonumber \\
\left[\left(\frac{\epsilon_{1}}{\kappa_{1}}+i\frac{\sigma}{\omega\epsilon_{0}}\right)\sinh(\kappa_{2}D/2)+\frac{\epsilon_{2}}{\kappa_{2}}\cosh(\kappa_{2}D/2)\right]=0.\label{eq:sym-asym}
\end{eqnarray}
The first product term in Eq.~(\ref{eq:sym-asym}) corresponds to the
symmetric mode with the electric field $z$ dependence even with respect to $z=D/2$,
while the second term describes the antisymmetric mode, odd with respect to the central plane.

Another analytic limit is that of $\epsilon_{1}=\epsilon_{2}=\epsilon_{3}=\epsilon$
and $q\gg\omega\epsilon^{1/2}/c$ (non-retarded or plasmon approximation).
In this case, Eq.~(\ref{eq:sym-asym}) reduces to
\begin{equation}
2\epsilon e^{qD}+i\frac{q\sigma}{\omega\epsilon_0}(e^{qD}\pm1)=0\,,
\end{equation}
where upper and lower signs correspond to the first (symmetric) and
second (antisymmetric) terms in (\ref{eq:sym-asym}), respectively.
If we assume that the conductivity of graphene is given by Eq.~(\ref{eq_sigma_approx}),
it is possible to obtain a close form for the spectrum of the surface
waves,
\begin{equation}
\hbar^{2}\Omega_{p}^{2}\approx\frac{2\alpha E_{F}}{\epsilon}\hbar qc\left(1\pm e^{-qD}\right)\,.\label{eq_spectrum_DL}
\end{equation}
Note that for $qD\gtrsim0$ the spectrum given by Eq.~(\ref{eq_spectrum_DL})
becomes inaccurate since the condition $qD\gg\omega q\epsilon^{1/2}/c$
is violated. In this regime the spectrum can only be obtained numerically.
In Fig.~\ref{fig_Dlayer_spectrum} we give the numerical solution
of Eq.~(\ref{eq_eigen_DLayer}) and obtain the two branches of the
spectrum; clearly they both vanish at zero $q$. The optical (symmetric)
branch lies above the dispersion curve of a single graphene layer
and the acoustic branch is below it. This dependence can be obtained
analytically as well by inspection of Eqs.~(\ref{eq_spectrum_DL})
and (\ref{eq_W_Plasmon_graphene}).

The propagation of TM-waves in multilayer graphene-based structures,
under strong light illumination, was considered by Dubinov \textit{et
al.} \cite{Dubinov}.


\section{Spectrum of TE SPPs in graphene}

\label{sec_TE_spectrum}

A broadband polarizer based of on the propagation
of TE-surface waves on graphene has been demonstrated \cite{KianPing}.
It is, therefore, interesting to compute the properties of a TE surface-wave.
Let us find the spectrum of an $s-$polarized (TE) surface wave in
graphene. These type of waves do not exist in the traditional 2D electron
gas because the imaginary part of the conductivity is always positive.
We assume a solution of Maxwell's equations in the form
\begin{eqnarray}
\vec{B} & = & (B_{m,x},0,B_{m,z})e^{iqx}e^{-\kappa_{m}\vert z\vert}\,,\\
\vec{E} & = & (0,E_{m,y},0)e^{iqx}e^{-\kappa_{m}\vert z\vert}\,,
\end{eqnarray}
where $m=1,2$ refers to the media 1 and 2 (see Fig.~\ref{fig_bc}).
In the case under consideration the boundary conditions (\ref{eq:bcond-et})
and (\ref{eq:bcond-ht}) are represented as
\begin{eqnarray}
E_{1,y}=E_{2,y}\,,\\
B_{1,x}-B_{2,x}=\mu_{0}\sigma_{yy}E_{1,y}\,.
\label {bc-TE}
\end{eqnarray}
For this particular case, Maxwell's equations read
\begin{eqnarray}
\nonumber
(-1)^{m+1}\kappa_{m}E_{m,y}=-i\omega B_{m,x}\,,\\
\nonumber
iqE_{m,y}=i\omega B_{m,z}\,,\\
(-1)^{m+1}\kappa_{m}B_{m,x}-iqB_{m,z}=-ic^{-2}\epsilon_{m}\omega E_{m,y}\,,
\label {ME-TE}
\end{eqnarray}
and we obrain the relations:
\begin{eqnarray}
B_{m,x} & = & (-1)^{m+1}i\frac{\kappa_{m}}{\omega}E_{m,y}\,,\\
B_{m,z} & = & \frac{q}{\omega}E_{m,y}\,,\\
\kappa_{m}^{2} & = & q^{2}-\frac{\omega^{2}}{c^{2}}\epsilon_{m}\,;\ \ \ \ \ m=1,2\:.
\end{eqnarray}
From the first boundary condition it follows $E_{1,y}=E_{2,y}$ and the second boundary condition (\ref {bc-TE}) yields the spectrum of the TE-waves
\cite{Falko},
\begin{equation}
\kappa_{1}+\kappa_{2}-i\omega\mu_{0}\sigma_{yy}=0\,.
\label{eq_W_SPP_2D_TE}
\end{equation}
We note that Eq.~(\ref{eq_W_SPP_2D_TE}) has real solutions if and only if
 the imaginary part of the conductivity is negative (and the real
part of the conductivity is zero). For $\sigma^{\prime}>0$ the wave
is damped by the imaginary part of either $q$ or $\omega$, depending
on the excitation conditions. Since the imaginary part of the Drude conductivity
is positive, this contribution alone  cannot give rise to TE-waves.
On the other hand, the interband contribution to $\sigma^{\prime\prime}$
is negative. Therefore, when the two contributions, for a given frequency,
add to a negative number, TE SPPs can propagate in graphene.

Eq.~(\ref{eq_W_SPP_2D_TE}) cannot be solved analytically. However,
the third term is small because it is proportional to the fine structure
constant. Then, in the simple case where $\epsilon_{1}=\epsilon_{2}=\epsilon$,
the spectrum of the TE wave is essentially equal to $\hbar\omega\lesssim\hbar cq/\sqrt{\epsilon}$,
that is, the dispersion relation of a free wave in a dielectric. Only
close to the threshold for interband transitions, $\hbar\omega=2E_{F}$,
does the spectrum deviate considerably from this result.

\section{Excitation of SPPs by evanescent waves}
\label{sec_excitation_by_evanescent}

\subsection{ATR configuration}
\label{ATR_configuration}

\begin{figure}[!ht]
\begin{centering}
\includegraphics[clip,width=7cm]{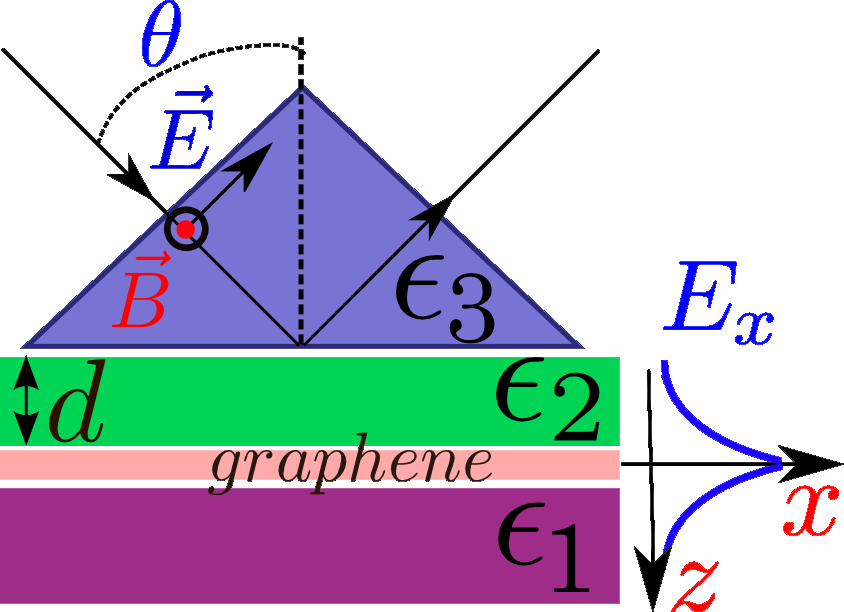} 

\par\end{centering}

\caption{Schematic representation of the experimental ATR setup needed to excite
surface plasmon-polaritons in graphene (Otto configuration). The
graphene layer is located between two dielectric media of relative
permittivity $\epsilon_{1}$ (considered semi-infinite) and $\epsilon_{2}$
(of thickness $d$). On top of the latter there is a prism of relative permittivity $\epsilon_{3}>(\epsilon_{1},\epsilon_{2})$
(which is a necessary condition for total internal reflection). The
incident angle of the incoming $p-$polarized wave is $\theta$, and the electric
and magnetic fields are $\vec{E}=(E_{x},0,E_{z})$ and $\vec{B}=(0,B_{y},0)$,
respectively.
}

\label{fig_plasmon_polariton}
\end{figure}

As discussed above, excitation of SPPs
on a flat metal surface is not possible by direct illumination (excepting the case of a periodically modulated conductivity; see Sec. \ref{sec_rippled_polaritonic}).
It is because  the wave vector of the
SPP, at a given frequency, is much larger than that of the impinging radiation. One way out
is excitation by evanescent waves. We assume an attenuated total reflection (ATR) configuration of the
form depicted in Fig. \ref{fig_plasmon_polariton}.
Electromagnetic radiation coming from the prism impinges into the
interface 3-2 at an angle $\theta$ larger than the critical angle
for total internal reflection, {$\theta>\arcsin \left [\sqrt{\mbox{max}(\epsilon_{1},\epsilon_{2})/\epsilon_{3}}\right ]$}.
So, only evanescent waves can exist in the layer with $\epsilon _2$ (of thickness
$d$) and in the half-space $\epsilon _1$. In this configuration it becomes possible to couple the incident wave to the SPPs
in graphene because the dielectric constant of the prism is larger
than those of the dielectrics cladding graphene.


\subsection{Fields in the ATR regime}
\label{subsec_TR_fields}

\begin{figure}[!ht]
\begin{centering}
\includegraphics[clip,width=9cm]{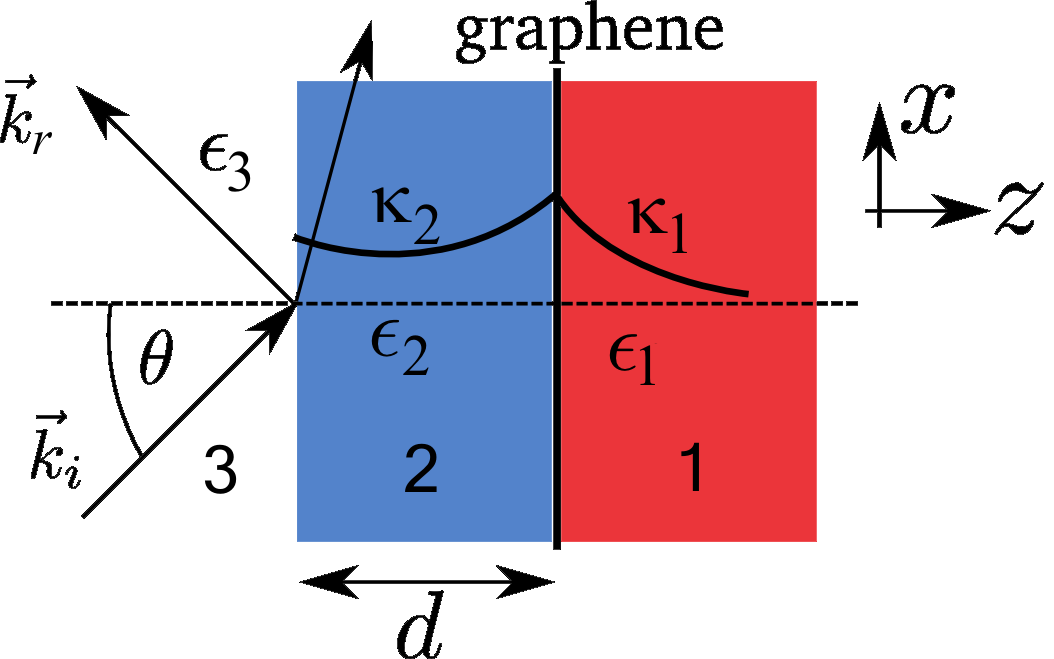} 
 \par\end{centering}
\caption{Three dielectrics with {$\epsilon_{3}>\epsilon_{1}$,$\epsilon_{2}$}.
Between the dielectrics {1 and 2} there is a graphene sheet.}
\label{fig_EPL}
\end{figure}

We start by deriving the fields in the media 1, 2 and 3 (refer to Fig.~\ref{fig_EPL}).
The incoming wave vector is {$\vec{k}_{i}=(k\sin\theta,0,k\cos\theta)$}
and the reflected one at the interface is {$\vec{k}_{r}=(k\sin\theta,0,-k\cos\theta)$}.
Due to translational invariance, the component {$q=k\sin\theta$}
is conserved at all interfaces.

We study the case of TM-waves, that is, the fields have the form
$\vec{B}_{m}=(0,B_{{m,y}},0)$ and $\vec{E}_{m}=(E_{{m,x}},0,E_{{m,z}})$ ($m=1-3$). If we further assume
that $\vec{B}_{3}=\vec{B}^{{(i,r)}}e^{i\vec{k}_{i,r}\cdot\vec{r}}$
and $\vec{E}_{3}=\vec{E}^{{(i,r)}}e^{i\vec{k}_{i,r}\cdot\vec{r}}$,
it follows from the Maxwell's equations that
\begin{eqnarray}
k & = & \omega\varepsilon_{3}^{1/2}/c\,,\\
B_{{y}}^{{(i,r)}} & = & {\pm}\frac{\epsilon_{3}\omega}{{c^{2}}k_{z}}E_{{x}}^{{(i,r)}}\,,\label{eq:byex3}\\
E_{z}^{{(i,r)}} & = & {\mp}\frac{q}{k_{z}}E_{x}^{{(i,r)}}\,.\label{eq:ezex3}
\end{eqnarray}
{Here $k_{z}=k\cos\theta$. In Eqs.~(\ref{eq:byex3})-(\ref{eq:ezex3})
the upper and lower signs correspond to the incident and reflected waves,
respectively.} On the other hand, if we assume that $\vec{B}_{m}=\vec{B}_{m}^{{(\pm)}}e^{iqx}e^{\pm\kappa_{m}z}$
and $\vec{E}_{m}=\vec{E}_{m}^{{(\pm)}}e^{iqx}e^{\pm\kappa_{m}z}$,
a situation corresponding to evanescent or exponentially growing waves
{[}simple generalization of Eqs.~(\ref{eq:specE-field})-(\ref{eq:specH-field}){]},
we find
\begin{eqnarray}
B_{m,y}^{{(\pm)}} & = & {\pm}i\frac{\omega\epsilon_{m}}{{c^{2}\kappa_{m}}}E_{m,x}^{{(\pm)}}\,,\label{eq:byex-m}\\
E_{m,z}^{{(\pm)}} & = & \mp i\frac{{q}}{{\kappa_{m}}}E_{m,x}^{{(\pm)}}\,,\label{eq:ezex-m}
\end{eqnarray}
and the dispersion relation is equivalent to (\ref{eq:disp}). Since $q$ is conserved, we have:
\begin{equation}
\kappa_{2}^{2}=\frac{\omega^{2}}{c^{2}}\epsilon_{3}\sin^{2}\theta-\frac{\omega^{2}}{c^{2}}\epsilon_{2}>0\Leftrightarrow\sin\theta>\left(\frac{\epsilon_{2}}{\epsilon_{3}}\right)^{1/2}\,.
\end{equation}
The last condition means that evanescent waves can occur in the
medium 2 if and only if $\epsilon_{3}>\epsilon_{2}$.
For the geometry of Fig. \ref{fig_EPL}, the boundary conditions at the interfaces $z=d$ and $z=0$ read as
\begin{eqnarray}
E_{m,x} & = & E_{m+1,x}\,,\\
B_{m,y} & = & B_{m+1,y}-\mu_{0}\sigma\delta_{m,1}E_{m,x}\,,
\end{eqnarray}
respectively, where $\delta_{m,m^{\prime}}$ is the Kronecker symbol.

\subsection{Total reflection}

\label{subsec_total_internal_reflection}

Let us assume that $d\rightarrow \infty$ and the graphene sheet and the medium 1 have no effect on the reflection at the interface 2-3. We shall show that, for $\theta_{c}=\arcsin{(\epsilon_{2}/\epsilon_{3})^{1/2}}$, the reflection
is total, even though there is an evanescent wave in the medium 2. For this
particular case the boundary conditions are written as $E_{x}^{(i)}+E_{x}^{(r)}=E_{2,x}^{(-)}$
and $B_{y}^{(i)}+B_{y}^{(r)}=B_{2,y}^{(-)}$, where the subscripts $i$
and $r$ stand for incident and reflected waves as before. Explicitly, the second boundary condition reads
\begin{eqnarray}
\frac{\epsilon_{3}}{k_{z}}E_{x}^{(i)}-\frac{\epsilon_{3}}{k_{z}}E_{x}^{(r)} & = & -i\frac{\epsilon_{2}}{\kappa_{2}}E_{2,x}^{(-)}\,,
\end{eqnarray}
with the following solution,
\begin{equation}
r\equiv \frac{E_{x}^{(r)}}{E_{x}^{(i)}}=\frac{\epsilon_{3}\kappa_{2}+i\epsilon_{2}k_{z}}{\epsilon_{3}\kappa_{2}-i\epsilon_{2}k_{z}}=e^{2i\alpha}\,,
\label{eq_r_total_reflect}
\end{equation}
where we have defined $\alpha=\arctan(\epsilon_{2}k_{z}/\epsilon_{3}\kappa_{2})$.
Note that for grazing incidence $\alpha=0$, while for $\theta=\theta_{c}$
we have $\kappa_{2}=0$ and $\alpha=\pi/2$.  We thus conclude that
total reflection ($\vert r \vert ^2=1$) occurs for $\theta \geq \theta_{c}$. The amplitude
of the evanescent field reads
\begin{equation}
\frac{E_{2,x}^{(-)}}{E_{x}^{(i)}}=\frac{2\epsilon_{3}\kappa_{2}}{\epsilon_{3}\kappa_{2}-i\epsilon_{2}k_{z}}=\frac{2\epsilon_{3}\kappa_{2}}{\epsilon_{3}\kappa_{2}-i\epsilon_{2}k\cos\theta}\,,\label{eq_amp_evanescent}
\end{equation}
where $\kappa_{2}=k\sqrt{\sin^{2}\theta-\epsilon_{2}/\epsilon_{3}}<k$.
We show in the next section that the presence of a graphene layer,
as in Fig.~\ref{fig_EPL}, can frustrate the total reflection in a dramatic
way.

\subsection{Otto configuration: $p-$polarized wave}
\label{subsec_Otto}

\subsubsection{Reflection coefficient }

We shall now compute the reflection coefficient in
the situation represented in Fig.~\ref{fig_EPL}, where we have three
dielectrics, a graphene layer at the 2-3 interface, and $\theta>\theta_{c}$.
We will see that in this case $\vert r\vert^{2}$ can be smaller than
unity (down to zero) due to the excitation of SPPs waves in the
graphene layer.

In the dielectric 3 we have both incoming and reflected waves, while in the dielectric 2 we can have both evanescent and exponentially
growing solutions because this region has a finite thickness $d$. In the dielectric 1 only evanescent waves can occur.

The fields in the three regions are:

Region {3}
\begin{eqnarray}
B_{{3},y} & = & B_{y}^{{(i)}}e^{i\vec{k}_{i}\cdot\vec{r}}+B_{y}^{{(r)}}e^{i\vec{k}_{r}\cdot\vec{r}}\,,\\
\vec{E}_{3} & = & \left(E_{x}^{{(i)}},0,E_{z}^{{(i)}}\right)e^{i\vec{k}_{i}\cdot\vec{r}}+\left(E_{x}^{{(r)}},0,E_{z}^{{(r)}}\right)e^{i\vec{k}_{r}\cdot\vec{r}}\,.
\end{eqnarray}

Region 2
\begin{eqnarray}
B_{2,y} & = & B_{{2,y}}^{{(+)}}e^{i\vec{k}_{2}^{{(+)}}\cdot\vec{r}}+B_{{2,y}}^{{(-)}}e^{i\vec{k}_{2}^{{(-)}}\cdot\vec{r}}\,,\\
\vec{E}_{2} & = & \left(E_{{2,x}}^{{(+)}},0,E_{{2,z}}^{{(+)}}\right)e^{i\vec{k}_{2}^{{(+)}}\cdot\vec{r}}+\left(E_{{2,x}}^{{(-)}},0,E_{{2,z}}^{{(-)}}\right)e^{i\vec{k}_{2}^{{(-)}}\cdot\vec{r}}\,,
\end{eqnarray}

Region {1}
\begin{eqnarray}
B_{1,y} & = & B_{{1,y}}^{{(-)}}e^{i\vec{k}_{1}^{{(-)}}\cdot\vec{r}}\,,\\
\vec{E}_{1} & = & \left(E_{{1,x}}^{{(-)}},0,E_{{1,z}}^{{(-)}}\right)e^{i\vec{k}_{1}^{{(-)}}\cdot\vec{r}}\,,
\end{eqnarray}
where $\vec{k}_{m,\pm}=(k\sin\theta,0,\mp i\kappa_{m})$, $m=1,2$.
The boundary conditions at the interfaces ${z}=0$ and ${z}=d$ are:
\begin{eqnarray}
E_{x}^{{(i)}}+E_{x}^{{(r)}} & = & E_{{2,x}}^{{(+)}}+E_{{2,x}}^{{(-)}}\,,\label{eq:0ef}\\
B_{y}^{{(i)}}+B_{y}^{{(r)}} & = & B_{{2,y}}^{{(+)}}+B_{{2,y}}^{{(-)}}\,.\label{eq:0mf}
\end{eqnarray}
\begin{eqnarray}
E_{{1,x}}^{{(-)}}e^{-\kappa_{1}d} & = & E_{{2,x}}^{{(+)}}e^{\kappa_{2}d}+E_{{2,x}}^{{(-)}}e^{-\kappa_{2}d}\,,\label{eq:def}\\
B_{{1,y}}^{{(-)}}e^{-\kappa_{1}d} & = & B_{{2,y}}^{{(+)}}e^{\kappa_{2}d}+B_{{2,y}}^{{(-)}}e^{-\kappa_{2}d}-\mu_{0}\sigma_{xx}E_{{1,x}}^{{(-)}}e^{-\kappa_{1}d}\,.\label{eq:dmf}
\end{eqnarray}
Explicitly, using relations (\ref{eq:byex3}) and (\ref{eq:byex-m}),
we can rewrite Eqs.~(\ref{eq:0mf}) and (\ref{eq:dmf}) in terms
of the $x$-component of the electric field,

$z=0$:
\begin{eqnarray}
-\frac{\epsilon_{3}}{k_{z}}E_{x}^{{(i)}}+\frac{\epsilon_{3}}{k_{z}}E_{x}^{{(r)}} & = & -i\frac{\epsilon_{2}}{\kappa_{2}}E_{{2,x}}^{{(+)}}+i\frac{\epsilon_{2}}{\kappa_{2}}E_{{2,x}}^{{(-)}}\,;
\label{eq:0mf-new}
\end{eqnarray}

\begin{figure}[!ht]
\begin{centering}
\includegraphics[clip,width=8cm]{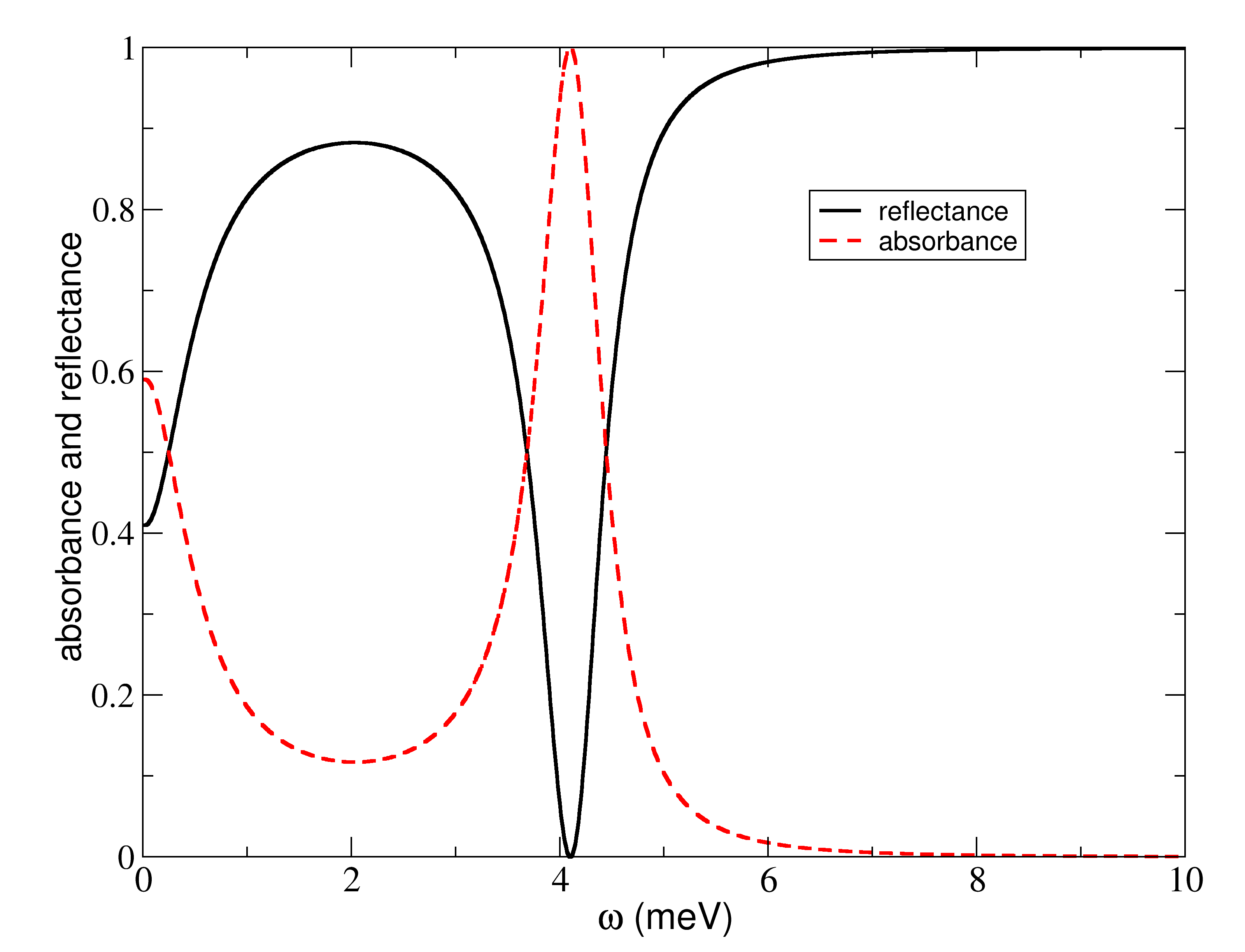} 

\par\end{centering}

\caption{Absorbance and reflectance spectra of graphene in the Otto configuration.
The parameters are: $E_{F}=0.45$~eV, $\Gamma=0.1$~meV, $\epsilon_{1}$=14,
$\epsilon_{2}=1$, $\epsilon_{3}=5$, $d=2.51$~$\mu$m, $\theta_{c}=36.7^{{\rm {o}}}$,
and $\theta_{i}=2.25\theta_{c}$. Solid black line: $\vert r\vert^{2}$;
Red dashed line: 1-$\vert r\vert^{2}$. In the calculation we have
used the Drude formula for the conductivity of graphene, Eq.~(\ref{eq_sigma_xx_Drude}).
}
\label{fig_reflectance}
\end{figure}

$z=d$:
\begin{eqnarray}
-\frac{\omega\epsilon_{2}}{\kappa_{2}}e^{\kappa_{2}d}E_{{2,x}}^{{(+)}}+\frac{\omega\epsilon_{2}}{\kappa_{2}}e^{-\kappa_{2}d}E_{{2,x}}^{{(-)}} & = & \left(\frac{\omega\epsilon_{1}}{\kappa_{1}}+i\frac{\sigma}{{\epsilon_{0}}}\right)e^{-\kappa_{1}d}E_{{1,x}}^{{(-)}}\,.
\label{eq:dmf-new}
\end{eqnarray}
After the elimination of $E_{{1,x}}^{{(-)}}$, the linear system of
equations defined by the boundary conditions can be written as
\begin{equation}
\Xi_{p}\left[\begin{array}{c}
E_{x}^{{(r)}}/E_{x}^{{(i)}}\\
E_{{2,x}}^{{(+)}}/E_{x}^{{(i)}}\\
E_{{2,x}}^{{(-)}}/E_{x}^{{(i)}}
\end{array}\right]=\Phi_{p},
\end{equation}
where
\begin{eqnarray}
\Xi_{p} & = & \left[\begin{array}{ccc}
-1 & 1 & 1\\
\kappa_{2}\epsilon_{3} & ik_{z}\epsilon_{2} & -ik_{z}\epsilon_{2}\\
0 & e^{\kappa_{2}d}(\kappa_{2}\Lambda_{1}^{(+)}+\chi_{1}) & e^{-\kappa_{2}d}(\kappa_{2}\Lambda_{1}^{(+)}-\chi_{1})
\end{array}\right],\label{eq:Xi-p}\\
\Phi_{p} & = & \left[\begin{array}{c}
1\\
\kappa_{2}\epsilon_{3}\\
0
\end{array}\right],\label{eq:Phi-p}\\
\chi_{m} & = & \kappa_{m}\omega\epsilon_{m+1}\,,\\
\Lambda_{m}^{(\pm)} & = & (\omega\epsilon_{m}\pm i\sigma\kappa_{m}/\epsilon_{0})\,.\label{eq:Lambda_pm}
\end{eqnarray}
The solution of the linear system yields:
\begin{eqnarray}
E_{x}^{{(r)}}/E_{x}^{{(i)}} & = & \frac{\kappa_{2}\epsilon_{3}\eta_{1}+ik_{z}\epsilon_{2}\eta_{2}}{\kappa_{2}\epsilon_{3}\eta_{1}-ik_{z}\epsilon_{2}\eta_{2}}\,,\label{eq_exact_r_Otto}\\
\frac{E_{{1,x}}^{{(-)}}e^{-\kappa_{1}d}}{E_{x}^{{(i)}}} & = & \frac{E_{2,x}^{{(+)}}}{E_{x}^{{(i)}}}e^{\kappa_{2}d}+\frac{E_{2,x}^{{(-)}}}{E_{x}^{{(i)}}}e^{-\kappa_{2}d}=\frac{2\omega\epsilon_{2}\kappa_{1}\kappa_{2}\epsilon_{3}}{\kappa_{2}\epsilon_{3}\eta_{1}-ik_{z}\epsilon_{2}\eta_{2}}\,,\label{eq:ez3}
\end{eqnarray}
where $\eta_{1}=\kappa_{2}\Lambda_{1}^{(+)}\sinh(\kappa_{2}d)+\cosh(\kappa_{2}d)\chi_{1}$
and $\eta_{2}=\kappa_{2}\Lambda_{1}^{(+)}\cosh(\kappa_{2}d)+\sinh(\kappa_{2}d)\chi_{1}$.
We note that, if the conductivity has an imaginary part only, then
the function $\Lambda_{1}^{(+)}$ is real and the reflection coefficient
is simply a phase. Therefore, the reflectance is equal to unity.
Thus, for detecting the excitation of a surface wave the conductivity
of  graphene must have a finite real part. In Fig.~\ref{fig_reflectance}
we represent the reflectance and the absorbance as functions of the
energy of the incoming ER. A sharp dip in the reflectance spectrum is
seen at an energy of about $\hbar\omega\approx4$ meV.

\begin{figure}[!htb]
\begin{centering}
\includegraphics[clip,width=8.5cm]{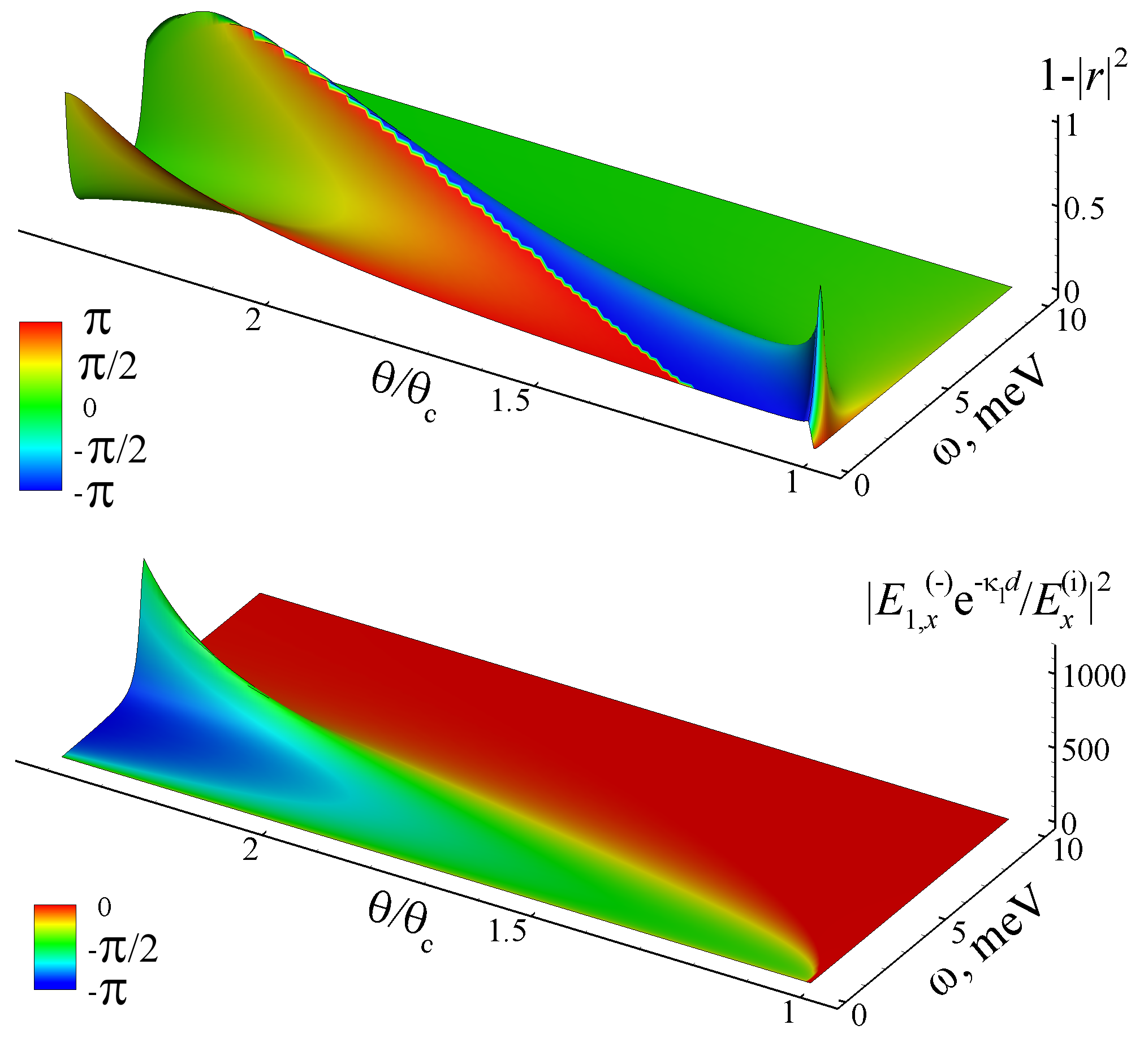} 

\par\end{centering}

\caption{Absorbance, 1-$\vert r\vert^{2}$, (top), and the electric field squared amplitude, $|E_{{1,x}}^{{(-)}}e^{-\kappa_{1}d}/E_{x}^{{(i)}}|^{2}$
at $z=d$ (bottom) {\it versus} photon energy and angle
of incidence. Phases of the reflection coefficient, $r$ (top) and the electric field (bottom)
are depicted by color. Other parameters as in Fig.~\ref{fig_reflectance}.}

\label{fig_absorvance_angle_omega}
\end{figure}

We also note a reduction of the reflectance close to zero energy.
We want to understand why the reflectance falls to zero at these
two energy values. First, we make a three-dimensional representation (Fig.~\ref{fig_absorvance_angle_omega}),
where the absorbance is represented as function of both the angle
of incidence and the incoming photon energy for a given width $d$.
Maximal absorbance (minimal reflectance) occurs close to zero and
4~meV, and is accompanied by a drastic change (from positive
to negative values) of the phase of the reflection coefficient as depicted in Fig.~\ref{fig_absorvance_angle_omega}.
The comparison of top and bottom panels in Fig.~\ref{fig_absorvance_angle_omega} shows that the resonance maximum of absorbance
 corresponds to maximal amplitude for excited SPPs and phase $\approx-\pi/2$.
The latter imposes the fulfillment of
the condition ${\rm Re}(E_{{1,x}}^{{(-)}}e^{-\kappa_{1}d})\approx0$ at resonance,
which, after taking into account Eq.~(\ref{eq:ez3}), leads to the
following constraint:
\begin{equation}
{\rm Re}(\eta_{1})=\left(\frac{\epsilon_{1}}{\kappa_{1}}-\frac{\sigma^{\prime\prime}}{\omega\epsilon_{0}}\right)\sinh(\kappa_{2}d)+\frac{\epsilon_{2}}{\kappa_{2}}\cosh(\kappa_{2}d)=0\:.
\label{eq:max-eq}
\end{equation}
Eq.~(\ref{eq:max-eq}) determines the frequency onset $\omega_{\mbox{min}}$,
and its main idea can be interpreted in the following manner. In the
limit $d\to\infty$ this equation can be obtained by the substitution
of the relation (known as ATR scanline),
\begin{equation}
k_{\mbox{\tiny SPP}}=q=k\sin\theta=\frac{\omega}{c}\sqrt{\epsilon_{3}}\sin\theta\,,\label{eq_ATR}
\end{equation}
into the SPP dispersion relation (\ref{eq_W_SPP_2D_numerical}). Rearranging
Eq.~(\ref{eq_ATR}), the energy of the photon reads
\begin{equation}
\hbar\omega_{\gamma}=\frac{q\hbar c}{\sqrt{\epsilon_{3}}\sin\theta}\,.\label{eq_gamma_spectrum_Otto}
\end{equation}
It becomes possible to excite SPPs when the in-plane momentum of
the photon (\ref{eq_ATR}) coincides with the momentum of the SPPs
at the same frequency. Under this condition, the value of the reflectance
at minimum can be represented as
\begin{eqnarray}
R_{\textrm{min}}=\frac{\omega\epsilon_{3}\kappa_{1}^{2}\kappa_{2}^{2}\frac{\sigma^{\prime}}{\epsilon_{0}}-k_{z}\left[(\omega\epsilon_{1}-\frac{\sigma^{\prime\prime}}{\epsilon_{0}}\kappa_{1})^{2}\kappa_{2}^{2}-(\omega\epsilon_{2}\kappa_{1})^{2}\right]}{\omega\epsilon_{3}\kappa_{1}^{2}\kappa_{2}^{2}\frac{\sigma^{\prime}}{\epsilon_{0}}+k_{z}\left[(\omega\epsilon_{1}-\frac{\sigma^{\prime\prime}}{\epsilon_{0}}\kappa_{1})^{2}\kappa_{2}^{2}-(\omega\epsilon_{2}\kappa_{1})^{2}\right]}.\label{eq:rmin}
\end{eqnarray}
Notice that the expression (\ref{eq:rmin}) must be evaluated at $\omega=\omega_{\textrm{min}}$.

\begin{figure}[!htb]
\begin{centering}
\includegraphics[clip,width=7cm]{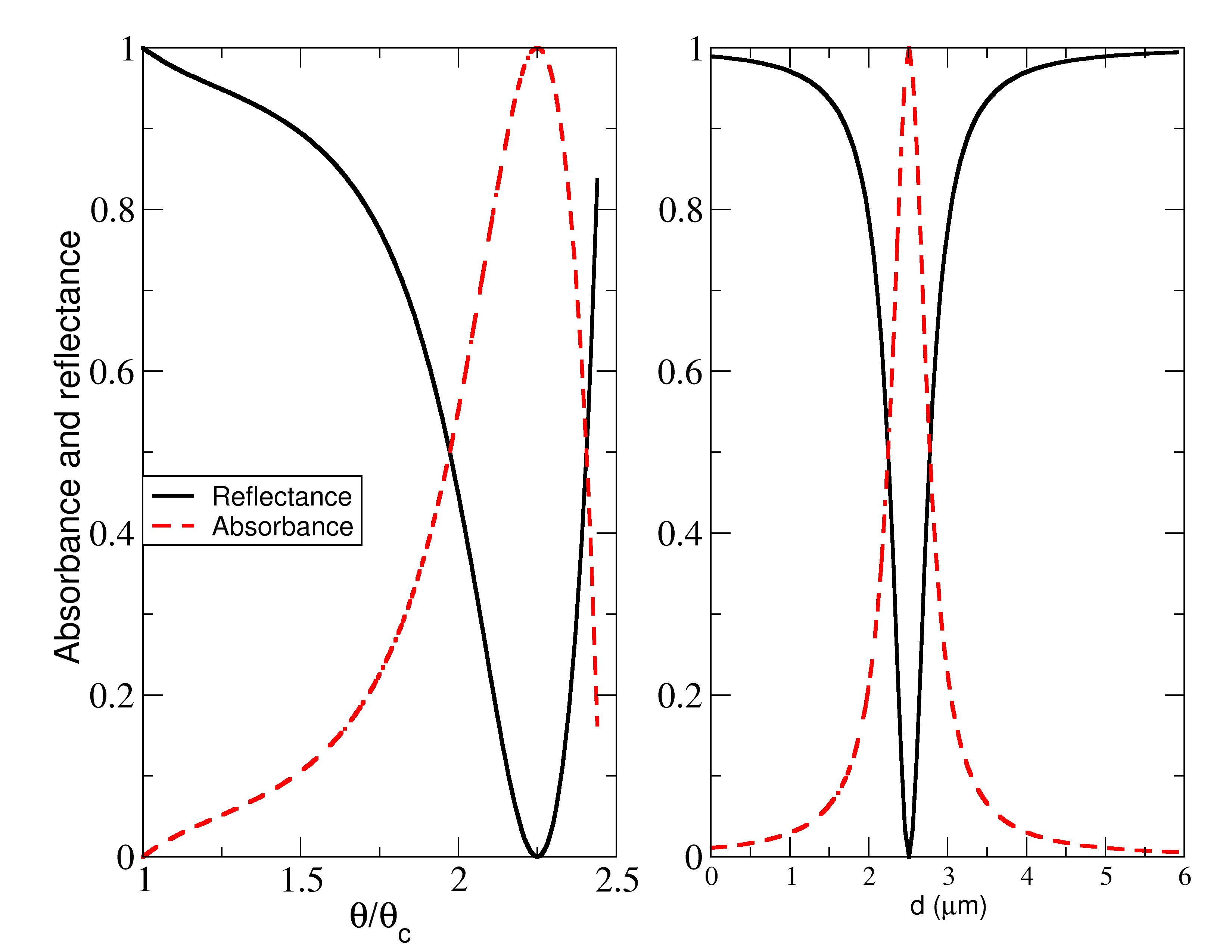}

\par\end{centering}

\caption{Absorbance and reflectance spectra. Left: reflectance $\vert r\vert^{2}$
(solid line), and absorbance $1-\vert r\vert^{2}$ (dashed line),
as function of the angle of incidence $\theta/\theta_{c}$. Right:
reflectance $\vert r\vert^{2}$ (solid line), and absorbance $1-\vert r\vert^{2}$
(dashed line), as function of the spacer thickness, $d$. In both panels
the photon energy is $\hbar\omega=4.1$~meV and other parameters are
as in Fig.~\ref{fig_reflectance}.}

\label{fig_absorvance_d_and_theta}
\end{figure}

In Fig.~\ref{fig_absorvance_d_and_theta} the reflectance
and the absorbance are plotted as functions of the incident angle $\theta$ and
the spacer thickness $d$.
We see that there are some optimal values of $\theta$ and $d$ for which
the reflectance is almost zero. They correspond to an efficient
excitation of SPPs in the graphene sheet.

\begin{figure}[!ht]
\begin{centering}
\includegraphics[clip,width=8.5cm]{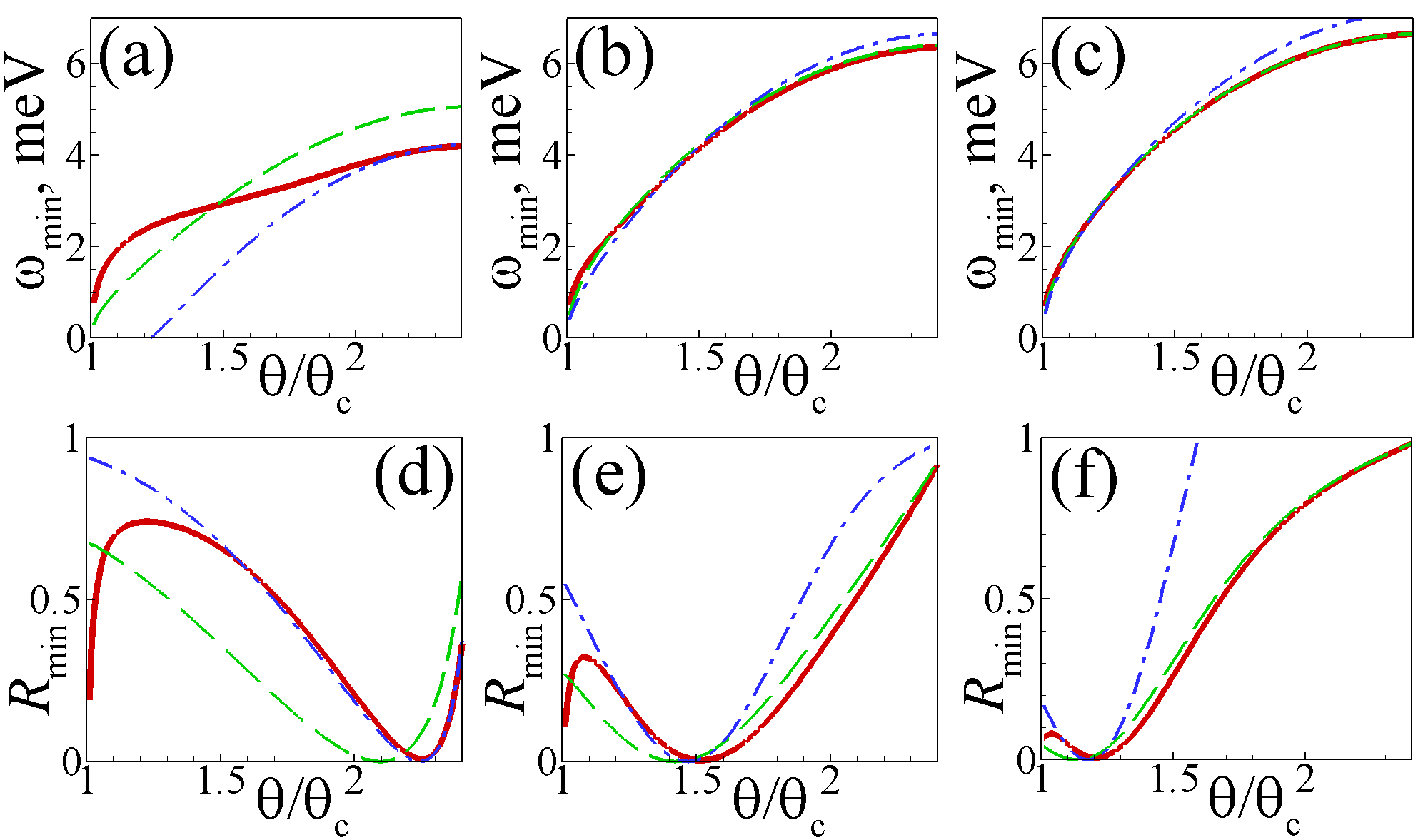} 

\par\end{centering}

\caption{Dependence of the resonant frequency $\omega_{\textrm{min}}$ (a--c),
and reflectance at resonance $R_{\textrm{min}}$ (d--f) upon angle
of incidence $\theta$, obtained from numerical calculation {[}red
solid line{]}, approximations (\ref{eq:max-sm}) {[}blue dashed-dotted
line{]} and (\ref{eq:max-big}) {[}green dashed line{]}. The parameters are: $d=2.51$~$\mu$m (a,d), $d=7.53$~$\mu$m (b,e), and $d=12.55$~$\mu$m
(c,f). Other parameters as in Fig.~\ref{fig_reflectance}.}

\label{fig:atr_appr}
\end{figure}


\subsubsection{The limit $\kappa_{2}d\ll1$ }

Putting $\kappa_{2}d=0$ in Eq.~(\ref{eq_exact_r_Otto}), we find:
\begin{eqnarray}
\frac{E_{x}^{{(r)}}}{E_{x}^{{(i)}}} & \approx & \frac{\kappa_{1}\omega\epsilon_{3}+ik_{z}\Lambda_{1}^{(+)}}{\kappa_{1}\omega\epsilon_{3}-ik_{z}\Lambda_{1}^{(+)}}\nonumber \\
 & = & \frac{\xi -k_{z}\sigma'\kappa_{1}/\epsilon_{0}}{\xi ^{\ast}+k_{z}\sigma'\kappa_{1}/\epsilon_{0}}\,,
 \label{eq_r_for_ll1}
\end{eqnarray}
where $\xi $ is defined as
\begin{equation}
\xi =\omega\epsilon_{3}+ik_{z}\omega\left(\frac{\epsilon_{1}}{\kappa_{1}}-\frac{\sigma''}{\epsilon_{0}\omega}\right)\,.
\end{equation}
The above equation is the total internal reflection
(\ref{eq_r_total_reflect}), modified by the presence of graphene.
The smallest value of $E_{x}^{{(r)}}/E_{x}^{{(i)}}$ is obtained for $\mbox {Im} \xi =0$, that is, for
\begin{equation}
\frac{\epsilon_{1}}{\kappa_{1}}-\frac{\sigma''}{\epsilon_{0}\omega}=0\,,\label{eq_pp_2q3}
\end{equation}
which coincides with the dispersion relation of SPP waves with the wavenumber
$\kappa_{1}$ in a medium of dielectric permittivity $\epsilon_{1}/2$.
We recall, however, that the fulfilment of the condition (\ref{eq_pp_2q3}) by itself
does not imply an efficient excitation of SPPs because this type of
electromagnetic waves cannot be excited by direct illumination. We
note that Eq.~(\ref{eq_pp_2q3}) fixes a relation between $\omega$
and $\theta$, which are no longer independent variables. Solving,
for example, Eq.~(\ref{eq_pp_2q3}) for $\sin\theta$, Eq.~(\ref{eq_r_for_ll1})
can be regarded as a function of $\omega$ only.

For finite $\kappa _{2}d\ll1$ Eq.~(\ref{eq:max-eq}) becomes
\begin{equation}
\omega_{\textrm{min}}=\frac{\sqrt{\epsilon_{3}\sin^{2}\theta-\epsilon_{1}}}{\epsilon_{1}d}\left[\frac{4\alpha E_{F}d}{\hbar}-\frac{c\epsilon_{2}}{\epsilon_{3}\sin^{2}\theta-\epsilon_{2}}\right].
\label{eq:max-sm}
\end{equation}
The main consequence of Eq.~(\ref{eq:max-sm}) is the existence of
a low-angle cutoff, that is, SPPs can be excited only for incidence angles higher than a certain $\Theta _{\min }$,
\begin{equation}
\sin^{2}\Theta_{\textrm{min}}=\frac{\epsilon_{2}}{\epsilon_{3}}\left[1+\frac{\hbar c}{4\alpha E_{F}d}\right]\:.
\label{eq:min-theta}
\end{equation}
This expression was obtained from Eq.~(\ref{eq:max-sm}) by setting
$\omega_{\textrm{min}}=0$. Decreasing the thickness of the gap between
the prism and graphene, $d$, leads to an increase of $\Theta_{\textrm{min}}$.
When the right-hand side of Eq.~(\ref{eq:min-theta}) becomes larger than unity, SPPs no longer can be excited
in the full range of incidence angles $\theta$.

The validity of 
approximation (\ref{eq:max-sm}) is demonstrated explicitly in Fig.~\ref{fig:atr_appr}(a).
For small $d$, the approximation (\ref{eq:max-sm}) is valid for $\theta\gtrsim\Theta_{\textrm{min}}\approx1.23\theta_{c}$.
At the same time, for $\theta\gtrsim2.15\theta_{c}$ one obtains a
good agreement between the approximation (\ref{eq:max-sm}) {[}blue
dashed-dotted line{]} and exact results {[}red solid line{]} both
for $\omega_{\textrm{min}}$ {[}Fig.~\ref{fig:atr_appr}(a){]} and
for $R_{\textrm{min}}$ {[}Fig.~\ref{fig:atr_appr}(d){]}. Finally,
increasing $d$ limits the validity of the approximation (\ref{eq:max-sm})
to small $\theta$ only {[}see Figs.~\ref{fig:atr_appr}(b),~\ref{fig:atr_appr}(c),~\ref{fig:atr_appr}(e),
and \ref{fig:atr_appr}(f){]}.


\subsubsection{The limit $\kappa_{2}d\gg1$ }

In the opposite limit of $\kappa _{2}d\to\infty$, we obtain for Eq.~(\ref{eq_exact_r_Otto})
\begin{eqnarray}
\frac{E_{x}^{{(r)}}}{E_{x}^{{(i)}}}\rightarrow\frac{i\kappa_{2}\epsilon_{3}-k_{z}\epsilon_{2}}{i\kappa_{2}\epsilon_{3}+k_{z}\epsilon_{2}}\,,
\end{eqnarray}
which is equivalent to Eq.~(\ref{eq_r_total_reflect}), as expected, because graphene is far away from the 2-3 interface. Solution of Eq.~(\ref{eq:max-eq}) by
means of perturbation theory yields:
\begin{eqnarray}
\omega_{\textrm{min}}=\frac{4\alpha E_{F}}{\hbar}\left[\frac{\epsilon_{1}}{\sqrt{\epsilon_{3}\sin^{2}\theta-\epsilon_{1}}}+\frac{\epsilon_{2} \beta }{\sqrt{\epsilon_{3}\sin^{2}\theta-\epsilon_{2}}}\right]^{-1},\label{eq:max-big}
\end{eqnarray}
where
\begin{eqnarray}
\nonumber
\beta = \tanh^{-1}\left(\omega_{0}\frac{d}{c}\sqrt{\epsilon_{3}\sin^{2}\theta-\epsilon_{2}}\right)\:,\\
\omega_{0}=\frac{4\alpha E_{F}}{\hbar}\left[\frac{\epsilon_{1}}{\sqrt{\epsilon_{3}\sin^{2}\theta-\epsilon_{1}}}+\frac{\epsilon_{2}}{\sqrt{\epsilon_{3}\sin^{2}\theta-\epsilon_{2}}}\right]^{-1}.\label{eq:max-big_aux}
\end{eqnarray}
Comparison of panels (a-c) in Fig.~\ref{fig:atr_appr} demonstrates that
increasing $d$ leads to a better correspondence between the approximation
(\ref{eq:max-big}) {[}green dashed lines{]} and the exact result
{[}red solid lines{]}.


\subsubsection{Transfer matrix method }

We will show now how the same problem of interaction of light with a graphene-based planar structure can be solved using the transfer matrix
method~\cite {Born}. Obviously, the result is the same, however, the method
is suitable for systems based on graphene multi-layers, such as that
of Fig.~\ref{fig_double_layer}. Referring to Fig.~\ref{fig_EPL},
the boundary conditions 
(\ref{eq:0ef}) and (\ref{eq:0mf-new}) at the 3-2 interface can be written in the matrix
form as
\begin{eqnarray}
\left[\begin{array}{c}
E_{x}^{(i)}\\
E_{x}^{(r)}
\end{array}\right] & = & \frac{1}{2i\epsilon_{3}\kappa_{2}}\left[\begin{array}{cc}
\epsilon_{3} & k_{z}\\
\epsilon_{3} & -k_{z}
\end{array}\right]\left[\begin{array}{cc}
i\kappa_{2} & i\kappa_{2}\\
-\epsilon_{2} & \epsilon_{2}
\end{array}\right]\left[\begin{array}{c}
E_{2,x}^{(+)}\\
E_{2,x}^{(-)}
\end{array}\right]\nonumber \\
 & \equiv & M^{3\rightarrow2}\left[\begin{array}{c}
E_{2,x}^{(+)}\\
E_{2,x}^{(-)}
\end{array}\right]\,.\label{eq:M3to2}
\end{eqnarray}
The determinant of $M^{3\Rightarrow2}$ is $\mbox{det }M^{3\Rightarrow2}=ik_{z}\epsilon_{2}/(\kappa_{2}\epsilon_{3})\,.$
Similarly, the boundary conditions at the 2-1 interface can be written as
\begin{eqnarray}
\left[\begin{array}{c}
E_{2,x}^{(+)}\\
E_{2,x}^{(-)}
\end{array}\right] & = & \left[\begin{array}{cc}
e^{-\kappa_{2}d} & 0\\
0 & e^{\kappa_{2}d}
\end{array}\right]\frac{1}{2\chi_{1}}\left[\begin{array}{cc}
\chi_{1} & 1\\
\chi_{1} & -1
\end{array}\right]\nonumber \\
 & \times & \left[\begin{array}{cc}
1 & 1\\
\kappa_{2}\Lambda_{1}^{(-)} & -\kappa_{2}\Lambda_{1}^{(+)}
\end{array}\right]\left[\begin{array}{c}
E_{1,x}^{(+)}e^{\kappa_{1}d}\\
E_{1,x}^{(-)}e^{-\kappa_{1}d}
\end{array}\right]\nonumber \\
 & \equiv & M^{2\Rightarrow1}\left[\begin{array}{c}
E_{1,x}^{(+)}\\
E_{1,x}^{(-)}
\end{array}\right],\label{eq:M2to1}
\end{eqnarray}
where $\Lambda_{1}^{(\pm)}$ has been defined in (\ref{eq:Lambda_pm}).
The determinant of $M^{2\Rightarrow1}$ is, $\mbox{det }M^{2\Rightarrow1}=\kappa_{2}\epsilon_{1}/(\kappa_{1}\epsilon_{2})\,.$
The full transfer matrix is $M=M_{3\Rightarrow2}M_{2\Rightarrow1}$
and its determinant reads:
\begin{equation}
\mbox{det }M=i\frac{k_{z}\epsilon_{1}}{\kappa_{1}\epsilon_{3}}\,.
\end{equation}
The reflection coefficient is given by
\begin{equation}
r\equiv\frac{E_{x}^{(r)}}{E_{x}^{(i)}}=\frac{M_{22}}{M_{12}}\,,\label{eq:tran-mat-ref}
\end{equation}
or explicitly,
\begin{eqnarray}
\frac{E_{x}^{(r)}}{E_{x}^{(i)}} & = & \frac{\kappa_{2}\cosh(\kappa_{2}d)(\chi_{1}\epsilon_{3}+ik_{z}\epsilon_{2}\Lambda_{1}^{(+)})+\sinh(\kappa_{2}d)(\kappa_{2}^{2}\epsilon_{3}\Lambda_{1}^{(+)}+ik_{z}\epsilon_{2}\chi_{1})}{\kappa_{2}\cosh(\kappa_{2}d)(\chi_{1}\epsilon_{3}-ik_{z}\epsilon_{2}\Lambda_{1}^{(+)})+\sinh(\kappa_{2}d)(\kappa_{2}^{2}\epsilon_{3}\Lambda_{1}^{(+)}-ik_{z}\epsilon_{2}\chi_{1})}\,,\nonumber \\
\label{eq_r_t_matrix}
\end{eqnarray}
where $M_{ij}$ are the elements of the matrix $M$.
 Notice, that Eq.(\ref{eq:tran-mat-ref}) 
can be obtained from Eqs.(\ref{eq:M2to1}) and (\ref{eq:M3to2}), 
by putting $E_{1,x}^{(+)}\equiv 0$ which is necessary to guarantee that  the 
field is finite at z
$\rightarrow \infty$. 
After some algebra,  Eq.~(\ref{eq_r_t_matrix}) can be shown equivalent
to Eq.~(\ref{eq_exact_r_Otto}).


\subsection{Otto configuration: $s-$polarized wave}

\label{subsec_Otto_s} We shall now discuss the reflection of an $s-$polarized
wave in the geometry depicted in Fig.~\ref{fig_EPL}. The wave vectors have been defined in Sec. \ref{subsec_Otto}. The fields in the three regions
are given by:

Region 3
\begin{eqnarray}
\vec{B}_{3} & = & (B_{x}^{(i)},0,B_{z}^{(i)})e^{i\vec{k}_{i}\cdot\vec{r}}+(B_{x}^{(r)},0,B_{z}^{(r)})e^{i\vec{k}_{r}\cdot\vec{r}}\,,\\
E_{3,y} & = & E_{y}^{(i)}e^{i\vec{k}_{i}\cdot\vec{r}}+E_{y}^{(r)}e^{i\vec{k}_{r}\cdot\vec{r}}\,;
\end{eqnarray}

Region 2
\begin{eqnarray}
\vec{B}_{2} & = & (B_{2,x}^{(+)},0,B_{2,z}^{(+)})e^{i\vec{k}_{2}^{(+)}\cdot\vec{r}}+(B_{2,x}^{(-)},0,B_{2,z}^{(-)})e^{i\vec{k}_{2}^{(-)}\cdot\vec{r}}\,,\\
E_{2,y} & = & E_{2,y}^{(+)}e^{i\vec{k}_{2}^{(+)}\cdot\vec{r}}+E_{2,y}^{(-)}e^{i\vec{k}_{2}^{(-)}\cdot\vec{r}}\,;
\end{eqnarray}

Region 1
\begin{eqnarray}
\vec{B}_{1} & = & (B_{1,x}^{(-)},0,B_{1,z}^{(-)})e^{i\vec{k}_{1}^{(-)}\cdot\vec{r}}\,,\\
E_{1,y} & = & E_{1,y}^{(-)}e^{i\vec{k}_{1}^{(-)}\cdot\vec{r}}\,.
\end{eqnarray}
 The boundary conditions at $z=0,\:d$ are:
\begin{eqnarray}
E_{y}^{(i)}+E_{y}^{(r)}=E_{2,y}^{(+)}+E_{2,y}^{(-)}\:;\label{eq:0ef-s}\\
B_{x}^{(i)}+B_{x}^{(r)}=B_{2,x}^{(+)}+B_{2,x}^{(-)}\:.\label{eq:0mf-s}
\end{eqnarray}
\begin{eqnarray}
E_{1,y}^{(-)}e^{-\kappa_{1}d} & = & E_{2,y}^{(+)}e^{\kappa_{2}d}+E_{2,y}^{(-)}e^{-\kappa_{2}d}\,,\label{eq:def-s}\\
B_{1,x}^{(-)}e^{-\kappa_{1}d} & = & B_{2,x}^{(+)}e^{\kappa_{2}d}+B_{2,x}e^{-\kappa_{2}d}+\mu_{0}\sigma E_{1,y}^{(-)}e^{-\kappa_{1}d}\,.\label{eq:dmf-s}
\end{eqnarray}
We can express the magnetic field components in terms of the electric
field ones. To that end we use Maxwell's equations (\ref{eq:Bx}),
and (\ref{eq:Bz}), (\ref{eq:Ey}) and obtain
\begin{eqnarray}
B_{x}^{(i,r)} & = & \mp\frac{k_{z}}{\omega}E_{y}^{(i,r)}\,,\\
B_{z}^{(i,r)} & = & \frac{q}{\omega}E_{y}^{(i,r)}\,,\\
B_{m,x}^{(\pm)} & = & \pm i\frac{\kappa_{m}}{\omega}E_{m,y}^{(\pm)}\,,\\
B_{m,z}^{(\pm)} & = & \frac{q}{\omega}E_{m,y}^{(\pm)}\,.
\end{eqnarray}
These allow to write Eqs.~(\ref{eq:0mf-s}) and (\ref{eq:dmf-s})
as
\begin{equation}
ik_{z}E_{y}^{(i)}-ik_{z}E_{y}^{(r)} = \kappa_{2}E_{2,y}^{(+)}-\kappa_{2}E_{2,y}^{(-)}
\label{eq:0mf-s-new}
\end{equation}
and
\begin{equation}
-(\kappa_{1}-i\omega\mu_{0}\sigma)E_{1,y}^{(-)}e^{-\kappa_{1}d}  =  \kappa_{2}E_{2,y}^{(+)}e^{\kappa_{2}d}-\kappa_{2}E_{2,y}e^{-\kappa_{2}d}\,,
\label{eq:dmf-s-new}
\end{equation}
respectively.
The problem now reduces to the solution of the following system of linear equations:
\begin{equation}
\Xi_{s}\left[\begin{array}{c}
E_{y}^{(r)}/E_{y}^{(i)}\\
E_{2,y}^{(+)}/E_{y}^{(i)}\\
E_{2,y}^{(+)}/E_{y}^{(i)}
\end{array}
\right]=\Phi_{s}\,.
\label{eq:linear}
\end{equation}
Here
\begin{eqnarray}
\Xi_{s} & = & \left[\begin{array}{ccc}
-1 & 1 & 1\\
ik_{z} & \kappa_{2} & -\kappa_{2}\\
0 & (\Lambda_{s}^{(-)}+\kappa_{2})e^{\kappa_{2}d} & (\Lambda_{s}^{(-)}-\kappa_{2})e^{-\kappa_{2}d}
\end{array}\right],\label{eq:Xi-s}\\
\Phi_{s} & = & \left[\begin{array}{c}
1\\
ik_{z}\\
0
\end{array}\right]\,,\label{eq:Phi-s}\\
\Lambda_{s}^{(\pm)} & = & \kappa_{1}\pm i\mu_{0}\omega\sigma\,.
\end{eqnarray}
From the solution of \ref{eq:linear} it follows that the reflection coefficient is given by
\begin{equation}
r_{s}=\frac{E_{y}^{(r)}}{E_{y}^{(i)}}=-\frac{\kappa_{2}\eta_{1,s}+ik_{z}\eta_{2,s}}{\kappa_{2}\eta_{1,s}-ik_{z}\eta_{2,s}}\,.\label{eq_rs}
\end{equation}
where $\eta_{1,s}=\Lambda_{s}^{(-)}\cosh(\kappa_{2}d)+\kappa_{2}\sinh(\kappa_{2}d)$,
$\eta_{2,s}=\Lambda_{s}^{(-)}\sinh(\kappa_{2}d)+\kappa_{2}\cosh(\kappa_{2}d)$.
We note that for $d\rightarrow\infty$, $r$ has a pole for $\Lambda_{s}^{(-)}+\kappa_{2}=0$, which yields the TE-wave dispersion relation {[}compare with Eq.~(\ref{eq_W_SPP_2D_TE}){]}.
\begin{figure}[!ht]
\begin{centering}
\includegraphics[clip,width=9cm]{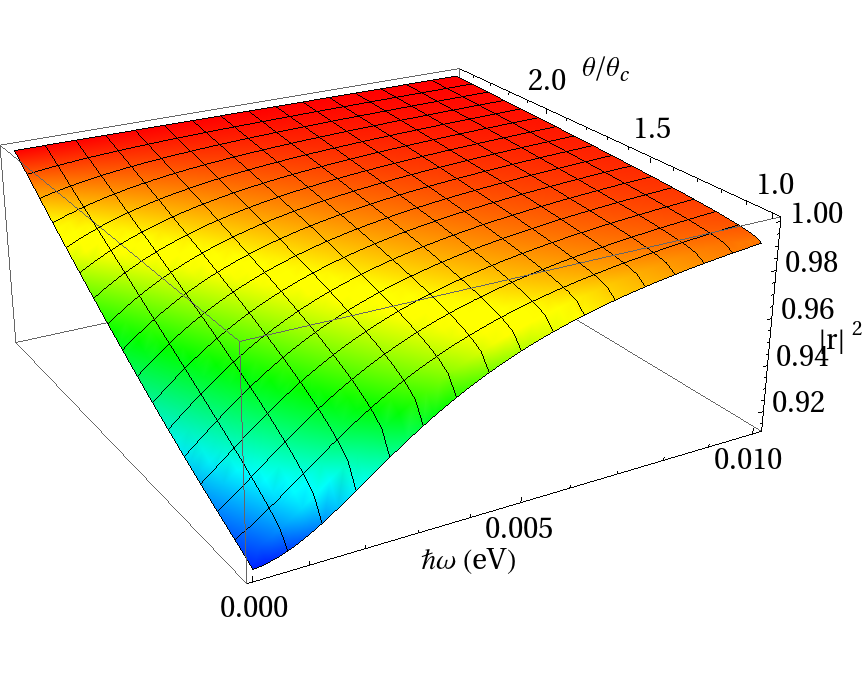}

\par\end{centering}

\caption{Reflectance of a $s-$polarized wave in the geometry of Fig.~\ref{fig_EPL}.
The parameters are the same of Fig.~\ref{fig_reflectance}.}

\label{fig_s_polarization_reflectance}
\end{figure}

In Fig.~\ref{fig_s_polarization_reflectance} the reflectance
of a $s-$polarized wave is plotted against the photon energy and the
angle of incidence. Clearly, the reflectance is close to unity everywhere excepting $\omega \rightarrow 0$.
Accordingly, the absorbance, $1-\vert r_{s}\vert^{2}$, is close to zero.
This behaviour contrasts with the case of $p-$polarized waves, where sharp SPP-related resonances occur. There are such dips in
the reflectance depicted in Fig.~\ref{fig_s_polarization_reflectance}
(note the vertical scale) because the conductivity has been modelled by the Drude formula (which has $\sigma^{\prime\prime}>0$). Hence, TE plasmon-polaritons in graphene are not captured
by the present calculation (c.f.~Sec.~\ref{sec_TE_spectrum}).


\subsection{Otto configuration: arbitrarily polarized waves}

\label{subsec_Otto_sp}
\begin{figure}[!ht]
\begin{centering}
\includegraphics[clip,width=9cm]{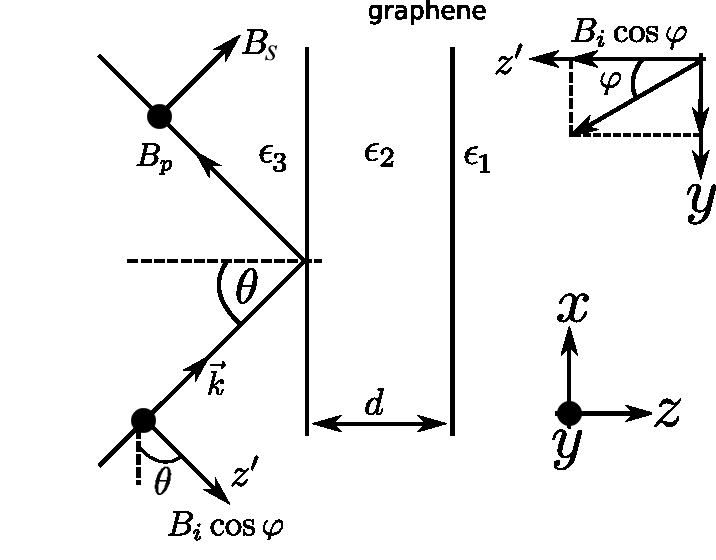}

\par\end{centering}

\caption{Geometry for the scattering of an electromagnetic wave containing
both $s-$ and $p-$polarization components.}

\label{fig_sp_scattering_geometry}
\end{figure}

In the general case of an arbitrary linearly polarized wave (see Fig.~\ref{fig_sp_scattering_geometry}), the incoming
magnetic field is,
\begin{equation}
\vec{B}_{i}=B_{i}(-\cos\varphi\cos\theta,\sin\varphi,\cos\varphi\sin\theta)e^{i\vec{k}_{i}\cdot\vec{r}}\,.
\end{equation}
Note that $\varphi=0$
corresponds to a purely $s-$polarized wave, whereas $\varphi=\pi/2$
corresponds to a purely $p-$polarized wave.

The incoming electric field is obtained from Maxwell's equations as
\begin{equation}
\vec{E}_{i}=-\frac{\omega}{k^{2}}
\vec{k}_i\times\vec{B}=
\frac{\omega}{k}B_{i}(\sin\varphi\cos\theta,\cos\varphi,
-\sin\varphi\sin\theta)e^{i\vec{k}_{i}\cdot\vec{r}}
\label{eq:Ei}
\end{equation}
and the reflected field can be written as
\begin{equation}
\vec{B}_{r}=(B_{s}\cos\theta,B_{p},B_{s}\sin\theta)e^{i\vec{k}_{r}\cdot\vec{r}}\,.
\end{equation}
Again, it follows from Maxwell's equations that the reflected electric
field is given by
\begin{equation}
\vec{E}_{r}=-\frac{\omega}{k^{2}}
\vec{k}_r
\times\vec{B}=\frac{\omega}{k}
(-B_{p}\cos\theta,B_{s},-B_{p}\sin\theta)e^{i\vec{k}_{r}\cdot\vec{r}}\,.\label{eq:Er}
\end{equation}

It becomes clear
that the arbitrarily polarized wave can be decomposed into $s-$ and
 $p-$polarized components, which are reflected independently.
We have already computed the reflection coefficients for $s-$ and
$p-$polarized waves. Thus, we have for the $s-$polarized component:
\begin{equation}
\frac{E_{y}^{(r)}}{E_{y}^{(i)}}=\frac{B_{s}}{B_{i}\cos\varphi}\:;\ \ \ \ \ \   \frac{B_{s}}{B_{i}}=\cos\varphi\frac{E_{y}^{(r)}}{E_{y}^{(i)}}\,.
\end{equation}
For $\varphi=\pi/2$, there is no $s-$polarized reflected
wave because the incoming ER is purely $p-$polarized. Notice
that $E_{y}^{(r)}/E_{y}^{(i)}$ has been defined in Eq.~(\ref{eq_rs}).

For the $p-$polarized component we have:
\begin{equation}
\frac{E_{x}^{(r)}}{E_{x}^{(i)}}=-\frac{B_{p}}{B_{i}\sin\varphi}\:;\ \ \ \ \ \   \frac{B_{p}}{B_{i}}=-\sin\varphi\frac{E_{x}^{(r)}}{E_{x}^{(i)}}\,,
\end{equation}
and $E_{x}^{(r)}/E_{x}^{(i)}$ has been defined in Eq.~(\ref{eq_exact_r_Otto}).
This result implies that for $\varphi=0$ there is no scattered $p-$polarized
wave because the incoming ER is purely $s-$polarized.

In the regime where the optical conductivity of graphene is dominated
by the Drude term, the excitation of a $s-$polarized surface waves
in graphene is not possible because the imaginary part of the conductivity
is positive (refer to Sec.~\ref{sec_TE_spectrum}). On the other
hand, choosing the parameters of the problem appropriately (refer
to Sec.~\ref{subsec_Otto}), it is possible to obtain a strong absorption
of the $p-$polarized component of the impinging ER. As a consequence,
the setup of Fig.~\ref{fig_plasmon_polariton} can work as a polarizer.

If we write $B_{s}/B_{i}=A_{s}e^{i\delta_{s}}$ and $B_{p}/B_{i}=A_{p}e^{i\delta_{p}}$,
the total reflected power is defined as
\begin{equation}
R_{T}=R_{s}+R_{p}=\vert A_{s}\vert^{2}+\vert A_{p}\vert^{2}\,,\label{eq_total_reflect_sp}
\end{equation}
and the angle of ellipticity, $\psi$, is obtained from \cite{Hecht}
\begin{equation}
\tan(2\psi)=\frac{2A_{p}A_{s}}{A_{s}^{2}-A_{p}^{2}}\cos(\delta_{p}-\delta_{s})\,.\label{eq_angle_psi}
\end{equation}
The angle $\psi$ vanishes when either $A_{s}$ or $A_{p}$ is equal
to zero, corresponding to a linearly polarized wave, or when $\delta_{p}-\delta_{s}=\pm\pi/2$,
corresponding to an elliptically polarized wave.
\begin{figure}[!ht]
\begin{centering}
\includegraphics[clip,width=9cm]{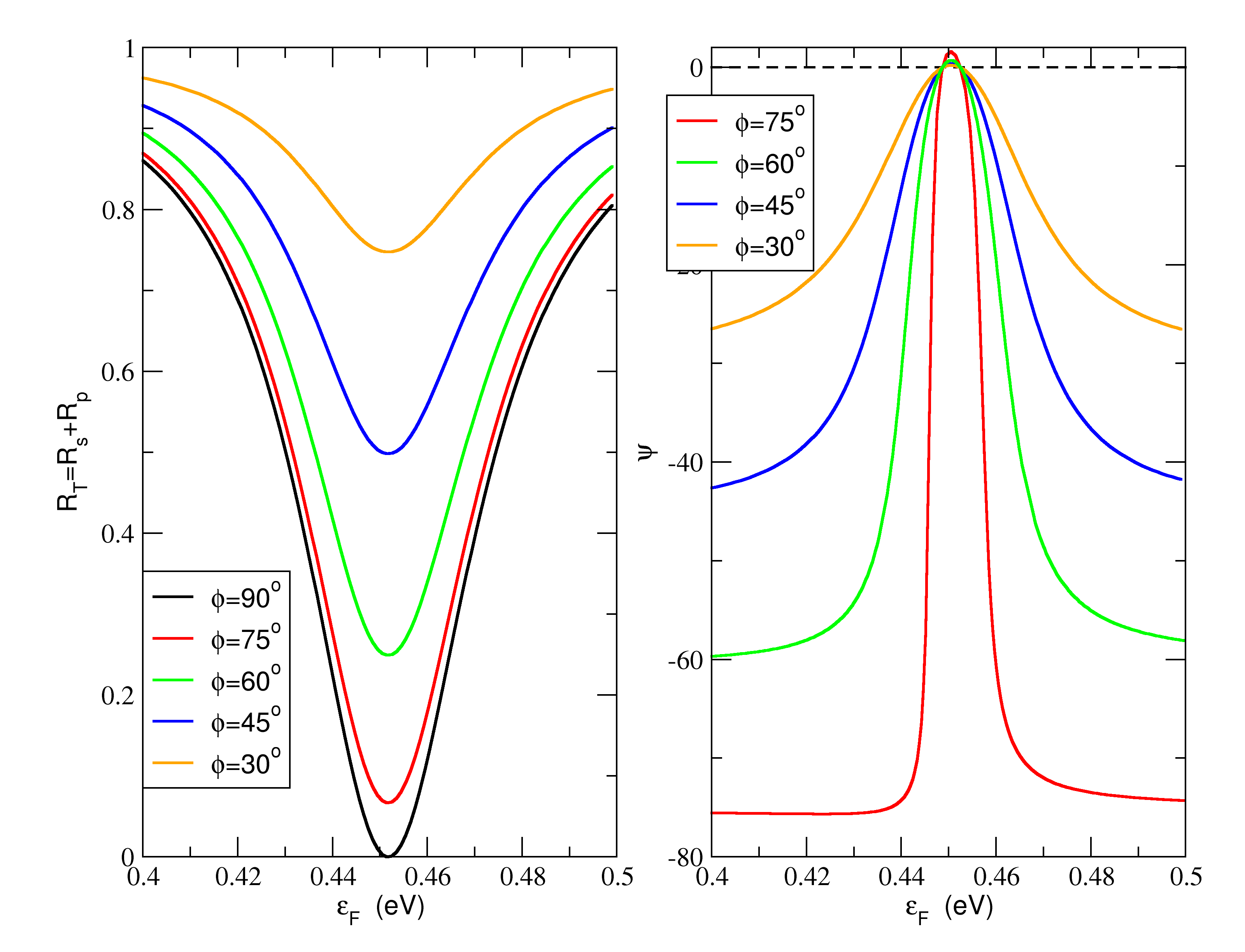} 

\par\end{centering}

\caption{Total reflectance $R_{T}$ and the angle of ellipticity $\psi$ as
a function of the Fermi energy $E_{F}$, for different angles
of polarization of the incident wave.
The reflectance of the $p-$polarized component
vanishes for $E_{F}=0.45$ eV. Other parameters as in Fig.~\ref{fig_reflectance}.
The angle of incidence is $\theta=1.95\theta_c$ and the energy of the
incoming light is $\hbar\omega=5.1$ meV.}
\label{fig_total_reflectance_s_and_p}
\end{figure}

In Fig.~\ref{fig_total_reflectance_s_and_p} we represent the total
reflectance $R_{T}$, given by Eq.~(\ref{eq_total_reflect_sp}),
and the angle of ellipticity $\psi$, as given by Eq.~(\ref{eq_angle_psi}),
as a function of the Fermi energy $E_{F}$, for fixed $\omega $ and $\theta $ and different angles
of polarization, $\varphi$, of the incident wave. When $\varphi=\pi/2$, the incoming wave is
purely $p-$polarized and the reflectance for $E_{F}=0.45$~eV is
zero (for the given $\omega $ and $\theta $). As $\varphi$ deviates from $\pi/2$, a finite reflectance for
$E_{F}=0.45$~eV appears due to the $s-$polarized component of the
incoming wave, as can be seen in the left panel
of Fig.~\ref{fig_total_reflectance_s_and_p}. In the right panel
of the same figure, the angle of ellipticity ($\psi$) is plotted. We
can see that there are two points for which $\psi$ is zero, they
correspond to $A_{p}=0$ at $E_{F}=0.45$~eV and to $\delta_{p}-\delta_{s}=\pm\pi/2$
at a sightly different energy.


\subsection{Otto configuration in a static magnetic field}

We shall now consider the Otto configuration of Fig.~\ref{fig_EPL} in
the presence of a static magnetic field $\vec{B}_{0}=(0,0,B_{0})$.
We take an arbitrarily polarized incident wave,
\begin{equation}
\vec{E}_{i}(\vec{r},t)=\left[(E_{x}^{(i)},0,E_{z}^{(i)})+(0,E_{y}^{(i)},0)\right]e^{i\vec{k}_{i}\cdot\vec{r}}\,,\label{eq:E_in}
\end{equation}
where we have explicitly separated the $p-$ and $s-$polarized
components. The starting point to solve the light reflection
problem is to write the ER field in regions $m=1,2,3$ in the most
general way. We obtain,

i) Region 3:
\begin{eqnarray}
\vec{E}_{3} & = & (E_{x}^{(i)},E_{y}^{(i)},E_{z}^{(i)})e^{i\vec{k}_{i}\cdot\vec{r}}+(E_{x}^{(r)},E_{y}^{(r)},E_{z}^{(r)})e^{i\vec{k}_{r}\cdot\vec{r}},\label{eq:E_region_3}
\end{eqnarray}

ii) Region 2:

\begin{equation}
\vec{E}_{2}=(E_{2,x}^{(+)},E_{2,y}^{(+)},E_{2,z}^{(+)})e^{i\vec{k}_{2}^{(+)}\cdot\vec{r}}+(E_{2,x}^{(-)},E_{2,y}^{(-)},E_{2,z}^{(-)})e^{i\vec{k}_{2}^{(-)}\cdot\vec{r}}\,,\label{eq:E_region_2}
\end{equation}

iii) Region 1:

\begin{equation}
\vec{E}_{1}=(E_{1,x}^{(-)},E_{1,y}^{(-)},E_{1,z}^{(-)})e^{i\vec{k}_{1}^{(-)}\cdot\vec{r}}\,,\label{eq:E_region_1}
\end{equation}
where the wavevectors $\vec{k}_{i,r}$, $\vec{k}_{m}^{(\pm)}$
have been defined in Sec.~\ref{subsec_Otto}. Similar equations apply
to the time-varying part of the magnetic field $\vec{B}$. 

The next step is to write the boundary conditions at each interface.
At $z=0$, the magnetic field is continuous, as well as the tangential
components of the electric field. The boundary conditions have the
same form as Eqs.~(\ref{eq:0ef})-(\ref{eq:0mf}), and (\ref{eq:0ef-s})-(\ref{eq:0mf-s}),
and shall not be reproduced again. The boundary conditions at the second
interface ($z=d$) must take into account the discontinuity of the
magnetic field across the graphene layer due to the existence of
a surface current $\vec{J}_{s}$ {[}see Eq.(\ref{eq:bcond-ht}){]},
\begin{equation}
\left.\left(\vec{B}_{1}-\vec{B}_{2}\right)\right|_{z=d}=\mu_{0}\vec{J}_{s}\times\hat{n}=\mu_{0}\hat{\sigma}(\omega,B_{0})\vec{E}_{\parallel}(z=d)\times\hat{n}\,,\label{eq:discontinuity_magnetic_field-1}
\end{equation}
with $\vec{E}_{\parallel}$ denoting the in-plane components
of the electric field with respect to the graphene sheet. The magneto-optical
conductivity tensor of graphene $\hat{\sigma}(\omega,B_{0})$ was
discussed earlier, in Sec.~3. The boundary conditions relating the
electric field amplitudes coincide with those given earlier, namely,
Eqs.~(\ref{eq:def}) and (\ref{eq:def-s}), whereas for the magnetic
field amplitudes we obtain the following set of conditions:
\begin{eqnarray}
B_{1,y}^{(-)}-e^{\kappa_{1}d}
\left(B_{2,y}^{(+)}e^{\kappa_{2}d}+B_{2,y}^{(-)}e^{-\kappa_{2}d}\right)=\nonumber\\
-\mu_{0}\left(\sigma_{xx}E_{1,x}^{(-)}+\sigma_{xy}E_{1,y}^{(-)}\right)\,,\\
\label{eq:BC_2_2}
B_{1,x}^{(-)}-e^{\kappa_{1}d}
\left(B_{2,x}^{(+)}e^{\kappa_{2}d}+B_{2,x}^{(-)}e^{-\kappa_{2}d}\right)=\nonumber\\
\mu_{0}\left(\sigma_{yx}E_{1,x}^{(-)}+\sigma_{yy}E_{1,y}^{(-)}\right)\,.
\label{eq:BC_2_3}
\end{eqnarray}
As before, we express the boundary conditions using the electric field
components only. Using Eqs.~(\ref{eq:0mf-new})-(\ref{eq:0mf-s-new}),
we easily obtain:
\begin{eqnarray}
\nonumber
\frac{\omega\epsilon_{1}}{\kappa_{1}}E_{1,x}^{(-)}+\frac{\omega\epsilon_{2}
e^{\kappa_{1}d}}{\kappa_{2}}
\left(E_{2,x}^{(+)}e^{\kappa_{2}d}-E_{2,x}^{(-)}e^{-\kappa_{2}d}\right)\\
= -\frac{i}{\epsilon_{0}}\left(\sigma_{xx}E_{1,x}^{(-)}+\sigma_{xy}E_{1,y}^{(-)}\right)\,;
\label{eq:BC_2_1_simp}\\
\nonumber
\kappa_{1}E_{1,y}^{(-)}+\kappa_{2}
\left(E_{2,y}^{(+)}e^{\kappa_{2}d}-E_{2,y}^{(-)}e^{-\kappa_{2}d}\right)\\
= i\omega\mu_{0}\left(\sigma_{yx}E_{1,x}^{(-)}+\sigma_{yy}E_{1,y}^{(-)}\right)\,.\label{eq:BC_2_2_simp}
\end{eqnarray}
Combining Eqs.~(\ref{eq:0ef}), (\ref{eq:0mf-new}), (\ref{eq:0ef-s}),
(\ref{eq:0mf-s-new}) and (\ref{eq:BC_2_1_simp})-(\ref{eq:BC_2_2_simp}),
and defining the following vectors,
\begin{equation}
e_{p}\equiv(E_{x}^{(r)}/E_{x}^{(i)},E_{2,x}^{(+)}/E_{x}^{(i)},E_{2,x}^{(-)}/E_{x}^{(i)})\,,\label{eq:eAC}
\end{equation}
\begin{equation}
e_{s}\equiv(E_{y}^{(r)}/E_{x}^{(i)},E_{2,y}^{(+)}/E_{x}^{(i)},E_{2,y}^{(-)}/E_{x}^{(i)})\,,\label{eq:eBD}
\end{equation}
a closed system of equations for the ER amplitudes is obtained according
to
\begin{equation}
\left(\begin{array}{cc}
\Xi_{p} & \Xi_{ps}\\
\Xi_{sp} & \Xi_{s}
\end{array}\right)\left(\begin{array}{c}
e_{p}\\
e_{s}
\end{array}\right)=\left(\begin{array}{c}
\Phi_{p}\\
f_{s}\Phi_{s}
\end{array}\right)\,,\label{eq:system}
\end{equation}
where $f_{s}\ge0$ denotes the relative fraction of the incoming electric
field stored in the $s-$polarized component, $f_{s}=E_{y,i}/E_{x,i}$,
and $\Xi_{p}$, $\Phi_{p}$, $\Xi_{s}$, $\Phi_{s}$ have been defined in
Eqs.~(\ref{eq:Xi-p}), (\ref{eq:Phi-p}), (\ref{eq:Xi-s}), and (\ref{eq:Phi-s}),
respectively, while other $\Xi$-matrices are are given by
\begin{eqnarray}
\Xi_{ps} & = & \left(\begin{array}{ccc}
0 & 0 & 0\\
0 & 0 & 0\\
0 & i\frac{\sigma_{xy}}{\epsilon_{0}}\kappa_{1}\kappa_{2} & i\frac{\sigma_{xy}}{\epsilon_{0}}\kappa_{1}\kappa_{2}
\end{array}\right)\,,\label{eq:Xi-ps}\\
\Xi_{sp} & = & \left(\begin{array}{ccc}
0 & 0 & 0\\
0 & 0 & 0\\
0 & -i\omega\mu_{0}\sigma_{yx} & -i\omega\mu_{0}\sigma_{yx}
\end{array}\right)\,.\label{eq:Xi-sp}
\end{eqnarray}
\begin{figure}
\centering{}\includegraphics[width=0.7\textwidth]{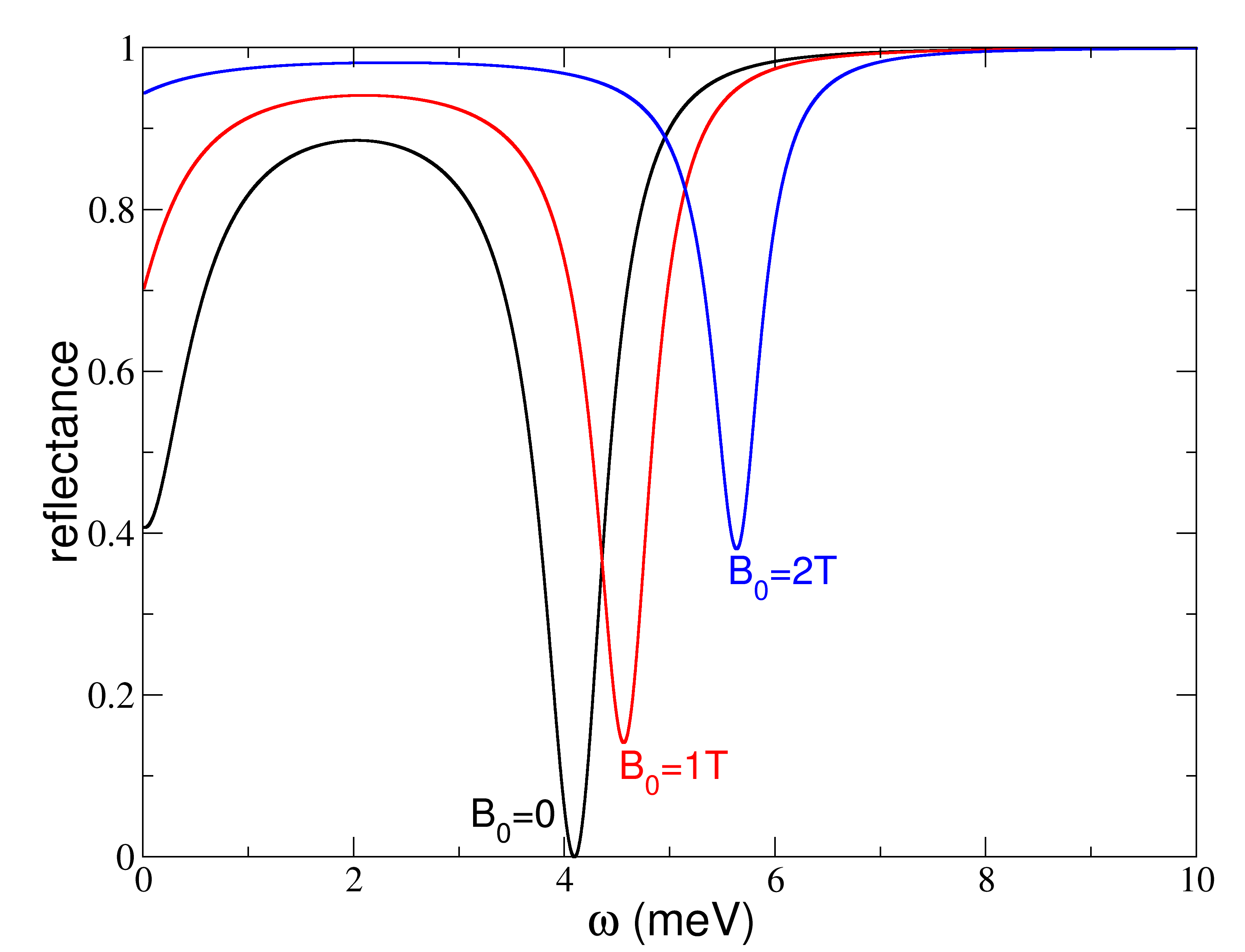}\caption{\label{fig:Otto_refletance_with_field}The reflectance as function
of the frequency for $p-$polarized incident light in the presence of
a static external magnetic field. Other parameters as in Fig.~\ref{fig_reflectance}.}
\end{figure}

It should be stressed that in Sec. \ref{subsec_Otto_sp} the $s-$ and
$p-$polarization components were independent, while in the case under consideration
(with external magnetic field) they are coupled through
the matrices $\Xi_{sp}$ and $\Xi_{ps}$. The optical properties of
the the system in Otto configuration, in the presence of an external magnetic field are fully
determined once the system of equations (\ref{eq:system}) is
solved. For instance, the reflectance is obtained according to
\begin{equation}
R=\frac{|E_{x}^{(r)}/E_{x}^{(i)}|^{2}+\cos^{2}\theta|E_{y}^{(r)}/E_{x}^{(i)}|^{2}}{1+f_{s}^{2}\cos^{2}\theta}\,,\label{eq:Reflectivity_general}
\end{equation}
and reduces to the familiar form $R=|E_{x}^{(r)}/E_{x}^{(i)}|^{2}$
in the particular case of a pure $p-$polarized incident wave ($f_{s}=0$).
Note that, by expressing the incident and reflected electric fields
in the form similar to Eqs.(\ref{eq:Ei}), and (\ref{eq:Er}),
\begin{eqnarray}
\vec{E}_{3}=E_{i}(\sin\varphi\cos\theta,\cos\varphi,-\sin\varphi\sin\theta)e^{i\vec{k}_{i}\cdot\vec{r}}+\nonumber \\
(-E_{p}\cos\theta,E_{s},-E_{p}\sin\theta)e^{i\vec{k}_{r}\cdot\vec{r}}\,,
\end{eqnarray}
the reflectance assumes the simple form $R=(|E_{p}|^{2}+|E_{s}|^{2})/|E_{i}|^{2}$,
and we have $f_{s}=\tan^{-1}\varphi\cos^{-1}\theta$.

In Fig.~\ref{fig:Otto_refletance_with_field} we plot the reflectance
as a function of $\omega$ for $p-$polarization, obtained by solving numerically
(\ref{eq:system}). We focus on the parameters considered in the absence
of magnetic field (refer to Fig.~\ref{fig_absorvance_d_and_theta}).
The suppression of the reflectance within specific frequency intervals,
interpreted as the excitation of SPPs for the combined layered structure
in zero-field (see Sec.~6.3), is still seen to occur. However, for
$B_{0}>0$ the dips are less pronounced and shifted
relative to the zero-field case. The dips in the reflectance spectrum
for non-zero magnetic field are related to a special type of surface
plasmonic waves known as magneto-plasmon polaritons, resulting from
the hybridization between plasmonic waves and cyclotronic modes. In
the semi-classical regime, the energy of a magneto-plasmon polariton
mode, in the non-retarded approximation $\omega\ll cq$ \cite{ConfinedMPP,Roldan2009}, is given by
\begin{equation}
\Omega(B)\simeq\sqrt{\left[\Omega(0)\right]^{2}+[\omega_{c}(B)]^{2}}\,,\label{eq:MPP}
\end{equation}
where $\Omega(0)$ is the SPP dispersion in the absence of a magnetic
field for graphene cladded by two dielectrics of permittivity
$\epsilon_{2}$ and $\epsilon_{1}$ {[}see Eq.~(\ref{eq_PP_spectrum_Otto}){]},
and the cyclotron frequency $\omega_{c}$ is given by Eq.~(\ref{eq:cyclotron_freq}).
A similar formula holds for the 2D electron gas \cite{Chui1974,Kukushkin2006}.
Away from the semi-classical regime (i.e., for low Fermi energy or high
magnetic fields), the quantization of the electronic spectrum must
be taken into account. In the quantized regime, the physics of graphene
becomes quite exotic, with interband transitions between Landau levels
in the valence and conduction bands of graphene controlling the characteristics
of the magneto-plasmon polaritons (e.g.,~their confinement and propagation)
and giving rise to quasi-transverse electric modes in specific frequency
windows. The reader can refer to Ref.~\cite{ConfinedMPP} for further
details.

The formula (\ref{eq:MPP}) can be used to estimate the shift of the
resonance positions due to the plasmon hybridization with cyclotronic modes. It
predicts a shift $\Omega(B)-\Omega(0)$ of roughly $0.56$~meV for
$B=1$~T, and $1.72$~meV for $B=2$~T, which agrees qualitatively
with the exact numerical results presented in Fig.~\ref{fig:Otto_refletance_with_field}.
We remark that the absolute position of the reflectance dips cannot
be obtained from Eq.~(\ref{eq:MPP}), since this result has been
derived for a simpler geometry consisting of graphene sandwiched between
two dielectrics (see also discussion in Sec.~6.3). An interesting
feature of the hybrid plasmon-cyclotronic modes is the absence of a clear
transverse-magnetic character (i.e.,~contrary to SPPs, magneto-plasmon
polaritons have non-zero magnetic field longitudinal components, $B_{x}$
and $B_{z}$ \cite{ConfinedMPP}). Strictly speaking, these modes
cannot be excited by means of pure $p-$polarized light, hence explaining
why the reflectance does not drop all the way to zero at the magneto-plasmon
polariton frequency (see Fig.~\ref{fig:Otto_refletance_with_field}).
For illumination with a TM wave, complete supression of the
reflectance can only occur when pure SPP (p-polarized) modes are excited
(that is, at zero external magnetic field).


\subsection{Transfer matrix method for $N-$layer problem }

\label{sec_scattering_TMatrix_N_layers}
\begin{figure}[!htb]
\begin{centering}
\includegraphics[clip,width=9cm]{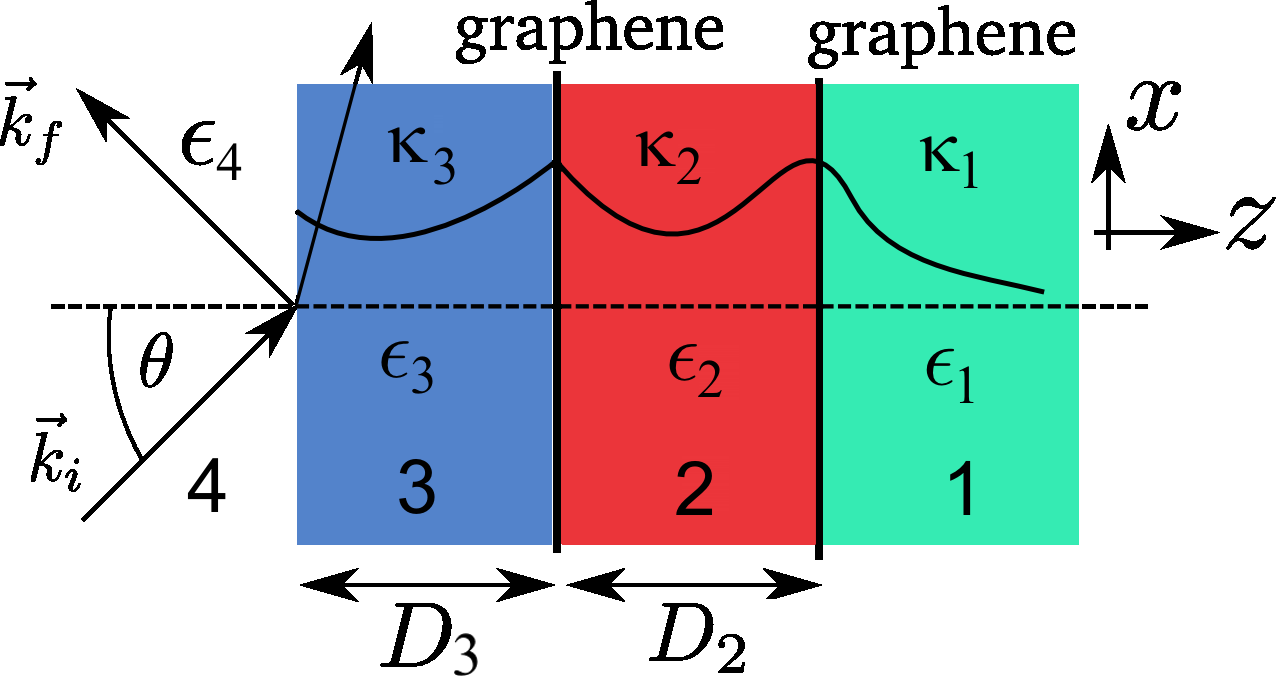} 

\par\end{centering}

\caption{Configuration for a double graphene-layer system. There are three
interfaces: $4-3$, $3-2$, and $2-1$.}

\label{fig_double_layer}
\end{figure}

We have computed the TM SPP spectrum for graphene double layer in Sec.~\ref{subsec_transfer_double_layer}. It is instructive
to apply the transfer matrix method to a more general problem
considering an array of $N$ graphene interfaces.
It allows
for the calculation of the reflection coefficient in
 the ATR configuration whose zeros correspond to the
SPP modes (as we will see below, for a propagating wave
the {\it poles} of the reflection coefficient can be related to the SPP modes).
Figure~\ref{fig_double_layer} depicts the particular case of $N=2$. Thus, the structure
under consideration contains $N+2$ dielectric layers, first and $N+2$-th
layers being semi-infinite with dielectric constants $\epsilon_{1}$
and $\epsilon_{N+2}$. At the same time, other layers are characterized
by width $D_{m}$ and dielectric constants $\epsilon_{m}$, $m=2,...,N+1$.
$N$ graphene layers are arranged at interfaces between $m$-th and
$m-1$-th layers ($m=2,...,N+1$). The final solution can be put
in the form:
\begin{eqnarray}
\left[\begin{array}{c}
E_{x}^{(i)}\\
E_{x}^{(r)}
\end{array}\right] & = & M\left[\begin{array}{c}
E_{1,x}^{(+)}\\
E_{1,x}^{(-)}
\end{array}\right]\,,
\end{eqnarray}
where the total transfer matrix $M=\prod_{m=N+1}^{1}M^{m+1\Rightarrow m}$
and matrices $M^{m+1\Rightarrow m}$ can be obtained as generalization
of the matrices in Eqs. (\ref{eq:M3to2})-(\ref{eq:M2to1}), namely
\begin{eqnarray}
M^{N+2\Rightarrow N+1} & = & \frac{-i}{2\epsilon_{N+2}\kappa_{N+1}}\left[\begin{array}{cc}
\epsilon_{N+2} & k_{z}\\
\epsilon_{N+2} & -k_{z}
\end{array}\right]\left[\begin{array}{cc}
i\kappa_{N+1} & i\kappa_{N+1}\\
-\epsilon_{N+1} & \epsilon_{N+1}
\end{array}\right]\,,\\
M^{m+1\Rightarrow m} & = & \left[\begin{array}{cc}
e^{-\kappa_{m+1}D_{m+1}} & 0\\
0 & e^{\kappa_{m+1}D_{m+1}}
\end{array}\right]\frac{1}{2\chi_{m}}\left[\begin{array}{cc}
\chi_{m} & 1\\
\chi_{m} & -1
\end{array}\right]\nonumber \\
 & \times & \left[\begin{array}{cc}
1 & 1\\
\kappa_{m+1}\Lambda_{m}^{(-)} & -\kappa_{m+1}\Lambda_{m}^{(+)}
\end{array}\right]\,.
\end{eqnarray}
The determinant of the total transfer matrix is given by
\begin{equation}
\mbox{det}(M)=i\frac{k_{z}\epsilon_{1}}{\kappa_{1}\epsilon_{N+2}}\,
\end{equation}
and the reflectance of the structure can be evaluated from the matrix elements [Eq.~(\ref{eq:tran-mat-ref})].

\begin{figure}[!htb]
\begin{centering}
\includegraphics[clip,width=6cm]{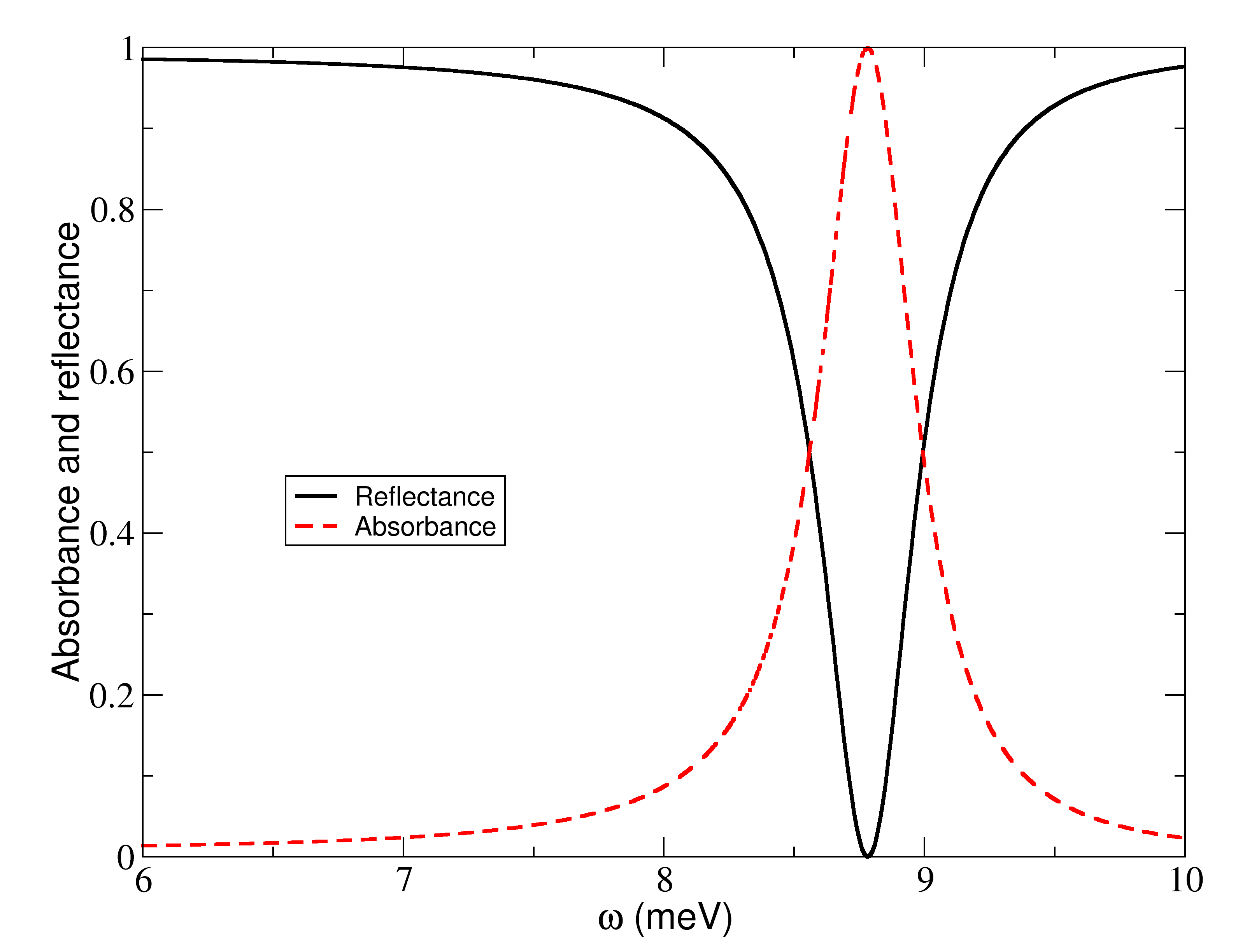}
\hspace{0.1cm} \includegraphics[clip,width=7cm]{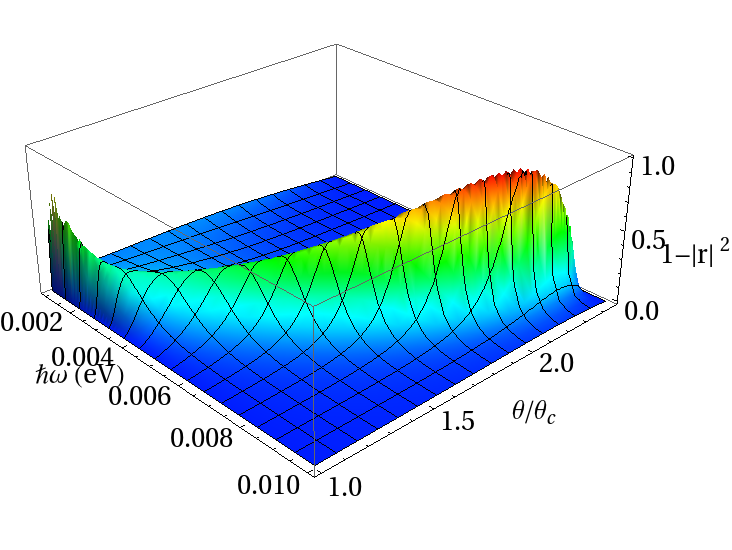}

\par\end{centering}

\caption{Top: Absorbance (dashed line) and reflectance (solid line) {\it versus} 
photon energy of double-layer graphene for $\theta=2.1\theta_{c}$. Bottom: reflectance
as function of both the energy $\hbar\omega$ and the angle of incidence,
$\theta/\theta_{c}$. The parameters are: $E_{F}=0.45$~eV, $\Gamma=0.1$~meV,
$\epsilon_{4}$=14, $\epsilon_{3}=1$, $\epsilon_{2}=\epsilon_{1}=5$,
$D_{3}=2.51$ $\mu$m, $D_{2}=D_{3}$, and $\theta_{c}=36.7^{{\rm {o}}}$.
}

\label{fig_absorvance_d_and_theta_DLayer}
\end{figure}

In Fig.~\ref{fig_absorvance_d_and_theta_DLayer} we present the reflectance
and the absorbance of double-layer graphene ($N=2$) as functions of both the
energy and the incident angle. The most notable effect is the shift
of the resonant energy from $\hbar\omega\approx4$~meV to $\hbar\omega\approx8.5$~meV (more than 100\%!). This is due to the hybridization
of the plasmon-polarion bands of the two graphene sheets when they
come closer to each other. It is possible to control the position of
the resonance by tuning the distance $D_{2}$.

\section{SPP excitation by incident wave at a metallic contact}
\label{sec_metallic contact}


\subsection{Green's function for ER scattering problem }

\label{sec_Green's function}

\begin{figure}[!htb]
\begin{centering}
\includegraphics[clip,width=7cm]{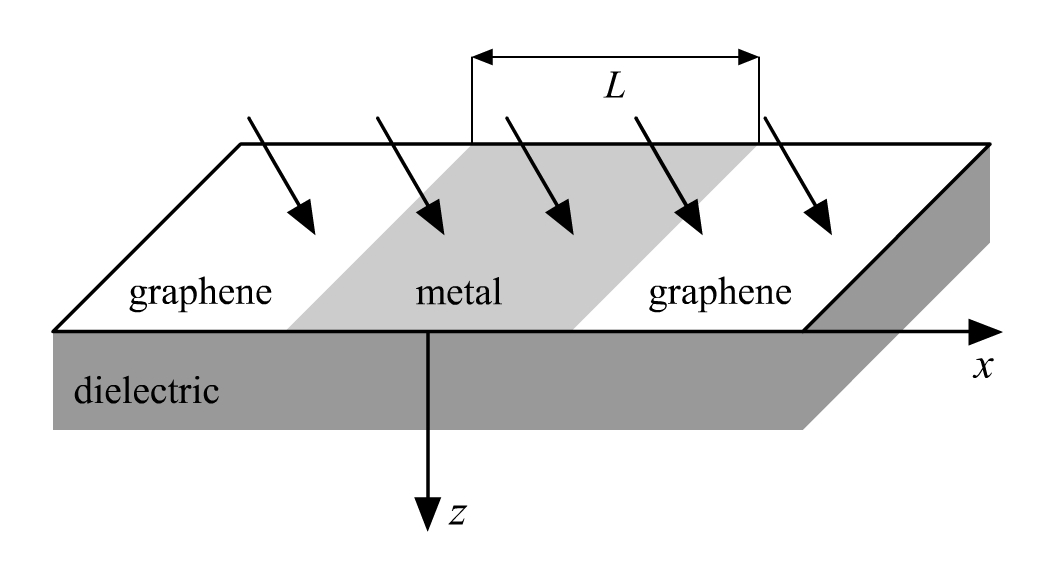}

\par\end{centering}

\caption{Graphene sheet on a dielectric substrate with a metal contact (stripe of width $L$) on top of it.
}

\label{fig_graphene_with_stripe}
\end{figure}

Here we shall consider the problem of scattering of electromagnetic radiation by a very thin metallic stripe (of width $L$) on top of a graphene sheet (see Fig. \ref{fig_graphene_with_stripe}).
It has been solved numerically by Satou and Mikhailov \cite {Satou_Mikhailov}, here we shall show how an approximate solution can be obtained analytically using Green's function method \cite {Morse}.
The scattering problem for a plane $p-$polarized wave leads to the two-dimensional Helmholtz equation \cite {Born},
\begin{equation}
\nabla _{x,z}^2 B_y + \epsilon (z) \frac {\omega ^2}{c^2} B_y=0\,,
\label{Helm}
\end{equation}
where $\nabla _{x,z}^2$ is the Laplace operator in the $xz$ plane, $\epsilon (z) =\epsilon_2$ for $z<0$ and $\epsilon (z) =\epsilon_1$ for $z>0$. Considering an incident wave coming from $z=-\infty$, the solution of Eq. (\ref{Helm}) for $z<0$ can be written as
\begin{equation}
B_y(x,z<0)=2B_i\cos (k_z z)e^{iq x}+B_{\mbox {scat}}\:.
\label{total_field-}
\end{equation}

The first term in Eq. (\ref {total_field-}) represents total reflection (as if the whole plane $z=0$ were covered with a perfect metal) and the second one (that we call scattered field) includes the contribution of all parts of the interface that are not covered by the metal (but covered with graphene). As before, here $k_z=k\cos \theta$, $q=k\sin \theta$ and $k=\omega \epsilon_2^{1/2}/c$.
The scattered field can be presented in terms of Green's function $G_{\vec{k}}(\vec{r};\vec{r}^\prime)$,
\begin{eqnarray}
\nonumber
B_{\mbox {scat}}=\int _{-\infty}^{\infty} \left \{ G_{\vec{k}}(\vec{r};x^\prime,z^\prime=0)\times \right .\\
 \left . \left [\Theta (x^\prime - L/2) + \Theta (-x^\prime -L/2)\right ]  \left.\frac {\partial B_y}{\partial z^\prime}\right\vert _{z^\prime =0^-}\right \} dx^\prime \:,
\label{Scat_field-}
\end{eqnarray}
which is given by \cite {Morse}
\begin{equation}
G_{\vec{k}}(\vec{r};\vec{r}^\prime)=\frac{i}{4} \left [ H_0^{(1)}(kR)+H_0^{(1)}(kR^\prime)  \right ]\,
\label{Green}
\end{equation}
with $R=\sqrt {(x-x^\prime)^2+(z-z^\prime)^2}$ and $R^\prime=\sqrt {(x-x^\prime)^2+(z+z^\prime)^2}$. Here $H_0^{(1)}$ denotes the Hankel function of the first kind and zero order.
The Green's function (\ref{Green}) is the solution of the equation
\begin{equation}
\nabla _{x,z}^2 G_{\vec{k}}(\vec{r};\vec{r}^\prime) + \epsilon_2 \frac {\omega ^2}{c^2} G_{\vec{k}}(\vec{r};\vec{r}^\prime)=-\delta(\vec{r}-\vec{r}^\prime)\,,
\label{Helm-Green}
\end{equation}
and takes into account the effects of a field source located at $\vec{r}^\prime=(x^\prime,z^\prime)$ and of an image source at $\vec{r}^\prime=(x^\prime,-z^\prime)$.
The Heaviside functions $\Theta(x)$ in Eq. (\ref {Scat_field-}) explicitly take into account the fact that the sources of the scattered field are located at $\vert x \vert \ge L/2$.

Similar to Eq. (\ref {total_field-}), the field in the substrate ($z>0$) can be written as
\begin{eqnarray}
\nonumber
B_y(x,z>0)=-\frac i{2}\int _{-\infty}^{\infty} \left \{ H_0^{(1)}([\epsilon_1/\epsilon_2]^{1/2}k\sqrt {(x-x^\prime)^2 + z^2}) \times \right .\\
 \left . \left [\Theta (x^\prime - L/2) + \Theta (-x^\prime -L/2)\right ] \left.\frac {\partial B_y}{\partial z^\prime} \right\vert _{z^\prime =0^+} \right \}dx^\prime \:.
\label{Scat_field+}
\end{eqnarray}


\subsection{Integral equation}
\label{sec_Integral equation}

Matching conditions at $z=0$, for $\vert x \vert \ge L/2$ read:
\begin{equation}
\frac {1}{\epsilon_2}\left. \frac {\partial B_y}{\partial z} \right\vert _{z =0^-}=\frac {1}{\epsilon_1} \left.\frac {\partial B_y}{\partial z} \right\vert _{z =0^+} \,
\label{matching1}
\end{equation}
and [see Eq. (\ref {bc-TM})]
\begin{equation}
B_y(x,0^-)-B_y(x,0^+)=\left.\frac {\sigma (\omega)}{i\varepsilon _0 \omega\epsilon_2}\frac {\partial B_y}{\partial z} \right\vert _{z =0^-} \,.
\label{matching2}
\end{equation}
Substituting Eqs. (\ref{total_field-}), (\ref{Scat_field-}), (\ref{Scat_field+}), and (\ref{matching1}) into (\ref{matching2}), we obtain:
\begin{eqnarray}
\nonumber
2i B_i e^{iq x}-\frac{1}{2\epsilon_2} \int _{\vert x^\prime \vert \ge L/2} g(x-x^\prime)\left.\frac{\partial B_y(x^\prime,z)}{\partial z} \right\vert _{z =0^-} dx^\prime \\
=\frac {\sigma (\omega)}{\varepsilon_0 \omega\epsilon_2} \left.\frac {\partial B_y(x,z)}{\partial z}\right\vert _{z =0^-}\,.
\label{matching3}
\end{eqnarray}
This is an integral equation for the magnetic field derivative (equal to $i\mu _0 \omega E_x$) on graphene, with
\begin{equation}
g(x-x^\prime)=\epsilon_2H_0^{(1)}(k\vert x-x^\prime \vert) + \epsilon_1 H_0^{(1)}([\epsilon_1/\epsilon_2]^{1/2}k\vert x-x^\prime \vert)\,,
\label{g_definition}
\end{equation}

Let us introduce an analytic function $f(x)$, such 
that it is equal to this derivative (with $\epsilon 2 B_l =1$) 
on uncovered parts of the graphene sheet,
\begin{equation}
f(x)=\left.\frac 1{\epsilon_2B_i} \frac {\partial B_y}{\partial z} \right\vert _{z =0^-} \:; \ \ \ \ \ \vert x^\prime \vert \ge L/2\,,
\label{f_definition}
\end{equation}
however, not necessarily vanishing on the metal stripe. It obeys the equation,
\begin{eqnarray}
2ie^{iq x}=\frac{\sigma (\omega)}{\varepsilon_0\omega}f(x)+\frac 12 \int _{\vert x^\prime \vert \ge L/2} g(x-x^\prime)f(x^\prime )dx^\prime \,,
\label{matching4}
\end{eqnarray}
that holds for $-\infty <x < \infty$. Multiplying Eq. (\ref{matching4}) by $\exp (iQx) $ and integrating we obtain:
\begin{eqnarray}
i 4\pi  \delta ({q +Q})=\frac{\sigma (\omega)}{\varepsilon_0\omega}\tilde f(Q)-i \mathcal{K} (Q)\int _{\vert x^\prime \vert \ge L/2} e^{iQx^\prime}f(x^\prime )dx^\prime \,,
\label{Int-eq}
\end{eqnarray}
where
$$
\tilde f(Q) = \int _{-\infty} ^{\infty } e^{iQx}f(x )dx
$$
is a Fourier transform of function $f(x)$, and
\begin{eqnarray}
\nonumber
\mathcal{K} (Q) =\frac i2\int _{-\infty} ^{\infty } e^{iQ(x-x^\prime)}g(x-x^\prime )d(x-x^\prime )=\frac {\epsilon_2}{\kappa_2(Q)}+ \frac {\epsilon_1}{\kappa_1(Q)}\,,
\label{kappa}
\end{eqnarray}
$\kappa_1(Q)$, $\kappa_2(Q)$ were defined in Eq.(\ref{eq:disp}) with replacing $q\Rightarrow Q$. The integral in the last term of Eq. (\ref{Int-eq}) can be written as
\begin{eqnarray}
\nonumber
\frac 1{2\pi} \int _{\vert x^\prime \vert \ge L/2} e^{iQx^\prime} \int _{-\infty} ^{\infty }\tilde f(Q^\prime ) e^{-iQ^\prime x^\prime} dQ^\prime dx^\prime \\
\nonumber
= \tilde f(Q) -\frac 1{\pi} \int _{-\infty} ^{\infty }\tilde f(Q^\prime ) \frac {\sin \left [(Q-Q^\prime)L/2 \right ]}{Q-Q^\prime}dQ^\prime
\,.
\end{eqnarray}
Therefore, Eq. (\ref{Int-eq}) reads as
\begin{eqnarray}
- 4\pi \delta ({q +Q})=i\frac{\sigma (\omega)}{\varepsilon_0\omega}\tilde f(Q)+\mathcal{K} (Q) \int _{-\infty} ^{\infty }\mathcal{L}({Q-Q^\prime})\tilde f(Q^\prime ) dQ^\prime \,.
\label{Int-eq2}
\end{eqnarray}
This is a Fredholm integral equation of the second kind \cite {Morse} with the kernel
\begin{equation}
\mathcal{L}({Q-Q^\prime}) =\delta({Q-Q^\prime})- \frac {\sin \left [(Q-Q^\prime)L/2 \right ]}{\pi (Q-Q^\prime)} \,.
\label{kernel}
\end{equation}

First, let us consider the trivial case of $L=0$ (no metallic stripe). Then $\mathcal{L}({Q-Q^\prime}) =\delta({Q-Q^\prime})$ and we have:
\begin{equation}
\nonumber
\tilde f_0(Q) = - \frac {4\pi\delta ({q +Q})}{\mathcal{K} (Q)+i\frac {\sigma (\omega)}{\varepsilon_0\omega}}
\end{equation}
and
\begin{equation}
f_0(x) = -\frac {2\exp ({iq x})}{\mathcal{K} (q)+i\frac{\sigma (\omega)}{\varepsilon _0\omega}} \,.
\label{f0_x}
\end{equation}
Substituting (\ref {f0_x}) into Eqs. (\ref {Scat_field-}) and (\ref {Scat_field+}), we obtain:
\begin{eqnarray}
\nonumber
B_y(x,z<0)=B_0e^{iq x}\left (e^{ik_z z}+r_y e^{-ik_z z}\right )\:;\\
B_y(x,z>0)= t_y B_ie^{i(q x-\kappa_1(q)z)}\:.
\label{total_field0}
\end{eqnarray}
with
\begin{eqnarray}
\nonumber
r_y= \frac {\mathcal{K} (q)+i\frac {\sigma (\omega)} {\epsilon _0\omega}-\frac {2\epsilon_2}{\kappa_2(q)} }{\mathcal{K} (q)+i\frac {\sigma (\omega)} {\epsilon _0\omega}}\:;\\
t_y = \frac {\epsilon_1}{\kappa_1(q)}\frac{2}{\mathcal{K} (q)+i\frac {\sigma (\omega)} {\varepsilon _0\omega}} \:,
\label{r_t}
\end{eqnarray}
Here we have taken 
into account, that $\kappa_2(q)=-ik_z$. 
We have used the following
relation for the Fourier transform of the Hankel function:
\begin{equation}
\int_{-\infty}^{\infty}dx H_0^{(1)}(k\sqrt {x^2 + z^2})\exp(-iQx)=
\frac{2\exp(i\sqrt{k^2-Q^2}x)}{\sqrt{k^2-Q^2}}.
\end{equation}

Of course, there is no real scattering in this case, just reflection, and the coefficients $r_y$ and $t_y$ could be found in a much simpler way but this simple example demonstrates how the method works.
Let us notice that, if we formally suppose that $\kappa_{1,2}(q)$ are real (respectively, $k_z$ is imaginary), the poles of the reflection and transmission coefficients (\ref {f0_x}) yield the dispersion relation of surface plasmon-polaritons in graphene. Indeed, it is easy to check that the equation
\begin{equation}
\mathcal{K} (q)+i\frac{\sigma (\omega)}{\varepsilon _0\omega}=0 \,
\label{disp_rel_100}
\end{equation}
is equivalent to Eq. (\ref{eq_W_SPP_2D}).


\subsection{Solution for $qL<<1$}
\label{sec_Solution}

\begin{figure}[!htb]
\begin{centering}
\includegraphics[clip,width=6cm]{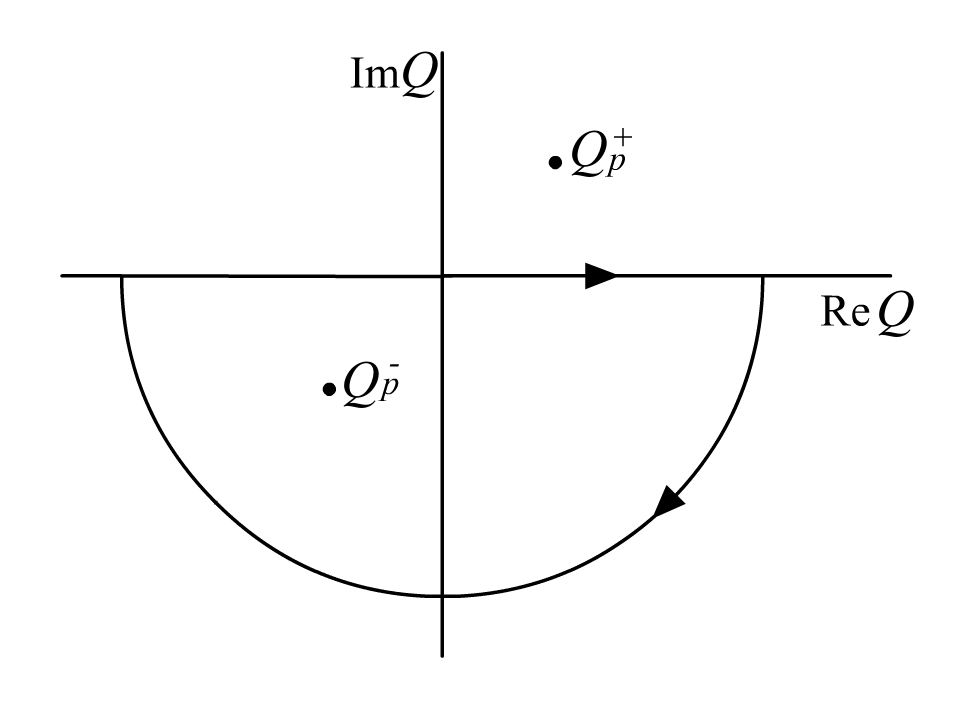}

\par\end{centering}

\caption{Contour for integral in Eq. \ref{f1_x} for $x>0$.
 For $x<0$ it has to be closed in the upper half-plane.
}

\label{contour}
\end{figure}

We shall now solve Eq. (\ref{Int-eq2}) for the case when $L$ is small, $QL<<1$, where $Q$ is any relevant wavenumber along $x$.
Then we put $\sin \left [(Q-Q^\prime)L/2 \right ]\approx (Q-Q^\prime)L/2$ and
\begin{equation}
\mathcal{L}({Q-Q^\prime}) =\delta({Q-Q^\prime})- \frac L{2\pi} \:.
\label{kernel1}
\end{equation}
By noting that
$$
\int _{-\infty} ^{\infty } \tilde f(Q) dq =2\pi f(Q)=const \:,
$$
Eq. (\ref{Int-eq2}) is solved to give
\begin{equation}
\nonumber
\tilde f(Q) =\tilde f_0(Q) + f(0)\frac {\mathcal{K} (Q)L}{\mathcal{K} (Q)+i\frac{\sigma (\omega)}{\varepsilon _0\omega}}
\end{equation}
and
\begin{equation}
f(x) =f_0(x) + f(0)\frac {L}{2\pi}\int _{-\infty} ^{\infty } \frac {\mathcal{K} (Q)}{\mathcal{K} (Q)+i\frac{\sigma (\omega)}{\varepsilon _0 \omega}}e^{-iQx}dQ\:,
\label{f1_x}
\end{equation}
where $f_0(x)$ is given by Eq. (\ref{f0_x}). The integral in Eq. (\ref{f1_x}) converges, since $\mathcal{K} (Q)\sim Q^{-1}$ for large $Q$, and can be calculated using Jordan's lemma and the contour shown in Fig. \ref {contour} (for $x>0$).
We denote by $Q_p^-$ the pole that lies in the lower half-plane and note that it is one of the two roots of Eq. (\ref{disp_rel_100}). In other words, $Q_p^-$ and $Q_p^+=-Q_p^-$ are two possible values of the SPP wavevector for a given frequency $\omega $. Calculation of the integral yields
\begin{eqnarray}
\frac 1{2\pi}\int _{-\infty} ^{\infty }\frac {\mathcal{K} (Q)}{\mathcal{K} (Q)+i\frac{\sigma (\omega)}{\varepsilon _0 \omega}}e^{-iQx}dQ =i\frac {\mathcal{K} (Q_p^-)}{\left.\frac{\partial\mathcal{K} (Q)}{\partial Q}\right|_{Q=Q_p^-}}e^{-iQ_p^-x}=\nonumber
\\-\frac{\sigma (\omega)}{\varepsilon _0 \omega}\frac {\kappa_1^3(Q_p)\kappa_2^3(Q_p)}{Q_p^-\left [\epsilon_2 \kappa_1^3(Q_p)+\epsilon_1 \kappa_2^3(Q_p) \right ]} e^{-iQ_p^- x}\:,\nonumber
\end{eqnarray}
and we obtain:
\begin{eqnarray}
f(x) =f_0(x) + f(0) L \mathcal{Q}e^{iQ_p^+ x}\:;\\
\mathcal{Q}=\frac{\sigma (\omega)}{\varepsilon _0 \omega}\frac {\kappa_1^3(Q_p)\kappa_2^3(Q_p)}{Q_p^+\left [\epsilon_2 \kappa_1^3(Q_p)+\epsilon_1 \kappa_2^3(Q_p) \right ]}\:.
\label{f1_x_2}
\end{eqnarray}
By requiring self-consistency of Eq. (\ref {f1_x_2}), we obtain 
$$
f(0)=-\frac{1}{1-iL\mathcal{Q}}\frac {2}{\mathcal{K} (q)+\frac{\sigma (\omega)}{\varepsilon _0\omega}} 
$$ and the final result is:
\begin{eqnarray}
f(x) =-\frac {2}{\mathcal{K} (q)+i\frac{\sigma (\omega)}{\varepsilon _0\omega}}
\left [\exp ({iq x}) + \frac {L\mathcal{Q}} {1-L\mathcal{Q}}e^{iQ_p^+ x}\right ] \:.
\label{f1_x_3}
\end{eqnarray}
\begin{figure}[!htb]
\begin{centering}
\includegraphics[clip,width=9cm]{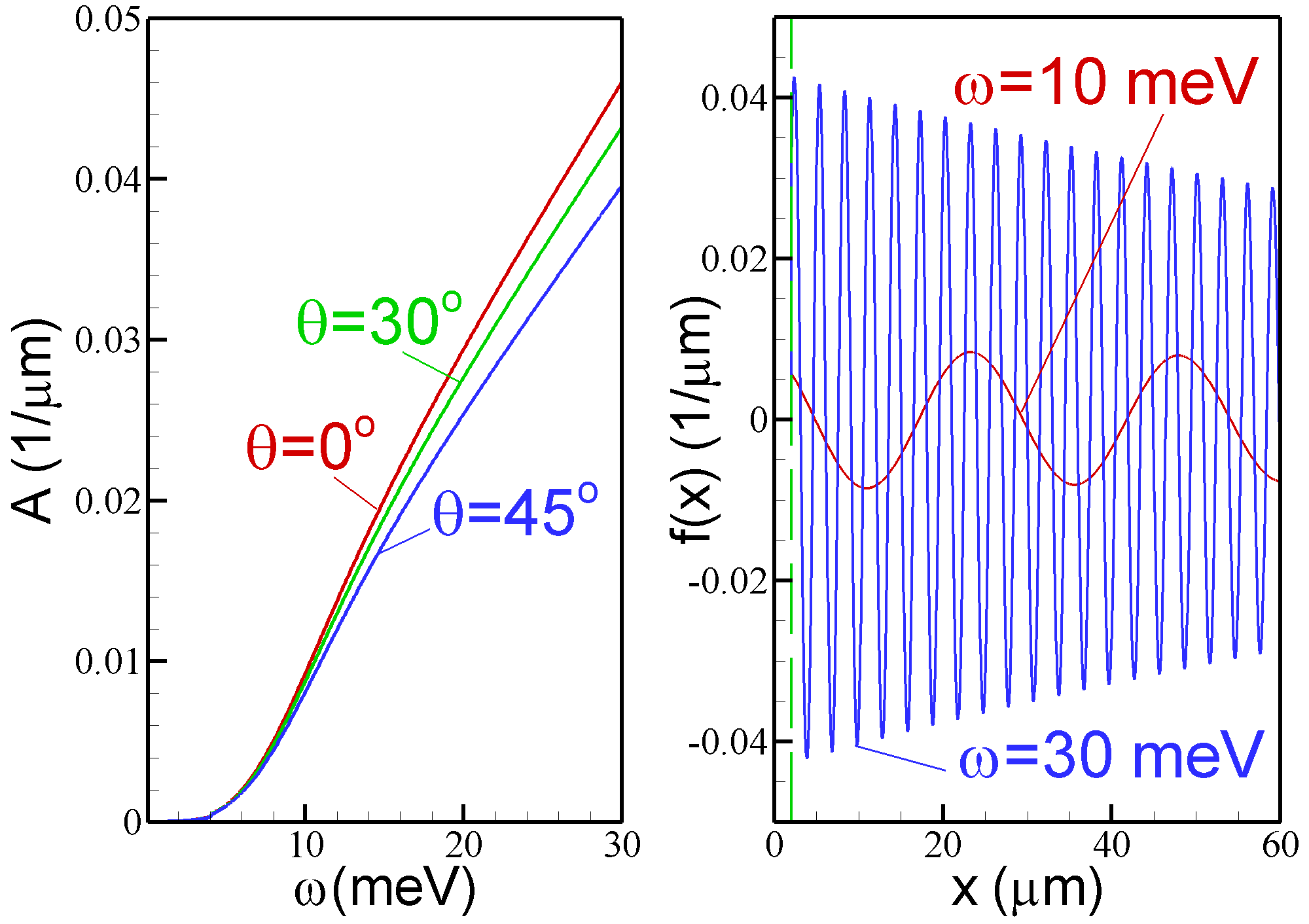}

\par\end{centering}

\caption{SPP excitation by incident ER at metallic stripe.
Left: Amplitude $A$ of the second term of the function $f(x)$ 
[Eq. (\ref{f1_x_3})] {\it versus} frequency $\omega$ for different
 angles of incidence as indicated. Right: Spatial profile of the generated SPP wave for two
 different frequencies indicated on the plot. The parameters are: $E_F=0.45$ eV,
 $\hbar\Gamma=0.1$ meV,  $\epsilon_2=1$, $\epsilon_1=5$, and $L=4$ $\mu$m.
}
\label{fig_fita}
\end{figure}

Equation (\ref{f1_x_3}) is valid for $x>0$, while for 
$x<0$ we have $Q_p^+ \rightarrow Q_p^-$.
The second term of this equation represents the SPP wave excited by shining ER radiation at the metallic contact. The two waves described by Eq. (\ref{f1_x_3}) can be referred to as forced and intrinsic oscillations, respectively, of the free 2D electron gas in graphene. The metallic stripe is a topological defect imposed on graphene and, as mentioned in the Introduction, can be used to launch SPP waves in this two-dimensional conductor. The amplitude of the SPP oscillations is proportional to the width of the contact ("defect power") and depends on the ER frequency as shown in Fig. Fig.~\ref{fig_fita}. Let us remind that the solution (\ref{f1_x_3}) is valid in the limit $\vert Q_p \vert L << 1$. As shown numerically by Satou and Mikhailov \cite {Satou_Mikhailov}, for larger $\vert Q_p \vert L $ the SPP amplitude becomes to oscillate as a function of frequency or contact width. The generated polaritons decay with the distance from the contact-stripe antenna because $Q_p$ is complex (see Fig.~\ref{fig_fita}). The scattered electromagnetic fields above and below the interface can be found by substituting (\ref{f1_x_3}) into Eqs. (\ref {Scat_field-}) and (\ref {Scat_field+}), leading to a rather complex dependence on $x$ and $z$ (or, in other words, on the scattering angle and distance from the contact).

\section{ER coupling to graphene with periodically modulated conductivity}
\label{sec_rippled}
\subsection{Polaritonic crystal}

So far, we have been considering graphene as a perfectly flat
membrane with a homogeneous optical conductivity. Now we shall relax
the latter assumption still keeping the former (i.e. graphene's flatness). There are different ways of implementing a position dependent conductivity in graphene: (i) patterning graphene micro-ribbons in an otherwise
homogeneous graphene sheet, (ii) inducing an inhomogeneous strain profile
in graphene, (iii) producing an inhomogeneous profile of adsorbed atoms or molecules on graphene sheet, and (iv) using patterned gates.
The last three mechanisms produce a position dependent
electronic density leading to a position dependent conductivity of graphene because of
the dependence of the Fermi energy on the density.

\begin{figure}[!htb]
\begin{centering}
\includegraphics[clip,width=7cm]{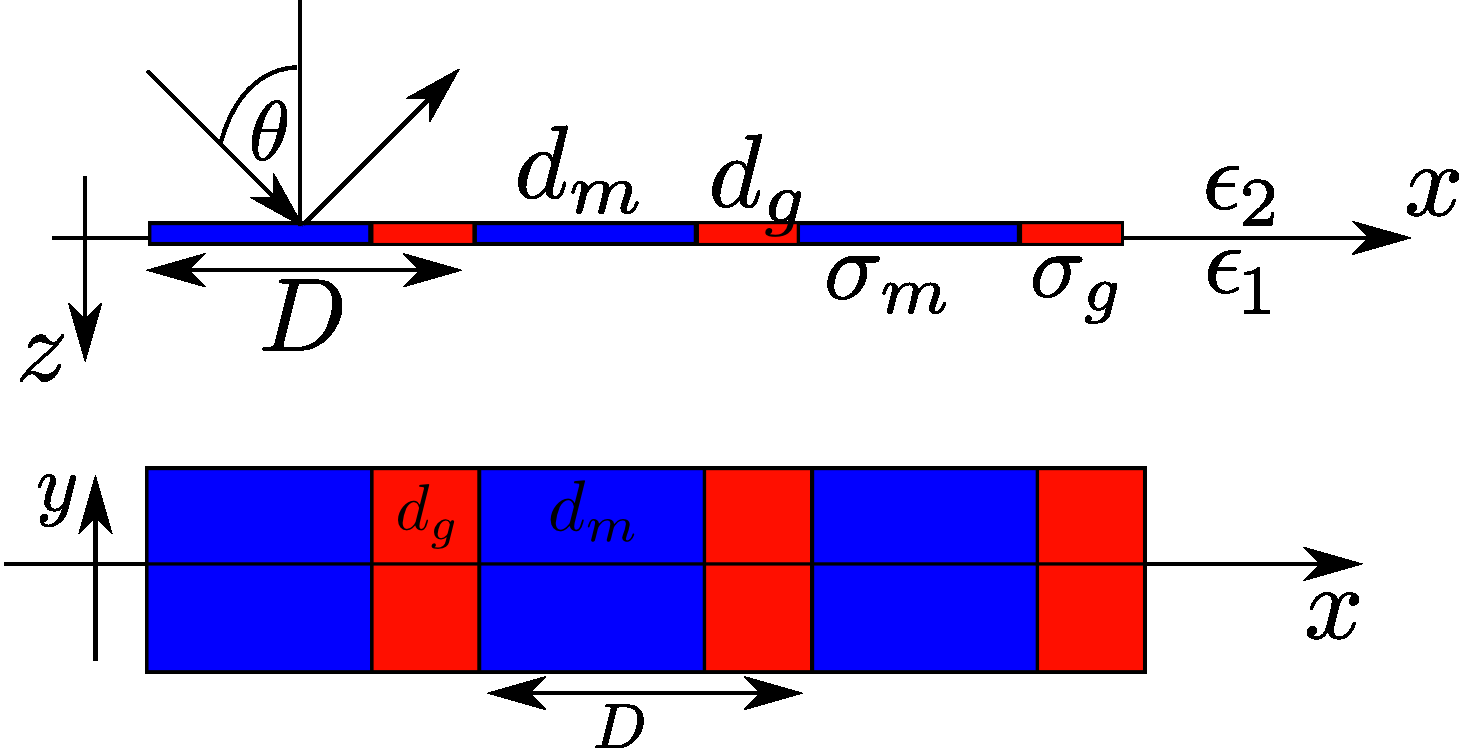} 

\par\end{centering}

\caption{System with a periodically modulated conductivity period $D$ (side and top views). }

\label{fig_periodic_cond}
\end{figure}

We assume a periodically modulated conductivity, $\sigma(x+D)=\sigma(x)$,
and introduce the one-dimensional (1D) reciprocal lattice wave vector, $G=2\pi/D$. The
system we have in mind is represented in Fig.~\ref{fig_periodic_cond}. It can be seen as an 1D polaritonic crystal \cite {nunoSPP,YuliyPRB}.

Since the system is periodic, the fields can be written in the
form of Fourier--Floquet series (in other words, they obey the Bloch theorem). In Fig. \ref{fig_periodic_cond} we have represented
an array of graphene micro-ribbons, of conductivity $\sigma_{g}$,
separated by regions of conductivity $\sigma_{m}$, which can be made
of either of a dielectric or a metal. However, we note that the formalism
we develop below applies to any profile of modulated conductivity.
Thus, Fig.~\ref{fig_periodic_cond} should be understood as a schematic
representation of one of the aforementioned possibilities.


\subsection{Formalism and results}

\label{sec_rippled_polaritonic}

First, we want to find the band structure of the TM-type surface plasmon--polaritons for a system of Fig.~\ref{fig_periodic_cond}. If it is composed of periodic
patches of graphene, the approach presented below fails because
it requires an infinite number of reciprocal lattice vectors for achieving
convergence. (The scattering problem we study later is, however, well defined.)

As before, we assume that the system supports surface waves propagating
along the graphene sheet and decaying away from it. We consider $p$-polarized
waves written in the form of Fourier--Floquet series,
\begin{eqnarray}
B_{m,y}(x,z) & = & \sum_{n}{\cal B}_{m,y||n}e^{i(q+nG)x}e^{-\kappa_{m||n}\vert z\vert}\,;\label{eq:}\\
E_{m,x}(x,z) & = & \sum_{n}{\cal E}_{m,x||n}e^{i(q+nG)x}e^{-\kappa_{m||n}\vert z\vert}\,;\label{eq:-1}\\
E_{m,z}(x,z) & = & \sum_{n}{\cal E}_{m,z||n}e^{i(q+nG)x}e^{-\kappa_{m||n}\vert z\vert}\,,
\end{eqnarray}
where the index $m$ labels the media above ($m=1$) and below ($m=2$) the $z=0$ plane,
and the sum runs over all integers. Since the system is
linear, each Fourier component is independent of the others. Then,
it follows from Maxwell's equations that
\begin{eqnarray}
\kappa_{m||n} & = & \sqrt{(q+nG)^{2}-\omega^{2}\epsilon_{m}/c^{2}}\,,\label{eq:lat-kappa}\\
{\cal B}_{m,y||n} & = & (-1)^{m}\frac{i\omega\epsilon_{m}}{c^{2}\kappa_{m||n}}{\cal E}_{m,x||n}\,,\label{eq:lat-byex}\\
{\cal E}_{m,z||n} & = & (-1)^{m+1}i\frac{q+nG}{\kappa_{m||n}}{\cal E}_{m,x||n}\,.\label{eq:lat-ezex}
\end{eqnarray}
Since the conductivity is periodic, it can be expanded in Fourier series
as
\begin{equation}
\sigma(x)=\sum_{l}e^{ilGx}\widetilde{\sigma}_{l}\,,
\end{equation}
where
\begin{equation}
\widetilde{\sigma}_{l}=\frac{1}{D}\int_{0}^{D}\sigma(x)e^{-ilGx}dx\,.
\end{equation}
The boundary conditions $E_{1,x}(x,0)-E_{2,x}(x,0)=0$, $B_{1,y}(x,0)-B_{2,y}(x,0)=-\mu_{0}\sigma(x)E_{1,x}(x,0)$
read:
\begin{eqnarray}
\sum_{n}({\cal E}_{1,x||n}-{\cal E}_{2,x||n})e^{inGx}=0\,; \label{eq:lat-bcond-ex}\\
\sum_{n}({\cal B}_{1,y||n}-{\cal B}_{2,y||n})e^{inGx}=-\mu_{0}\sum_{l,p}\widetilde{\sigma}_{l}{\cal E}_{1,x||p}e^{i(l+p)Gx}\,,\label{eq:lat-bcond-hy}
\end{eqnarray}
or, alernatively ($l+p=n\Leftrightarrow l=n-p$),
\begin{equation}
\sum_{n}({\cal B}_{1,y||n}-{\cal B}_{2,y||n})e^{inGx}=-\mu_{0}\sum_{n,p}\widetilde{\sigma}_{n-p}{\cal E}_{1,x||p}e^{inGx}\,.
\end{equation}
As a result, Eqs.~(\ref{eq:lat-bcond-ex})-(\ref{eq:lat-bcond-hy})
imply that
\begin{eqnarray}
{\cal E}_{1,x||n}-{\cal E}_{2,x||n}=0\,\\
{\cal B}_{1,y||n}-{\cal B}_{2,y||n}+\mu_{0}\sum_{p}\widetilde{\sigma}_{n-p}{\cal E}_{1,x||p}=0\,.
\end{eqnarray}
After some algebra, we obtain a non-linear eigenvalue problem,
\begin{eqnarray}
\frac{\epsilon_{1}}{\kappa_{1||n}}{\cal E}_{1,x||n}+\frac{\epsilon_{2}}{\kappa_{2||n}}{\cal E}_{1,x||n}+\frac{i}{\omega\epsilon_{0}}\sum_{p}\widetilde{\sigma}_{n-p}{\cal E}_{1,x||p}=0\,,
\label{eq_non_linear_eigenvalue}
\end{eqnarray}
for the frequency $\omega$.
If we consider only the Fourier component $p=n$ in the sum in Eq.(\ref{eq_non_linear_eigenvalue}), we obtain
\begin{equation}
\frac{\epsilon_{1}}{\kappa_{1||n}}+\frac{\epsilon_{2}}{\kappa_{2||n}}+\frac{i}{\omega\epsilon_{0}}\widetilde{\sigma}_{0}=0\,,
\end{equation}
which is just the equation for the SPP frequency of the wavevector $q+nG$ {[}compare with Eq.~(\ref{eq_W_SPP_2D}){]}.
The quantity $\widetilde{\sigma}_{0}$ is the average of the conductivity
in the primitive cell,
\begin{equation}
\widetilde{\sigma}_{0}=\frac{1}{D}\int_{0}^{D}\sigma(x)dx\,.
\end{equation}
The presence of harmonics with other Fourier components of $\sigma(x)$,
$\widetilde{\sigma}_{l}$ with $l\ne0$ in Eq.~(\ref{eq_non_linear_eigenvalue})
gives rise to the band-gap structure of the polaritonic spectrum, equivalent to the so called empty lattice
approximation of electrons in a periodic potential \cite {Ziman}.
It can be presented either in the extended scheme if we consider the SPP wavevector varying from $-\infty $ to $\infty $, or in the reduced scheme if we limit the wavevector to the first Brillouin zone, $-\frac \pi D \leq q <\frac \pi D$. 
In the latter case, folding the dispersion curve into the first 
Brillouin zone produces upper branches, i.e. the 
polaritonic crystal structure for the SPP frequencies.

We note that the non-linear eigenvalue problem can be transformed into
a linear one if we make the non-retarded approximation. As discussed
in Sec.~\ref{sec_TM_spectrum}, we do not expect to obtain an accurate
solution for the band at wave vectors close to zero. In the non-retarded
approximation we have $\kappa_{m||n}\approx\vert q+nG\vert$. Using Eq.~(\ref{eq_sigma_approx}) and writing
the spatial-dependent conductivity as
\begin{equation}
\sigma(x)=i\frac{\nu}{\omega}s(x)\,,
\end{equation}
the Fourier transform of $\sigma(x)$ can be written as
\begin{equation}
\widetilde{\sigma}_{l}=i\frac{\nu}{\omega}{\cal S}_{l}\,,
\label{eq:sigma_Fourier}
\end{equation}
where ${\cal S}_{l}$ is the $l$-th Fourier component of $s(x)$.
This allows to write Eq.~(\ref{eq_non_linear_eigenvalue}) as
\begin{equation}
\frac{\epsilon_{1}+\epsilon_{2}}{\vert q+nG\vert}{\cal E}_{1,x||n}-\frac{\nu}{\omega^{2}\epsilon_{0}}\sum_{p}{\cal S}_{n-p}{\cal E}_{1,x||p}=0\,,
\end{equation}
or as
\begin{equation}
\frac{\nu\vert q+nG\vert}{2\bar{\epsilon}\epsilon_{0}}\sum_{p}{\cal S}_{n-p}{\cal E}_{1,x||p}=\omega^{2}{\cal E}_{1,x||n}\,,\label{eq_linear_BS_polaritonic}
\end{equation}
which has the standard form of a linear eigenvalue problem. However, the band
structure derived from such procedure is not quantitatively accurate.

For obtaining the spectrum in the vicinity of $q=G/2=\pi/D$, we need to
take into account field harmonics ${\cal E}_{1,x||n}$ with $n=-1,0$
only (which correspond to the lowest gap in the spectrum). In this
case, Eq.~(\ref{eq_linear_BS_polaritonic}) reduces to two equations
only:
\begin{eqnarray}
\frac{\nu G}{4\bar{\epsilon}\epsilon_{0}}\left({\cal S}_{0}{\cal E}_{1,x||-1}+{\cal S}_{-1}{\cal E}_{1,x||0}\right) & = & \omega^{2}{\cal E}_{1,x||-1}\,,\\
\frac{\nu G}{4\bar{\epsilon}\epsilon_{0}}\left({\cal S}_{1}{\cal E}_{1,x||-1}+{\cal S}_{0}{\cal E}_{1,x||0}\right) & = & \omega^{2}{\cal E}_{1,x||0}\,,
\end{eqnarray}
which form a $2\times2$ eigenvalue problem. The eigenvalues are
\begin{equation}
\omega^{2}=\frac{\nu G}{4\bar{\epsilon}\epsilon_{0}}\left({\cal S}_{0}\pm\sqrt{{\cal S}_{1}{\cal S}_{-1}}\right)=\frac{\alpha E_{F}cG}{\bar{\epsilon}\hbar}\left({\cal S}_{0}\pm\sqrt{{\cal S}_{1}{\cal S}_{-1}}\right)\,.
\label{eq_eigen_linear}
\end{equation}
Equation (\ref{eq_eigen_linear}) describes the form of the spectrum in the vicinity of the gap at the Brillouin zone edge. As expected,
the value of the gap depends on the Fourier components $\widetilde{\sigma}_{\pm1}$
of the conductivity. The approximated description of the spectrum that
we have presented has to be checked against a full numerical calculation.
We have verified that for the case of an array of micro-ribbons this
simplified description fails because a large number of reciprocal
lattice vectors are necessary to describe the spectrum accurately.
In fact, this particular case is poorly convergent. This is a consequence
of the non-continuous nature of graphene in the micro-ribbon structure.

For further progress we need a model for $\sigma(x)$. We assume a
conductivity profile of the form
\begin{equation}
\sigma(x)=\sigma_{D}s(x)=\sigma_{D}[1-h\cos(2\pi x/D)]\,,\label{eq:lat-sig-cos}
\end{equation}
where $\sigma_{D}$ is given by Eq.~(\ref{eq_sigma_xx_Drude}). The
Fourier transform of $s(x)$ reads:
\begin{eqnarray}
{\cal S}_{0} & = & 1\,,\label{eq:lat-sig-cos-h0}\\
{\cal S}_{l} & = & -\frac{h}{2}(\delta_{l,1}+\delta_{l,-1})\,.\label{eq:lat-sig-cos-h1}
\end{eqnarray}
For this form of $s(x)$ the eigenvalue problem (\ref{eq_linear_BS_polaritonic})
has a fast convergence.
\begin{figure}[!htb]
\begin{centering}
\includegraphics[clip,width=8cm]{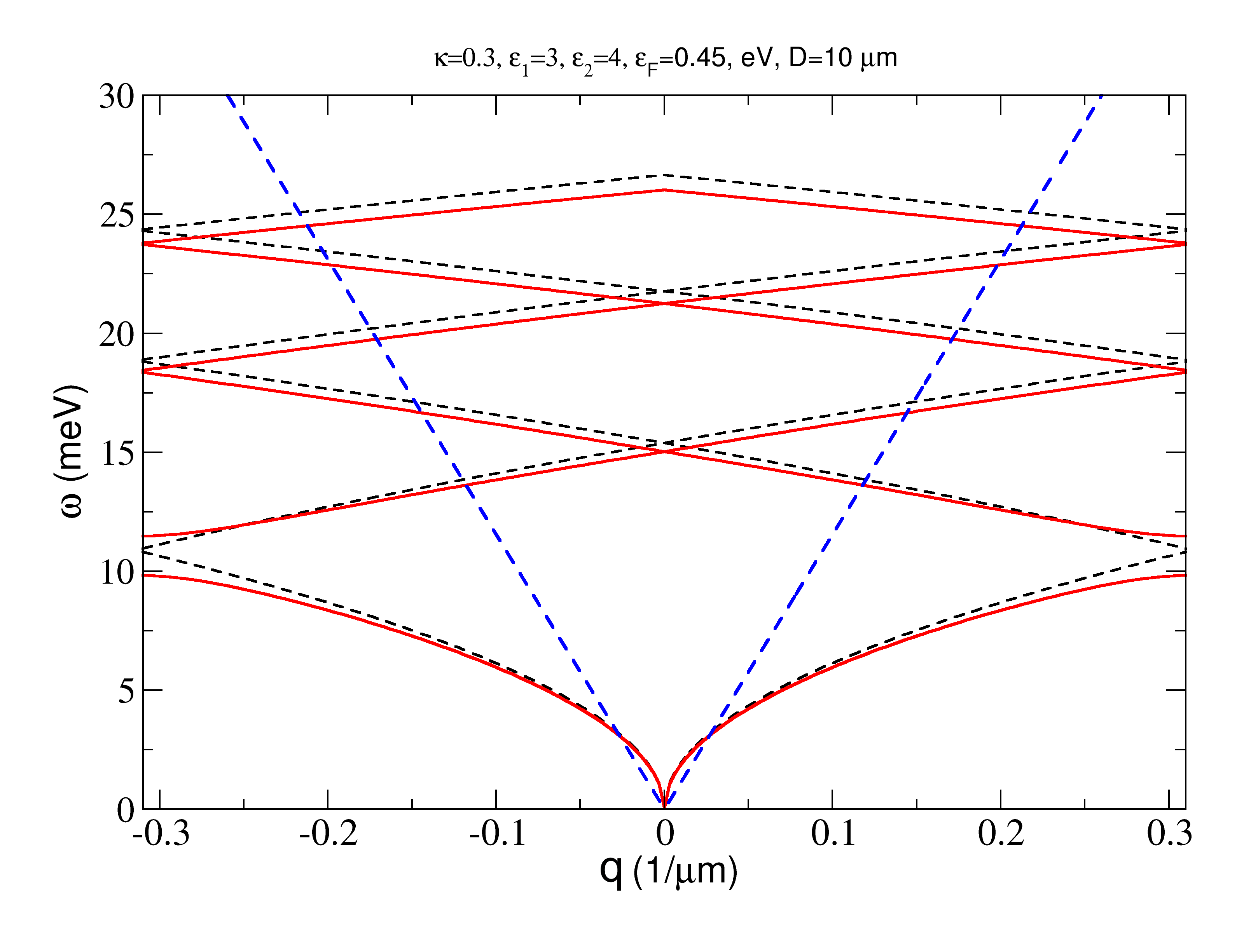} 
\vspace {0.2cm}
\includegraphics[clip,width=7.5cm]{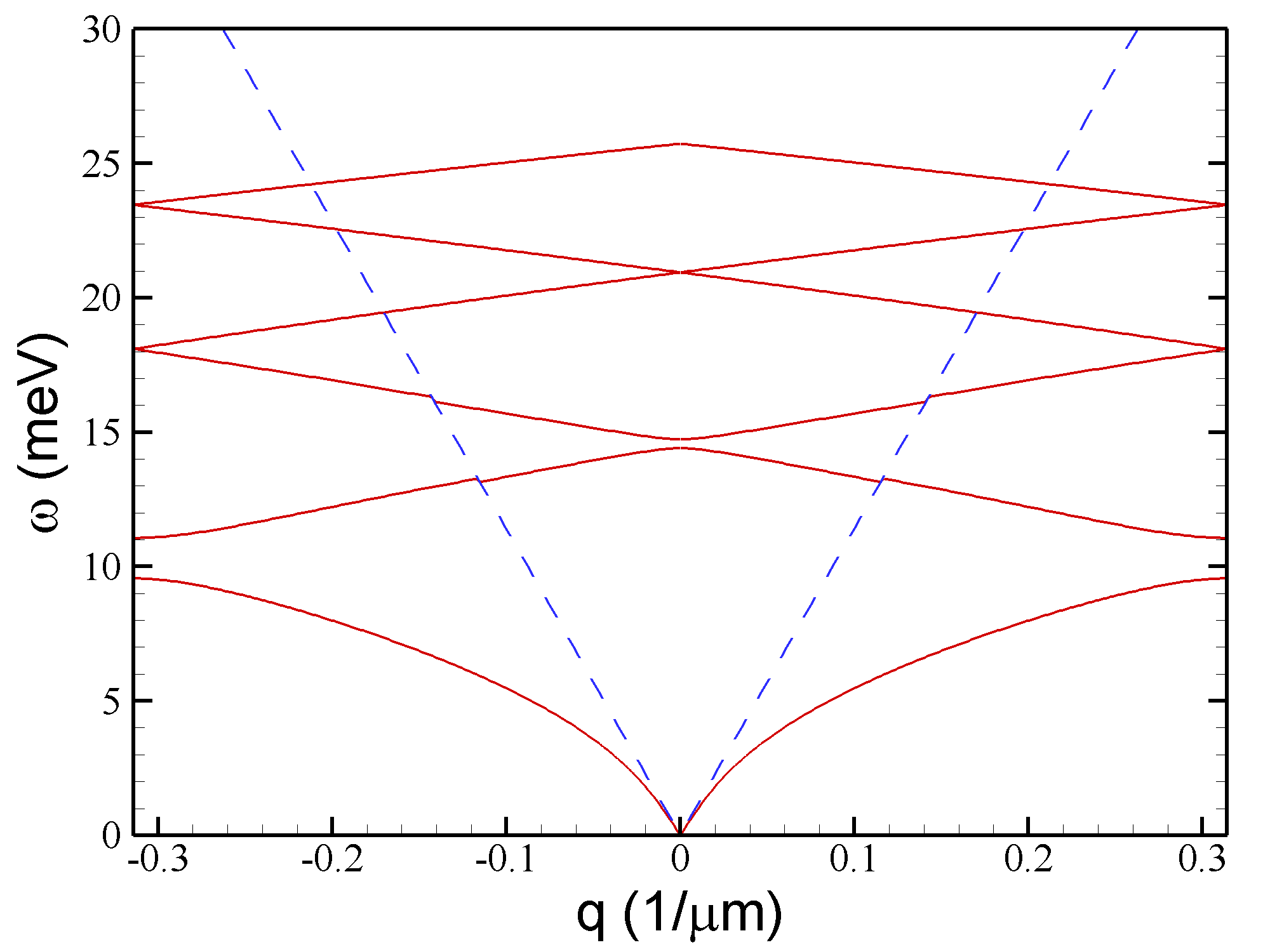} 
\par\end{centering}

\caption{Band structure of the polaritonic crystal obtained from the solution
of the linear eigenvalue problem (\ref{eq_linear_BS_polaritonic}), i.e. in the retarded approximation (upper panel) and by directly solving Eq. (\ref{eq_non_linear_eigenvalue}) (lower panel). Only
the first six bands are shown. 
The dashed curves in the upper panel correspond to the 
"empty lattice" approximation. Notice the anti-crossing of 
the bands $n=\pm 1$  in the vicinity of $q=0$ (near 15 mev). The parameters are $E_{F}=0.45$~eV,
$D=10$~$\mu$m, $\epsilon_{1}=3$, $\epsilon_{2}=4$, and $h=0.3$.
}

\label{fig_polaritonic_crystal}
\end{figure}

In the top plot of Fig.~\ref{fig_polaritonic_crystal} we present the band structure
of the polaritonic crystal produced by the conductivity modulation, computed
using the non-retarded approximation. The dashed black lines represent
the folding of the bare SPP dispersion curve of homogeneous graphene into the first
Brillouin zone. The solid lines represent the spectrum when the conductivity
is modulated. Clearly, there is a large gap between the first and the
second bands and smaller gaps between the other upper bands. The blue
dashed straight lines represent the light cone, $\hbar ck/\sqrt{\epsilon_{1}}$.
The states of the upper bands located within the cone can be excited by
shining light on the flat graphene sheet without the aid of a prism.

As it can be observed in Fig.~\ref{fig_polaritonic_crystal}, the band structure for is inaccurate $k\approx0$, the dispersion curve for the lowest band lies inside the light cone and the curves representing the upper branches do not show the expected anti-crossing.
These are artefacts produced by the non-retarded approximation used.
The band structure calculated taking into account the retardation effect \cite {YuliyPRB} is
free from these artefacts (see the lower plot of  Fig.~\ref{fig_polaritonic_crystal}).

\subsection{Scattering by a polaritonic crystal}
\label{sec_rippled_scatt}

\subsubsection{Formalism}

Having studied the emergence of the polaritonic band structure in a
system with periodically modulated conductivity, we now want to discuss the scattering
of light from a structure of the type represented in Fig.~\ref{fig_periodic_cond}.
We consider TM-polarized waves in the form
\begin{eqnarray}
B_{m,y}(x,z) & = & \delta_{m,2}B_{y}^{(i)}e^{iqx}e^{ik_{z}z}+\sum_{n}{\cal B}_{m,y||n}e^{i(q+nG)x}e^{-\kappa_{m||n}\vert z\vert}\,,\label{eq:lat-by}\\
E_{m,x}(x,z) & = & \delta_{m,2}E_{x}^{(i)}e^{iqx}e^{ik_{z}z}+\sum_{n}{\cal E}_{m,x||n}e^{i(q+nG)x}e^{-\kappa_{m||n}\vert z\vert}\,,\label{eq:lat-ex}\\
E_{m,z}(x,z) & = & \delta_{m,2}E_{z}^{(i)}e^{iqx}e^{ik_{z}z}+\sum_{n}{\cal E}_{m,z||n}e^{i(q+nG)x}e^{-\kappa_{m||n}\vert z\vert}\,.\label{eq:lat-ez}
\end{eqnarray}
where $q=k\sin\theta$, $k_z=k\cos\theta$, $k=\sqrt{\epsilon_{2}}(\omega/c)$.
The relations between the amplitudes of the reflected and transmitted
fields, ${\cal B}_{m,y||n}$, ${\cal E}_{m,x||n}$, and ${\cal E}_{m,z||n}$
are described by Eqs.(\ref{eq:lat-byex}) and (\ref{eq:lat-ezex}),
while those between the amplitudes of the incoming field are the same
as Eqs.~(\ref{eq:byex3}) and (\ref{eq:ezex3}), with $\epsilon_{3}$
being replaced by $\epsilon_{2}$.

Notice that in Eqs.~(\ref{eq:lat-by}), (\ref{eq:lat-ex}) and (\ref{eq:lat-ez})
the ratio $(q+nG)/k$ can be interpreted as the sinus of the scattering angle of the
Fourier mode $n$, that is,
\begin{equation}
e^{i(q+nG)x}=e^{ik\sin\theta_{2||n}x}\,,
\end{equation}
leading to
\begin{equation}
\theta_{2||n}=\arcsin[(q+nG)/k]\,.\label{eq:thetar-n-I}
\end{equation}
As a result, the usual condition for the Bragg scattering reads as
\begin{eqnarray}
\nonumber
k\sin\theta_{2||n}=k\sin\theta+n2\pi/D<k\:;\\
D(\sin\theta_{2||n}-\sin\theta)=n\lambda\,,
\end{eqnarray}
where $\lambda=2\pi/k$ is the wavelength of light in the top dielectric
with $\epsilon_{2}$. In fact, $(q+nG)^{2}-[\kappa_{2||n}]^{2}=k^{2}$
for both propagating and evanescent waves, but for diffraction orders ($n$) corresponding to propagating waves
\begin{equation}
\kappa_{2||n}=-ik\cos\theta_{2||n}.\label{eq:thetar-n-II}
\end{equation}
In a similar manner it is possible to introduce the scattering angles for transmitted waves,
$\theta_{1||n}$
\begin{eqnarray}
q+nG=(\omega/c)\epsilon_{1}^{1/2}\sin\theta_{1||n}\,,\label{eq:thetat-n-I}\\
\kappa_{1||n}=-i(\omega/c)\epsilon_{1}^{1/2}\cos\theta_{1||n}.\label{eq:thetat-n-II}
\end{eqnarray}
The boundary conditions imply:
\begin{eqnarray}
{\cal E}_{1,x||0} & = & {\cal E}_{2,x||0}+E_{x}^{(i)}\,,\\
{\cal E}_{1,x||n} & = & {\cal E}_{2,x||n}\hspace{0.3cm}\wedge\hspace{0.3cm}n\ne0\,,
\end{eqnarray}
and
\begin{eqnarray}
{\cal B}_{1,y||0}-{\cal B}_{2,y||0}-B_{y}^{(i)}=-\mu_{0}\sum_{p}\widetilde{\sigma}_{-p}{\cal E}_{1,x||p}\,,\\
{\cal B}_{1,y||n}-{\cal B}_{2,y||n}=-\mu_{0}\sum_{p}\widetilde{\sigma}_{n-p}{\cal E}_{1,x||p}\hspace{0.3cm}\wedge\hspace{0.3cm}n\ne0\,.
\end{eqnarray}
Using the relations between the fields, Eqs.~(\ref{eq:byex3}) and (\ref{eq:lat-byex}),
the set of boundary conditions reduces to
\begin{eqnarray}
\left(\frac{\epsilon_{1}}{\kappa_{1||0}}+\frac{\epsilon_{2}}{\kappa_{2||0}}\right){\cal E}_{1,x||0}+\frac{i}{\omega\epsilon_{0}}\sum_{p}\widetilde{\sigma}_{-p}{\cal E}_{1,x||p} & = & \frac{2i\epsilon_{2}}{k_{z}}E_{x}^{(i)}\,,\label{eq_linear_a}\\
\left(\frac{\epsilon_{1}}{\kappa_{1||n}}+\frac{\epsilon_{2}}{\kappa_{2||n}}\right){\cal E}_{1,x||n}+\frac{i}{\omega\epsilon_{0}}\sum_{p}\widetilde{\sigma}_{n-p}{\cal E}_{1,x||p} & = & 0\hspace{0.3cm}\wedge\hspace{0.3cm}n\ne0\,.\label{eq_linear_b}
\end{eqnarray}
We recall that
\begin{equation}
\kappa_{m||n}=\sqrt{(k\sin\theta+nG)^{2}-\omega^{2}\epsilon_{m}/c^{2}}
\end{equation}
for non-negative arguments of the square root, otherwise $\kappa_{m||n}$
is written as
\begin{equation}
\kappa_{m||n}=-i\sqrt{\omega^{2}\epsilon_{m}/c^{2}-(k\sin\theta+nG)^{2}}\,.
\end{equation}
The last equation imposes $\kappa_{1||0}=-ik_{z}$. The choice for
the sign of the square root is dictated by physical reasons: we must
have reflected waves for $z<0$ and transmitted waves for $z>0$ (so called Rayleigh conditions).

\subsubsection{Reflectance and transmittance efficiencies}

Let us compute the fraction of the ER energy carried by the different
diffracted orders. The time average of the Poynting vector, $\langle\vec{S}\rangle=\Re(\vec{E}\times\vec{H}^{\ast})/2$,
is given by (recall Sec.~\ref{sec_TE_TM})
\begin{equation}
\langle\vec{S}\rangle=\frac{1}{2\mu_{0}}\Re(-\hat{x}E_{z}B_{y}^{\ast}+\hat{y}E_{x}B_{y}^{\ast})\,.
\end{equation}
Thus, for diffraction order $n$ we have
\begin{eqnarray}
S_{z}^{(i)} & = & \frac{\omega\epsilon_{2}}{2\mu_{0}c^{2}k_{z}}\left|E_{x}^{(i)}\right|^{2}\,,\\
S_{2,z||n} & = & -\frac{\omega\epsilon_{2}}{2\mu_{0}c^{2}|\kappa_{2||n}|}\left|E_{2,x||n}\right|^{2}\,,\\
S_{1,z||n} & = & \frac{\omega\epsilon_{1}}{2\mu_{0}c^{2}|\kappa_{1||n}|}\left|E_{1,x||n}\right|^{2}
\end{eqnarray}
for the incident, reflected, and transmitted power per unit area (along
the $\hat{z}$ direction), respectively. Here we have assumed
that $\kappa_{m||n}$ is pure imaginary; if it is real, then the corresponding
order $n$ carries no energy due to evanescent character
of the corresponding wave. The negative sign of the reflected
power corresponds to the wave propagation in the negative direction along the
axis $\hat{z}$. Finally, the reflectance and transmittance efficiencies
are
\begin{eqnarray}
{\cal R}_{n} & = & -\frac{S_{2,z||n}}{S_{z}^{(i)}}=\frac{k_{z}}{|\kappa_{2||n}|}\left|r_{x||n}\right|^{2}\,,\\
{\cal T}_{n} & = & \frac{S_{1,z||n}}{S_{z}^{(i)}}=\frac{\epsilon_{1}k_{z}}{\epsilon_{2}|\kappa_{1||n}|}\left|t_{x||n}\right|^{2}\,,
\end{eqnarray}
respectively. The diffuse reflectance and diffuse transmittance
amplitudes of the order $n$ entering these relations are defined as
\begin{eqnarray}
r_{x||n}=\frac{{\cal E}_{2,x||n}}{E_{x}^{(i)}}\,,\\
t_{x||n}=\frac{{\cal E}_{1,x||n}}{E_{x}^{(i)}}\,.
\end{eqnarray}

The specular reflectance and transmittance ($n=0$ mode) are given by
\begin{equation}
{\cal R}_{0}=\vert r_{x||0}\vert^{2}\,,
\label{eq:specular_reflectance}
\end{equation}
and
\begin{equation}
{\cal T}_{0}=\left\vert t_{x||0}\right\vert ^{2}\frac{\epsilon_{1}}{\epsilon_{2}}\frac{\cos\theta}{\sqrt{\epsilon_{1}/\epsilon_{2}-\sin^{2}\theta}}\,.\label{eq:lat-t0}
\end{equation}
The last expression is valid for $\epsilon_{1}>\epsilon_{2}$ or for
$\theta<\theta_{c}=\arcsin(\sqrt{\epsilon_{1}/\epsilon_{2}})$, with
$\theta_{c}$ denoting the critical angle for total reflection at
the interface.

In general, for $n\ne0$, using the scattering angles for diffuse reflection and transmision modes, Eqs.
(\ref{eq:thetar-n-II}), (\ref{eq:thetat-n-II}), the diffuse reflectance
efficiency is
\begin{equation}
{\cal R}_{n}=\frac{\cos\theta}{\cos\theta_{2||n}}\left\vert r_{x||n}\right\vert ^{2}\,,
\label{eq:lat-rn}
\end{equation}
and the diffuse transmittance efficiency is given by
\begin{equation}
{\cal T}_{n}=\sqrt{\frac{\epsilon_{1}}{\epsilon_{2}}}\frac{\cos\theta}{\cos\theta_{1||n}}\left\vert t_{x||n}\right\vert ^{2}\,.
\label{eq:lat-tn}
\end{equation}
It should be noticed that Eqs.~(\ref{eq:lat-rn})
and (\ref{eq:lat-tn}) are valid for positive arguments of the square
root only (i.e. for propagating waves) and for real dielectric permitivitties
$\epsilon_{1}$ and $\epsilon_{2}$.  The absorbance is defined as
${\cal A}=1-\sum_{n}({\cal R}_{n}+{\cal T}_{n})$, where the sum over
$n$ is restricted to the diffraction orders corresponding to propagating waves only.

\subsubsection{Two special cases}
\label{subsec_scatt_special}

Let us consider a special case of vanishing conductivity ("graphene is absent").
Then Eqs.~(\ref{eq_linear_a}) and (\ref{eq_linear_b}) give
${\cal E}_{1,x||n}=0$ ($n\ne0$) (no diffuse reflectance and transmittance in this case) and
\begin{equation}
\frac{{\cal E}_{1,x||0}}{E_{x}^{(i)}}=2\left(1+\sqrt{\frac{\epsilon_{1}}{\epsilon_{2}}}\frac{\cos\theta}{\sqrt{1-\epsilon_{2}\sin^{2}\theta/\epsilon_{1}}}\right)^{-1}\,.
\end{equation}
The reflection coefficient is given by
\begin{equation}
 r_{x||0}=\frac{{\cal E}_{1,x||0}}{E_{x}^{(i)}}-1= \frac{\sqrt{1-\epsilon_{2}\sin^{2}\theta/\epsilon_{1}}-\sqrt{\epsilon_{1}/\epsilon_{2}}\cos\theta}{\sqrt{1-\epsilon_{2}\sin^{2}\theta/\epsilon_{1}}+\sqrt{\epsilon_{1}/\epsilon_{2}}\cos\theta}\,,\label{eq:lat-ref0-simp}
\end{equation}
which reproduces the well known result from elementary optics.

Another particular case is obtained when $d_{g}=D$. In this case,
only the $m=0$ Fourier component of $\sigma(x)$ survives. Thus,
we obtain:
\begin{equation}
\frac{{\cal E}_{1,x||0}}{E_{x}^{(i)}}\left(\frac{1}{\cos\theta}+\frac{\epsilon_{1}/\epsilon_{2}}{\sqrt{\epsilon_{1}/\epsilon_{2}-\sin^{2}\theta}}+\frac{\sigma_{D}}{\sqrt{\epsilon_{2}}\epsilon_{0}c}\right)=\frac{2}{\cos\theta}\,.\label{eq:lat-r0-simp}
\end{equation}
Again, using Eq. (\ref{eq:lat-t0}) and putting $\epsilon_{1}=\epsilon_{2}=1$
we obtain the well known result for the transmittance of free-standing graphene (for a TM wave),
\begin{equation}
{\cal T}_{0}=\left\vert \frac{2}{2+\sigma_{D}\cos\theta/\epsilon_{0}c}\right\vert ^{2}\,.\label{eq:lat-t0-simp}
\end{equation}

\subsubsection{Scattering by a grid of graphene micro-ribbons}

\label{subsec_scatt_grid}

After checking two special cases of the general expressions, we now
want to address an example where the full set of Eqs. (\ref{eq:specular_reflectance}) - (\ref{eq:lat-tn}) has to be
used. This particular problem was considered in Ref. \cite{Nikitin1}.
For an array of graphene ribbons, the Fourier coefficients in Eq. (\ref{eq:sigma_Fourier}) are given by ${\cal S}_{0}=d_{g}/D$ and
\begin{equation}
{\cal S}_{l}=\frac{1}{l\pi }\sin\frac{l\pi d_{g}}{D}e^{-il\pi d_{g}/D}\:;\ \ \ \ \ \ \ \ \ l\neq 0\,.
\end{equation}
\begin{figure}[!htb]
\begin{centering}
\includegraphics[clip,width=8cm]{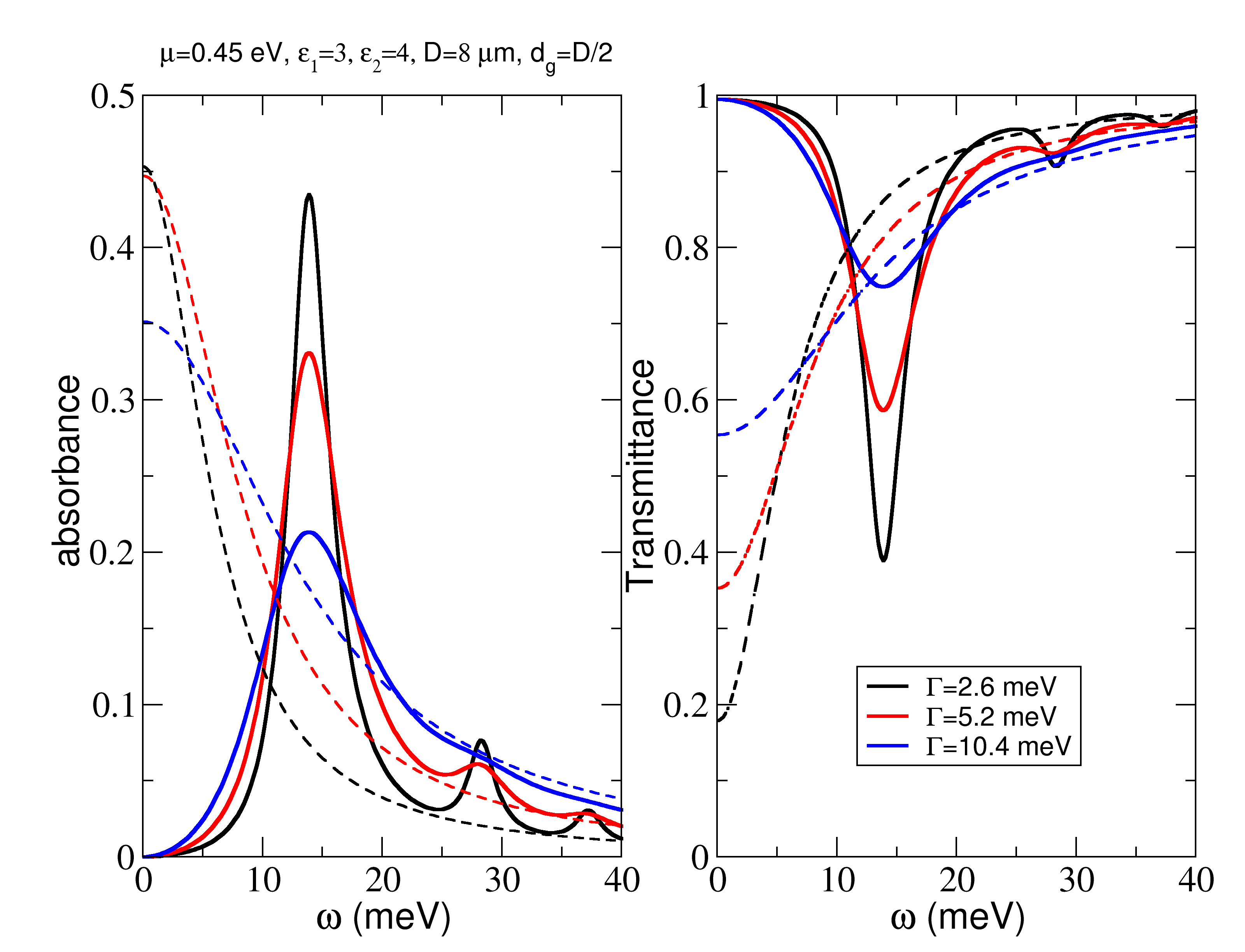} 

\par\end{centering}

\caption{Absorbance (left) and transmittance (right) of a grid of graphene
ribbons for different values of the broadening $\Gamma$. The dashed
lines are the absorbance/transmittance curves for pristine graphene
for the same $\Gamma$ values and chemical potential $E_{F}$. The parameters are: $E_{F}=0.45$~eV, $D=8$~$\mu$m, $d_{g}=D/2$,
$\epsilon_{1}=4$, and $\epsilon_{2}=3$.}

\label{fig_grid_absorb_gamma}
\end{figure}

In Fig.~\ref{fig_grid_absorb_gamma} we represent the absorbance
and the transmittance of a grid of graphene ribbons for different
values of the broadening $\Gamma$. This calculation required 600 reciprocal lattice vectors.
The same quantities for uniform (pristine)
graphene are represented by dashed curves. For small values of $\Gamma$,
both the absorbance and the transmittance spectra present a set of
resonances, with the most prominent one at low energies. On the other
hand, for large $\Gamma$, the weaker resonances are washed away and
the most prominent one becomes less pronounced. Compared to the case
of pristine graphene, the absorption is suppressed at low energies
whereas the transmittance is increased. The energy of the largest
resonance is approximately given by $\hbar\Omega_{p}(\pi/d_{g})$, suggesting a coupling of the impinging ER to the SPPs modes of graphene.
Unfortunately, the analysis is not so simple, because each ribbon
has a finite size and the electron confinement has to be taken into account.
The behaviour of the transmittance at low energies should also be noted. In
the  case of grid it tends to unity whereas in the pristine graphene case it attains
its lowest values. The grid transmits more at low energies although
the size of the gaps between the ribbons is much smaller than the wavelength of the incoming
radiation.

\begin{figure}[!htb]
\begin{centering}
\includegraphics[clip,width=8cm]{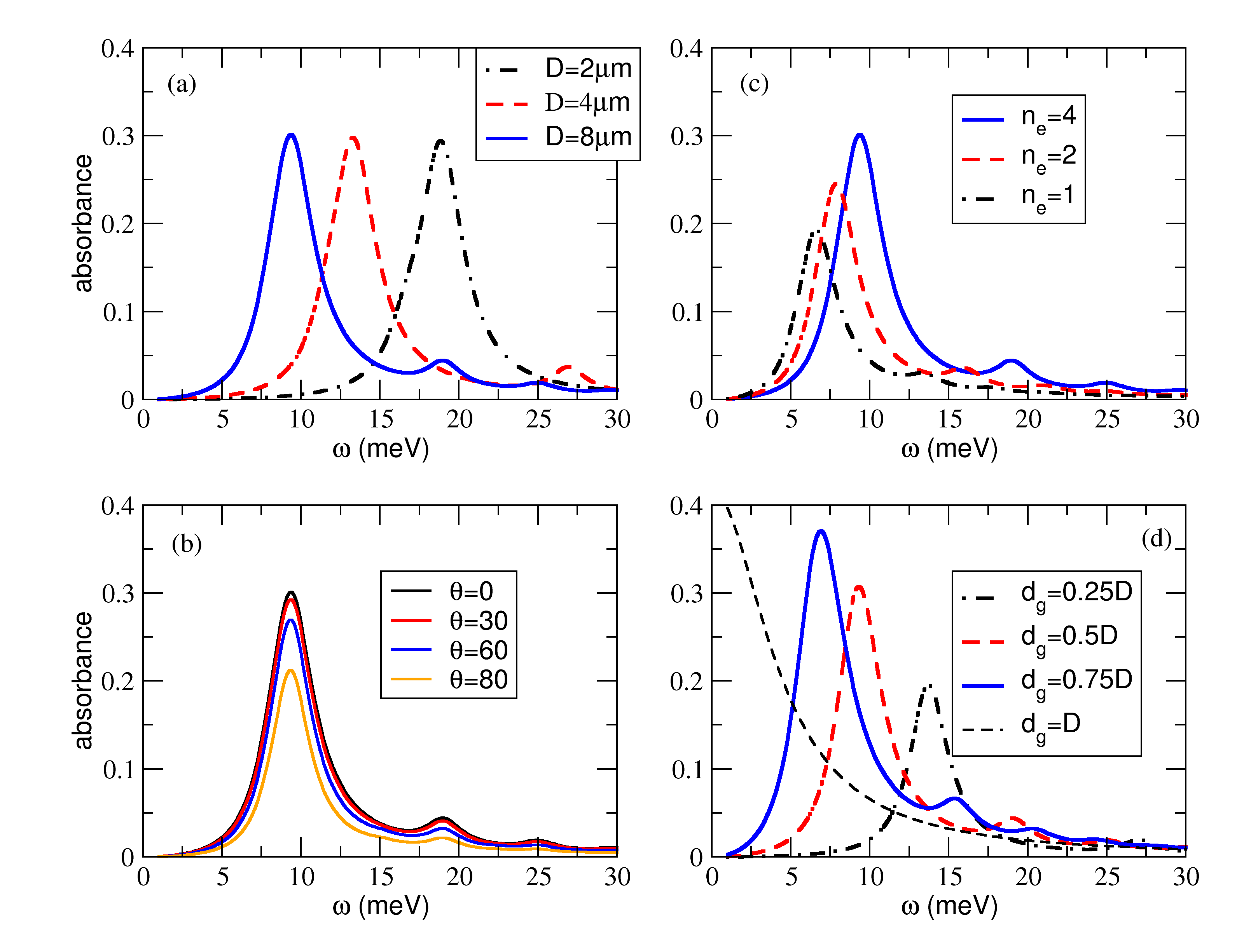} 

\par\end{centering}

\caption{Absorbance of a grid of graphene ribbons: (a) dependence on the lattice
parameter $D$; (b) dependence on the angle of incidence; (c) dependence
on the electronic density; (d) dependence on the width of the graphene
ribbon (keeping the width of the unit cell constant). The parameters for the reference curve
(solid blue) are: $E_{F}=0.23$~eV, $D=8$~$\mu$m, $\epsilon_{1}=5$,
$\epsilon_{2}=3$, and $\Gamma=2.6$~meV. In panel (c) $n_{e}$ is
given in units of $10^{12}$ cm$^{-2}$.}

\label{fig_absorb_grid}
\end{figure}

\begin{figure}[!htb]
\begin{centering}
\includegraphics[clip,width=8cm]{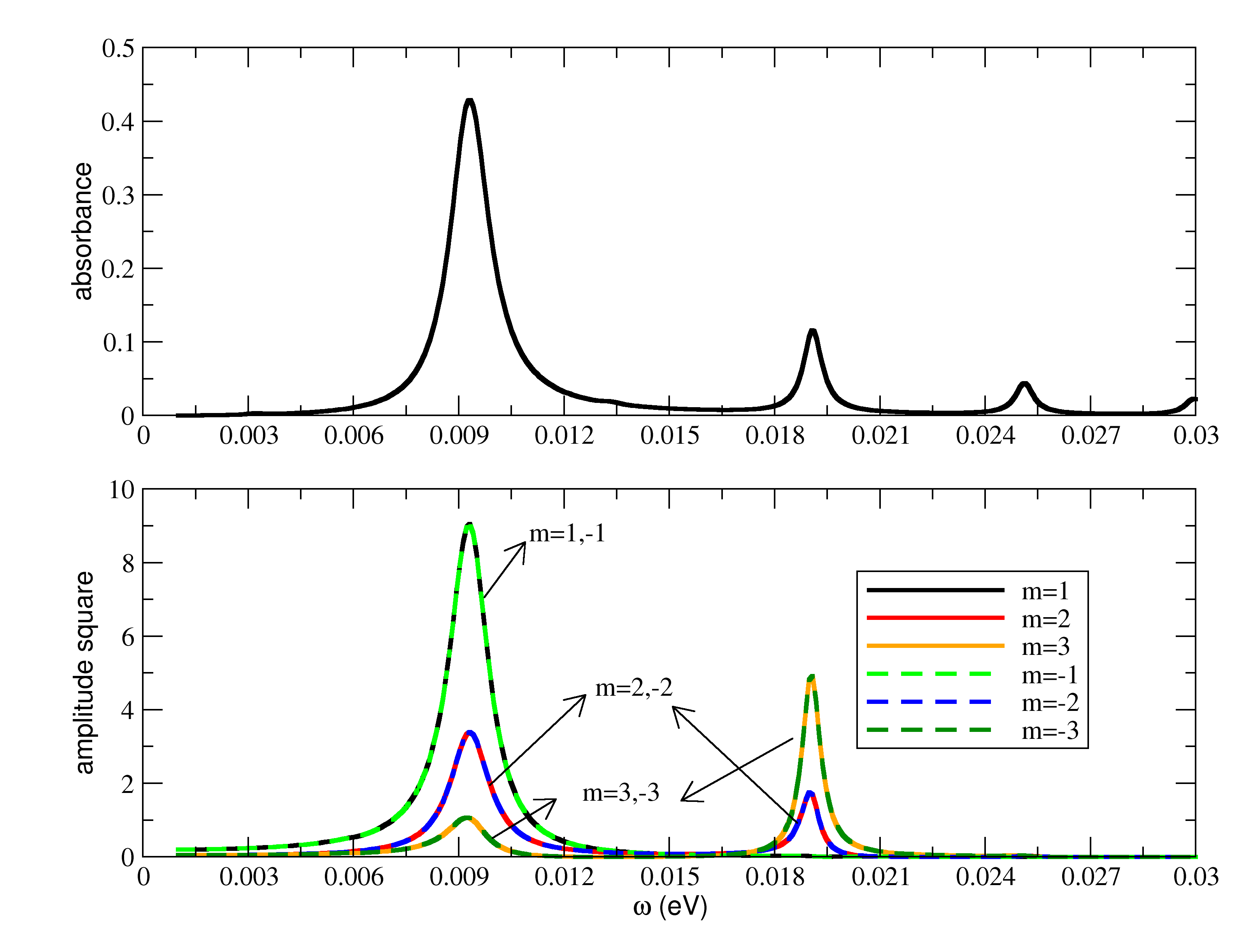} 

\par\end{centering}

\caption{Contribution of the first three harmonics to the absorbance curve.
The reference curve (top panel) has $\Gamma=0.6$~meV. Other parameters
as in Fig.~\ref{fig_absorb_grid}. In the bottom panel,
the quantity $\vert{\cal E}_{1,x||n}/E_{x}^{(i)}\vert^{2}$ is depicted.}

\label{fig_absorb_grid_Gmodes}
\end{figure}

In Fig.~\ref{fig_absorb_grid} we study the effect of different parameters of the problem on the absorbance curve. In panel
(a) we find the variation of absorbance with the parameter $D$, keeping
$d_{g}=D/2$. The shift of the absorbance maximum scales as
\begin{equation}
\omega_{\mbox{max}}\propto\sqrt{\frac{1}{D}}\,.\label{eq_realtion_1}
\end{equation}
In panel (b), the dependence of the absorbance on the angle of incidence,
$\theta$, is given. No appreciable effect is seen here, except close to
grazing incidence. We will see below that in the case of a continuous sheet
with a modulated conductivity the situation is quite different. In
panel (c), the dependence on the electronic density is shown (the values on the plot are given in units of $10^{12}$ cm$^{-2}$). The red shift scales with $n_{e}$ as
\begin{equation}
\omega_{\mbox{max}}\propto(n_{e})^{1/4}\,.\label{eq_realtion_2}
\end{equation}
Finally, in panel (d) we present the dependence of the absorbance
on $d_{g}$, keeping $D$ constant and equal to 8~$\mu$m. The
blue shift follows the scaling relation
\begin{equation}
\omega_{\mbox{max}}\propto\sqrt{\frac{1}{d_{g}}}\,.\label{eq_realtion_1b}
\end{equation}
In this latter panel the absorbance of an infinite graphene
sheet is also plotted. Clearly, there is an enhancement of absorption due to SPP
around certain specific frequencies, where the absorbance is higher for the grid of ribbons than for an infinite graphene sheet. Relations
(\ref{eq_realtion_1}), (\ref{eq_realtion_2}), and (\ref{eq_realtion_1b})
have the same functional form of dependence on $d_{g}$ and $n_{e}$ as the SPP frequency in the continuous system for the wavenumber $q=\pi/d_{g}$. Considering
the case of Fig.~\ref{fig_absorb_grid_Gmodes}, we predict a maximum at the frequency
\begin{equation}
\hbar\Omega_{p}\approx\sqrt{\frac{4\alpha}{\epsilon_{1}+\epsilon_{2}}E_{F}\hbar cq}=\sqrt{\frac{4}{137\times8}0.23\times0.2\frac{\pi}{8}}\simeq\mbox{8.1 meV}\,,
\end{equation}
a value close to the position of the first maximum in the absorbance
spectrum.

In order to understand the origin of the second maximum (and the
third, as well) seen in the absorbance spectrum, we plotted
in Fig.~\ref{fig_absorb_grid_Gmodes} the squared absolute value of the
amplitude ${\cal E}_{1,x||n}/E_{x}^{(i)}$ for $n=\pm1,\pm2$ and $\pm3$.
Clearly, the spectral weight associated with different harmonics is
centered at the maxima observed in the absorbance spectrum. This representation
gives a qualitative understanding of how the spectral weight associated
with the different SPP modes in the infinite sheet is redistributed to form
the absorbance spectrum in the periodic system. Different harmonics
contribute differently to these maxima, for example, none of
the depicted harmonics contribute to the maximum observed above $24$~meV, only those with $|n|\ge4$ contribute to it.


\subsubsection{Scattering from graphene sheet with cosine-modulated conductivity}

\label{subsec_scatt_cosine}

We next consider an example where the graphene sheet is continuous
(as opposed to the grid of ribbons) and possesses a periodically modulated optical conductivity.
As in Sec.~\ref{sec_rippled_polaritonic}, we assume a conductivity
profile of the form (\ref{eq:lat-sig-cos}), with the Fourier harmonics given by Eqs.~(\ref{eq:lat-sig-cos-h0}) and (\ref{eq:lat-sig-cos-h1}).
We shall see that, for angles of incidence different from zero, the
absorbance peak will split into two, associated with the lattice vectors
$G=\pm2\pi/D$. We give below a derivation for the energy of the peaks
as function of the angle of incidence. In the non-retarded approximation,
the energy of the plasmon-polaritons is given by Eq.~(\ref{eq_PP_spectrum_Otto}),
with $\bar{\epsilon}=(\epsilon_{1}+\epsilon_{2})/2$. For an ER wave incoming
at an angle of incidence $\theta$, the wavevector of the excited
SPP is given by
\begin{equation}
q=\vert k\sin\theta+nG\vert=\vert\frac{\omega\sqrt{\epsilon_{1}}}{c}\sin\theta-mG\vert\,.
\end{equation}
From Eq. (\ref {eq_W_SPP_2D_numerical}) for the SPP dispersion relation it follows that

\begin{equation}
q_{p}=\vert\frac{\omega\sqrt{\epsilon_{2}}}{c}\sin\theta+nG\vert=\frac{(\hbar\omega)^{2}}{4\alpha}\frac{\epsilon_{1}+\epsilon_{2}}{E_{F}\hbar c}\,.
\end{equation}
We assume that $G>\frac{\omega\sqrt{\epsilon_{2}}}{c}$. There are two possibilities:
\begin{enumerate}
\item $n=-1$

\begin{equation}
\frac{(\hbar\omega)^{2}}{4\alpha}\frac{\epsilon_{1}+\epsilon_{2}}{E_{F}\sqrt{\epsilon_{2}}}+\hbar\omega\sin\theta-G\hbar c/\sqrt{\epsilon_{2}}=0\,;
\end{equation}

\item $n=1$

\begin{equation}
\frac{(\hbar\omega)^{2}}{4\alpha}\frac{\epsilon_{1}+
\epsilon_{2}}{E_{F}\sqrt{\epsilon_{2}}}-
\hbar\omega\sin\theta-G\hbar c/\sqrt{\epsilon_{2}}=0\,.
\end{equation}

\end{enumerate}

Introducing the parameters
\begin{eqnarray}
a & = & \frac{1}{4\alpha E_{F}}\frac{\epsilon_{1}+\epsilon_{2}}{\sqrt{\epsilon_{2}}}\,,
\end{eqnarray}
and
\begin{eqnarray}
b & = & G\frac{c\hbar}{\sqrt{\epsilon_{2}}}\,,
\end{eqnarray}
the solution of the two equations for $\omega(\theta)$ reads as

\begin{eqnarray}
\hbar\omega & = & \pm\frac{\sin\theta}{2a}+\frac{1}{2a}
\sqrt{\sin^{2}\theta+4ab}\,,\label{eq_branch_minus_plus}
\end{eqnarray}
for $n=1$ and $n=-1$, respectively. Thus, we expect to observe a
peak splitting for $\theta>0$ at the left ($n=-1$) and at the right
($n=1$) of the single peak at $\theta=0$. We shall see that this
is indeed the case.

\begin{figure}[!htb]
\begin{centering}
\includegraphics[clip,width=9cm]{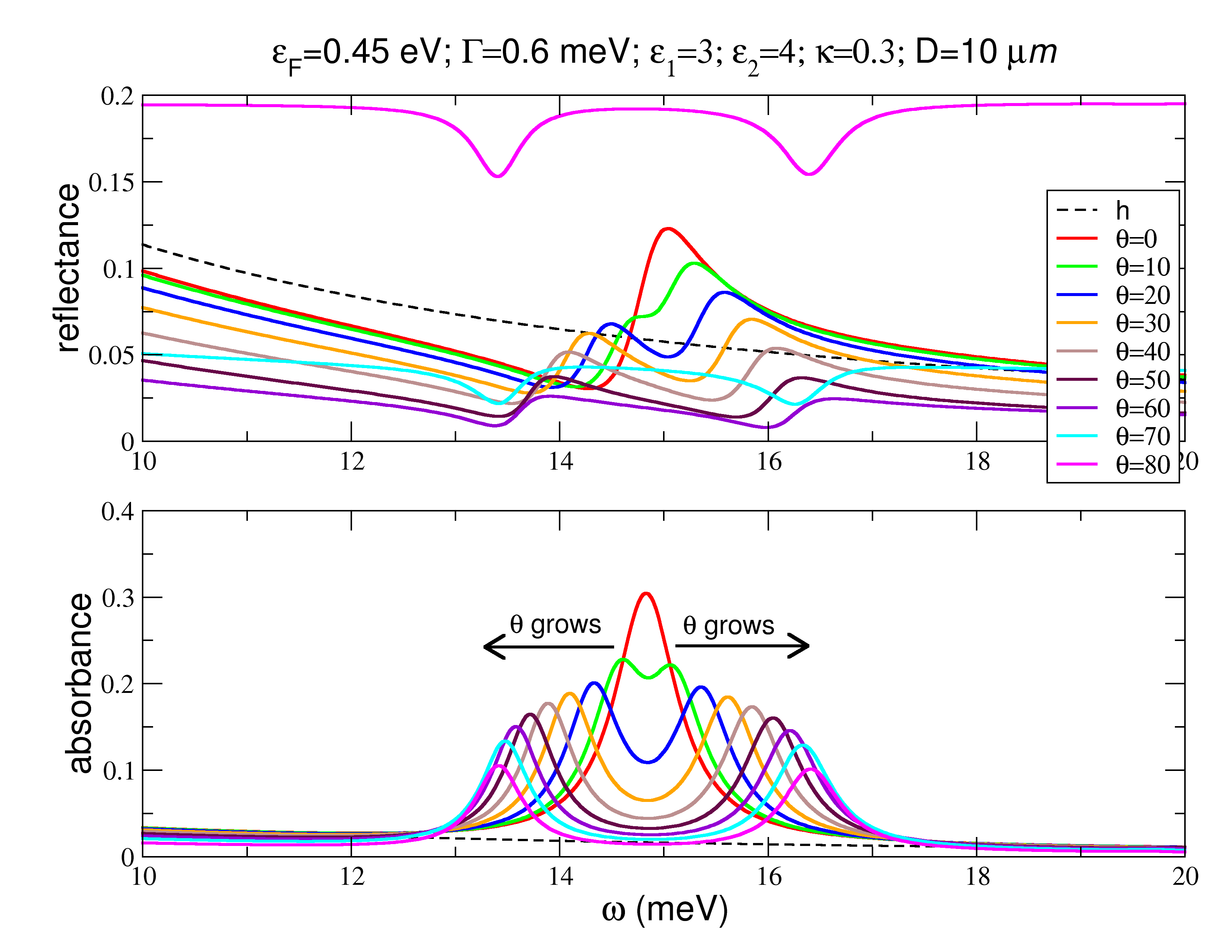} 

\par\end{centering}

\caption{Spectra of a graphene sheet with a cosine-modulated conductivity.
Reflectance (top) and absorbance (bottom) {\it versus} frequency for different angles of incidence.
 The parameters are: $E_{F}=0.452$~eV,
$D=10$~$\mu$m, $\epsilon_{1}=4$, $\epsilon_{2}=3$, $\Gamma=0.6$~meV,
and $h=0.3$. The Brewster angle for these dielectrics is 49.1$^{o}$.}

\label{fig_modulated_sigma}
\end{figure}

The linear system defined by Eqs.~(\ref{eq_linear_a}) and (\ref{eq_linear_b})
is solved numerically and the sums over $n$ are cut off at $n=-N,\ldots,0,\ldots,N$.
The numerical solution rapidly converges with $N$.
Results for the specular reflectance, ${\cal R}_{0}$, and for the
absorbance, ${\cal A}=1-{\cal R}_{0}-{\cal T}_{0}$, are given in
Fig.~\ref{fig_modulated_sigma}. For a modulated conductivity the
momentum of the SPPs is conserved up to a reciprocal lattice vector
$nG$, with $n=\pm1,\pm2,\ldots$, that is,

\begin{equation}
q_{p}=\vert k\sin\theta+nG\vert\,.\label{eq_q_SPP}
\end{equation}
In this case, even for normal incidence, it is possible to excite
SPPs. We stress that in the present case the Bragg scattering mechanism
expressed in Eq.~(\ref{eq_q_SPP}), allowing to overcome the momentum
mismatch between the propagating wave and SPPs, is induced entirely by the conductivity
modulation and it is not a consequence of a external grating. The
excitation of SPPs at normal incidence is illustrated in Fig.~\ref{fig_modulated_sigma}.
The dashed black curve represents the behaviour of the system for
a homogeneous conductivity and impinging ER at normal incidence; clearly
the curve is featureless. For the inhomogeneous case, a large enhancement
of the absorbance is seen around the energy given by Eq.~(\ref{eq_PP_spectrum_Otto})
with $q_{p}=2\pi/D$ (note that using this equation implies the extended band scheme). The position of the peak does
not coincide exactly with the number given by Eq.~(\ref{eq_PP_spectrum_Otto})
because this equation is not sensitive to the details of the band structure.
From Fig.~\ref{fig_polaritonic_crystal} we can observe that the bands
for $n=\pm1$ at the zone center have an energy of about 15 meV (the
energy for which the reflectance curve has a maximum for $\theta=0$).
As the angle of incidence approaches the Brewster angle, $\Theta_{B}$, for the two dielectrics, the reflectance decreases substantially. Note that the Brewster
angle of the system is not given exactly by the usual formula,
$\Theta_{B}=\arctan\sqrt{\epsilon_{1}/\epsilon_{2}}$, because of the presence
of graphene. For incidence angles above $\Theta_{B}$ the reflectance
develops two dips and can be larger than it would be for $\theta=0$
(see curve for $\theta=80^{o}$ in Fig.~\ref{fig_modulated_sigma}).

When the incoming beam deviates from normal incidence (i. e. $\theta\ne0$),
there is a peak splitting both in the reflectance and in the absorbance
curves, as predicted above. We would like to understand in qualitative
terms the behaviour of the splitting as function of $\theta$. Using Eq.~(\ref{eq_q_SPP}) in Eq.~(\ref{eq_PP_spectrum_Otto})
yields
\begin{equation}
\hbar\Omega_{p}=\sqrt{\frac{2\alpha}{\bar{\epsilon}}}E_{F}\hbar c\vert k\sin\theta+nG\vert\,.\label{eq_WAbsspp}
\end{equation}
Clearly, when $n>0$ and $\theta$ increases, the frequency shifts
toward higher energies. On the other hand, when $n<0$, the energy
decreases as $\theta$ increases. This behavior can be understood
from the analysis of Fig.~\ref{fig_polaritonic_crystal}. For $\theta=0$
the light line, Eq.~(\ref{eq_ATR}), is vertical and touches the
second band at the center of the Brillouin zone ($q=0$) where the branches
associated with $n=1$ and $n=-1$ almost touch each other (they do touch within the non-retarded approximation, see Fig. \ref {fig_polaritonic_crystal}). As $\theta$
grows, the slope of light line decreases and the branches with both
$n=1$ and $n=-1$ split away.
\begin{figure}[!htb]
\begin{centering}
\includegraphics[clip,width=9cm]{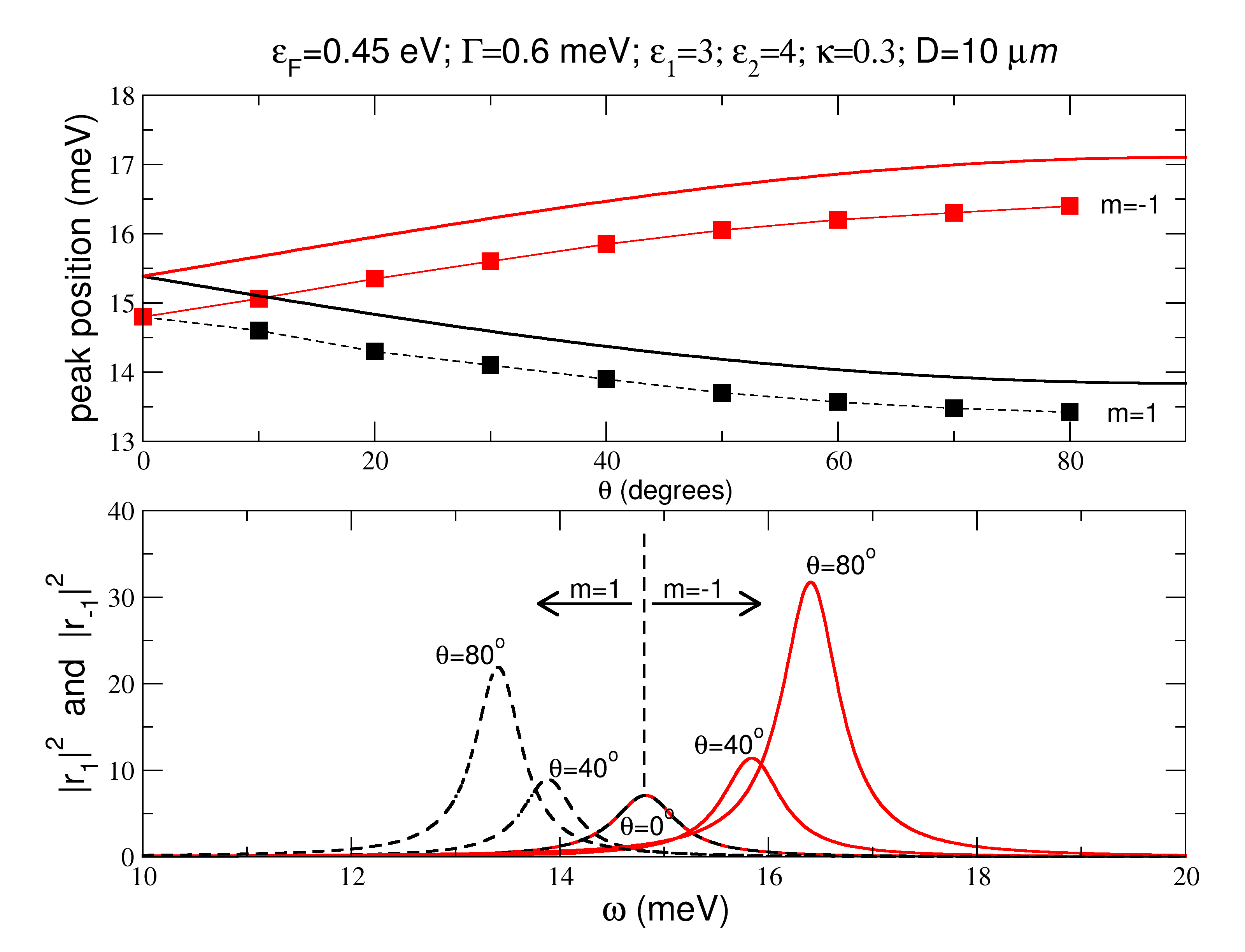} 

\par\end{centering}

\caption{Position of the absorbance peak as function of the angle of incidence.
 Top: dependence of the absorbance peak energy upon the
angle. The lines with squares are obtained from the bottom
panel of Fig.~\ref{fig_modulated_sigma}; the solid lines are the
two branches of Eq.~(\ref{eq_branch_minus_plus}). The bottom panel
shows the squared absolute values of the amplitudes of the modes associated
with the SPPs of momentum $\pm G$. The parameters are the same of
Fig.~\ref{fig_modulated_sigma}. }

\label{fig_peak_position}
\end{figure}

In Fig.~\ref{fig_peak_position} we plot the two branches of Eq.~(\ref{eq_branch_minus_plus})
and compare them with the positions of the absorbance peaks obtained
from Fig.~\ref{fig_modulated_sigma}. The agreement is only qualitative because, first, Eq.~(\ref{eq_branch_minus_plus})
is derived from a kinematic argument and, therefore, misses the dependence
on $h$ (and, eventually, some symmetries a particular problem may
have) and, secondly, we used the non-retarded approximation. However,
for small $h$ the agreement is quite good. Indeed, if we shift the solid curves in Fig.~\ref{fig_peak_position} by adding a constant,
they would fit the points (solid squares) obtained from Fig.~\ref{fig_modulated_sigma}.
In the bottom panel of Fig.~\ref{fig_peak_position} we plot $\vert r_{x||\pm1}\vert^{2}$
for different $\theta$ as a function of the energy. The energies of
the peaks of $\vert r_{x||\pm1}\vert^{2}$ are the same as those of the transmittance minima (absorbance maxima). This shows that the polaritons of the $n=\pm 1$ branches
with the wavevector $q_p\approx 0$ are responsible for the features in the reflectance, transmittance and absorbance spectra.


\section{Scattering of ER from corrugated graphene}

\label{sec_rayleigh}


\subsection{Setting the problem and definitions }
\label{sec_def}

In Secs.~\ref{sec_excitation_by_evanescent}, \ref{sec_metallic contact} and \ref{sec_rippled}
we have seen that resonant coupling between ER and SPPs can be achieved
either using the ATR scheme (Otto configuration), or due to a topological defect on graphene or its modulated conductivity.
Another possibility is the use of a dielectric grating \cite{ResonantCoupling}.
In what follows we discuss the coupling of ER to SPPs in graphene-based
gratings.

\begin{figure}[!ht]
\begin{centering}
\includegraphics[clip,width=10cm]{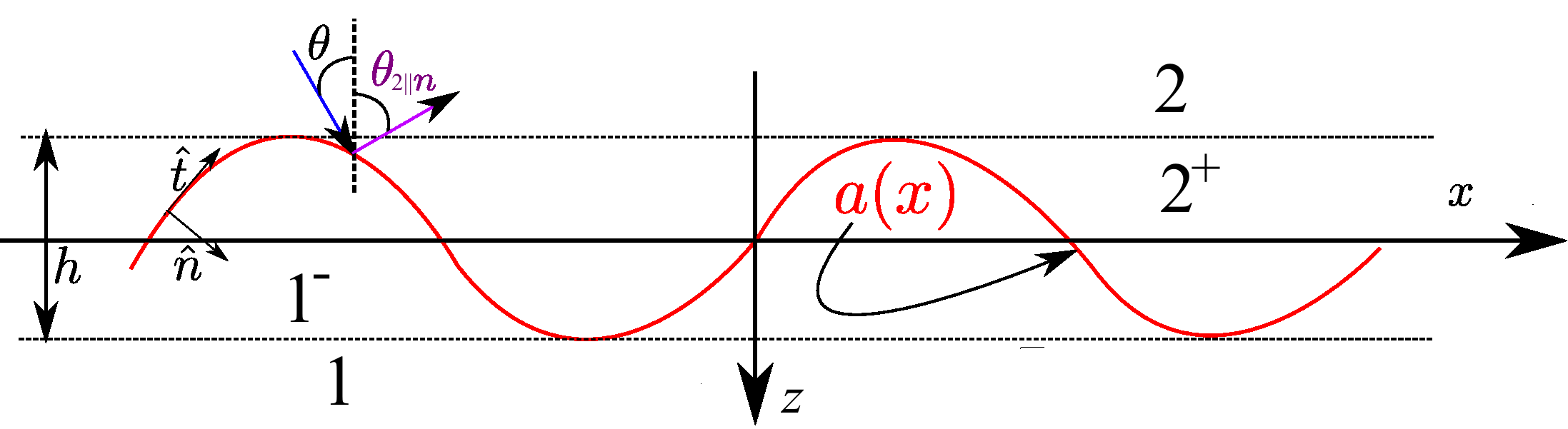}
\par\end{centering}

\caption{Grating geometry and different regions referred in the text. Graphene follows the profile
$a(x)$.}

\label{fig_grat}
\end{figure}

We want to solve the scattering problem of light impinging on a diffraction
grating, as schematically represented in Fig.~\ref{fig_grat}. The region 1 is
such that $z>\mbox{max}[a(x)]$, whereas the region 1$^{-}$ is such
that $a(x)<z<\mbox{max}[a(x)]$. A similar definition applies to the regions 2
and 2$^{+}$ when $z<\mbox{min}[a(x)]$ and $\mbox{min}[a(x)]<z<a(x)$,
respectively. In general, the calculation of the fields in the regions
2$^{+}$ and 1$^{-}$, i.e. $\mbox{min}[a(x)]<z<\mbox{max}[a(x)]$
is a challenging problem.

It will be assumed that in both regions 1 and 2 the dielectric functions,
$\epsilon_{1}$ and $\epsilon_{2}$, are constant. For a TM wave we
have Maxwell's equations (\ref{eq:By}), (\ref{eq:Ex}) and (\ref{eq:Ez}),
while the electromagnetic fields in regions 1 and 2 are the same as
represented by Eqs.~(\ref{eq:lat-by}), (\ref{eq:lat-ex}) and (\ref{eq:lat-ez}).
The Rayleigh-Fano approximation (also called Rayleigh hypothesis)
assumes that Rayleigh expansions (see below) are valid in the regions 1$^{-}$ and 2$^{+}$,
which is not true in general \cite{Toigo} (for example, the method fails for rectangular gratings). However, in the case of smooth grating profiles it can be proved that the solution of Maxwell's equations exists and is unique \cite {Chen-Friedman,Smirnov}, and can be approximated by linear combinations of reflected and transmitted waves in the regions 1$^{-}$ and 2$^{+}$ (see Ref. \cite {Smirnov} for details).



Since we are dealing with a corrugated surface, we need to defined tangent and normal vectors to the curve $z=a(x)$ (see Fig. \ref{fig_grat}).
The unit vector tangent to the curve $z=a(x)$ at point $x$ is given by
\begin{equation}
\hat{t}=\frac{1}{\sqrt{1+\left(\frac{da}{dx}\right)^{2}}}\left(\hat{x}+\frac{da}{dx}\hat{z}\right)\,,\quad|\hat{t}|=1.
\end{equation}
The normal unit vector to the curve is
\begin{equation}
\hat{n}=\frac{1}{\sqrt{1+\left(\frac{da}{dx}\right)^{2}}}\left(-\frac{da}{dx}\hat{x}+\hat{z}\right)\,,\quad|\hat{n}|=1;
\end{equation}
and clearly $\hat{n}\cdot\hat{t}=0$. Since the current density vector is tangent
to the graphene sheet, we must have $\vec{J}_{s}=\sigma\vec{E}_{t}=\sigma E_{t}\hat{t}$,
where $E_{t}$ is given by
\begin{equation}
E_{t}=\vec{E}\cdot\hat{t}=\frac{1}{\sqrt{1+\left(\frac{da}{dx}\right)^{2}}}\left(E_{x}+\frac{da}{dx}E_{z}\right)\:.
\end{equation}
One of the boundary conditions (\ref{eq:bcond-ht}) requires the determination
of the vector product $\vec{J}_{s}\times\hat{n}$:
\begin{eqnarray}
\vec{J}_{s}\times\hat{n}=-\frac{\sigma}{\sqrt{1+\left(\frac{da}{dx}\right)^{2}}}\left(E_{x}+\frac{da}{dx}E_{z}\right)\hat{z}\,,
\end{eqnarray}
and the respective boundary condition $B_{1,y}(x,z)\hat{y}-B_{2,y}\hat{y}=\mu_{0}\sigma\vec{J}\times\hat{n}$
reads
\begin{equation}
B_{1,y}(x,z)-B_{2,y}=-\frac{\mu_{0}\sigma}{\sqrt{1+\left(\frac{da}{dx}\right)^{2}}}\left(E_{x}+\frac{da}{dx}E_{z}\right)\,.\label{eq_BC1}
\end{equation}
The second boundary condition is $\vec{E}_{1,t}=\vec{E}_{2,t}$, which
can be written as
\begin{equation}
E_{1,x}(x,z)+\frac{da}{dx}E_{1,z}(x,z)=E_{2,x}(x,z)+\frac{da}{dx}E_{2,z}(x,z)\,.\label{eq_BC2}
\end{equation}
These boundary conditions are to be applied to a general profile $a(x)$.


\subsection{Rayleigh method by Toigo \textit{et al.}}

\label{sec_rayleigh_Toigo}

We give a representation of Rayleigh equations based on Ref.~\cite{Toigo}.
Using Eqs.~(\ref{eq:lat-by})-(\ref{eq:lat-ez}), the boundary conditions,
Eqs.~(\ref{eq_BC1})-(\ref{eq_BC2}), can be written in terms of
the $x$-component of the electric field as
\begin{eqnarray}
\epsilon_{1}\sum_{n}\frac{{\cal E}_{1,x||n}}{\kappa_{1||n}}e^{i[(q+nG)x-\kappa_{1||n}a(x)]}+\epsilon_{2}\sum_{n}\frac{{\cal E}_{2,x||n}}{\kappa_{2||n}}e^{i[(q+nG)x+\kappa_{2||n}a(x)]}-\nonumber \\
i\frac{\epsilon_{2}}{k_{z}}E_{x}^{(i)}e^{i[qx+k_{z}a(x)]}=-i\frac{\sigma}{\epsilon_{0}\omega{\sqrt{1+\left(\frac{da}{dx}\right)^{2}}}}\sum_{n}\left[1+i\frac{da}{dx}\frac{q+nG}{\kappa_{1||n}}\right]\times\nonumber \\
{\cal E}_{1,x||n}e^{i[(q+nG)x-\kappa_{1||n}a(x)]}\,;
\label{eq:BC1_toigoI}\\
\sum_{n}{\cal E}_{1,x||n}e^{i[(q+nG)x-\kappa_{1||n}a(x)]}\left[1+i\frac{da}{dx}\frac{q+nG}{\kappa_{1||n}}\right]=\nonumber \\
\sum_{n}{\cal E}_{2,x||n}e^{i[(q+nG)x+\kappa_{2||n}a(x)]}\left[1-i\frac{da}{dx}\frac{q+nG}{\kappa_{2||n}}\right]+\nonumber \\
\left[1-\frac{da}{dx}\frac{q}{k_{z}}\right]E_{x}^{(i)}e^{i[qx+k_{z}a(x)]}\,.\label{eq:BC2_toigoII}
\end{eqnarray}
Eqs.~(\ref{eq:BC1_toigoI}),(\ref{eq:BC2_toigoII}) were obtained using
Eqs.~(\ref{eq:byex3})-(\ref{eq:ezex3}) (replacing $\epsilon_{3}$ by $\epsilon_{2}$) and (\ref{eq:lat-byex})-(\ref{eq:lat-ezex}).
The boundary conditions are applied at $z=a(x)$. Multiplying
Eq.~(\ref{eq:BC1_toigoI}) by $e^{-i(q+pG)x}$ and integrating over
$\int_{0}^{D}dx/D$, we obtain
\begin{eqnarray}
\epsilon_{1}\sum_{n}\frac{{\cal E}_{1,x||n}}{\kappa_{1||n}}M_{1||p-n}+\epsilon_{2}\sum_{n}\frac{{\cal E}_{2,x||n}}{\kappa_{2||n}}M_{2||p-n}-i\frac{\epsilon_{2}}{k_{z}}E_{x}^{(i)}M_{p}^{(i)}=\nonumber \\
-i\frac{\sigma}{\epsilon_{0}\omega}\sum_{n}\left(L_{p-n}+i\frac{q+nG}{\kappa_{1||n}}N_{p-n}\right){\cal E}_{1,x||n}\,,
\end{eqnarray}
where
\begin{eqnarray}
M_{p}^{(i)} & = & \frac{1}{D}\int_{0}^{D}dxe^{-ipGx}e^{ik_{z}a(x)}\,,\\
M_{m||p-n} & = & \frac{1}{D}\int_{0}^{D}dxe^{-i(p-n)Gx}e^{(-1)^{m}\kappa_{m||n}a(x)}\,,\\
L_{p-n} & = & \frac{1}{D}\int_{0}^{D}dx\frac{e^{-i(p-n)Gx}}{\sqrt{1+\left(\frac{da}{dx}\right)^{2}}}e^{-\kappa_{1||n}a(x)}\approx M_{1||p-n}\,,\\
N_{p-n} & = & \frac{1}{D}\int_{0}^{D}dx\frac{da}{dx}\frac{e^{-i(p-n)Gx}}{\sqrt{1+\left(\frac{da}{dx}\right)^{2}}}e^{-\kappa_{1||n}a(x)}\approx\nonumber \\
 &  & -i\frac{(p-n)G}{\kappa_{1||n}}M_{1||p-n}\,.
\end{eqnarray}
After some manipulations we arrive at the following equation:
\begin{eqnarray}
\sum_{n}\left(\frac{\epsilon_{1}}{\kappa_{1||n}}+i\frac{\sigma}{\epsilon_{0}\omega}\frac{(q+nG)(q+pG)-\omega^{2}\epsilon_{1}/c^{2}}{\kappa_{1||n}^{2}}\right){\cal E}_{1,x||n}M_{1||p-n}=\nonumber \\
-\epsilon_{2}\sum_{n}\frac{{\cal E}_{2,x||n}}{\kappa_{2||n}}M_{2||p-n}+i\frac{\epsilon_{2}}{k_{z}}E_{x}^{(i)}M_{p}^{(i)}\,,
\label{eq_BC1_toigo}
\end{eqnarray}

In a similar manner, multiplying Eq.~(\ref{eq:BC2_toigoII}) $e^{-i(q+pG)x}$
and integrating over $\int_{0}^{D}dx/D$, we obtain:
\begin{eqnarray}
\sum_{n}\frac{(q+nG)(q+pG)-\omega^{2}\epsilon_{1}/c^{2}}{\kappa_{1||n}^{2}}{\cal E}_{1,x||n}M_{1||p-n}=\nonumber \\
\sum_{n}\frac{(q+nG)(q+pG)-\omega^{2}\epsilon_{2}/c^{2}}{\kappa_{2||n}^{2}}{\cal E}_{2,x||n}M_{2||p-n}+\nonumber\\
\sum_{n}\frac{\omega^{2}\epsilon_{1}/c^{2}-q(q+pG)}{k_{z}^{2}}E_{x}^{(i)}M_{p}^{(i)}\,.
\label{eq_BC2_toigo}
\end{eqnarray}
Solution of the boundary problem (\ref{eq_BC1_toigo}) and
(\ref{eq_BC2_toigo}) will allow for the calculation of the reflectance,
transmittance, and absorbance spectra.


\subsection{Three limiting cases}

\label{sec_Toigo_special_limits}
First, we will show that, if $a(x)=0$ and $E_{x}^{(i)}=0$, one obtains from Eqs.~(\ref{eq_BC1_toigo}) and
(\ref{eq_BC2_toigo}) the usual SPP spectrum folded into the first Brillouin zone.
When $a(x)=0$, we have $M_{m||p-n}=\delta_{p,n}$
and Eqs.~(\ref{eq_BC1_toigo}) and (\ref{eq_BC2_toigo}) reduce to
\begin{eqnarray}
\left(\frac{\epsilon_{1}}{\kappa_{1||p}}+i\frac{\sigma}{\epsilon_{0}\omega}\right){\cal E}_{1,x||p}=-\epsilon_{2}\frac{{\cal E}_{2,x||p}}{\kappa_{2||p}}\,;\qquad{\cal E}_{1,x||p}={\cal E}_{2,x||p}\,.
\end{eqnarray}
Solving for ${\cal E}_{2,x||p}$, we obtain
\begin{equation}
\frac{\epsilon_{1}}{\kappa_{1||p}}+\frac{\epsilon_{2}}{\kappa_{2||p}}+i\frac{\sigma}{\epsilon_{0}\omega}=0\,.\label{eq_SPP_Toigo}
\end{equation}
Since the SPP amplitude decays away from the graphene sheet, we consider
$\kappa_{m||n}$ in the form (\ref{eq:lat-kappa}) with $q\in[-\pi/D,\pi/D]$.
With this choice we recover Eq. (\ref{eq_W_SPP_2D}) folded into the
first Brillouin zone.

When $E_{x}^{(i)}$ is finite, we have a scattering problem. In this
case $M_{p}^{(i)}=\delta_{p,0}$ and Eqs.~(\ref{eq_BC1_toigo}) and
(\ref{eq:BC2_toigoII}) reduce to
\begin{eqnarray}
\left(\frac{\epsilon_{1}}{\kappa_{1||0}}+i\frac{\sigma}{\epsilon_{0}\omega}\right){\cal E}_{1,x||0}=-\epsilon_{2}\frac{{\cal E}_{2,x||0}}{\kappa_{2||0}}+i\frac{\epsilon_{2}}{k_{z}}E_{x}^{(i)}\,,\\
{\cal E}_{1,x||0}={\cal E}_{2,x||0}+E_{x}^{(i)}\,.\nonumber
\end{eqnarray}
In the limit $\sigma=0$ we obtain Eq.~(\ref{eq:lat-ref0-simp}),
the well-known reflectance amplitude from elementary optics. In the
case ${\sigma}\ne0$ and $\epsilon_{1}=\epsilon_{2}$ we obtain
Eq. (\ref{eq:lat-t0-simp}), which gives the transmittance amplitude for
graphene.


\subsection{A non-trivial example I: sine profile}

\label{subsec_sine_profile}

The last two examples are trivial for they refer to the limit of zero
curvature. We now consider a non-trivial case where graphene has a well defined
periodic corrugation. We assume a profile of the form
\begin{equation}
a(x)=h\sin(2\pi x/D)\,,\label{eq_sin_profile}
\end{equation}
from which follows $da/dx=(2\pi h/D)\cos(2\pi x/D)$. For this choice
of profile the Rayleigh hypothesis is exact. We have:
\begin{eqnarray}
M_{p}^{(i)} & = & J_{p}(k_{z}h)\,,\\
M_{m||p-n} & = & J_{p-n}[i(-1)^{m+1}\kappa_{m||n}h]=i^{p-n}I_{p-n}[(-1)^{m+1}\kappa_{m||n}h]\,,
\end{eqnarray}
where $J_{p}(z)$ and $I_{p}(z)$ are the usual and modified Bessel
functions of order $p$, respectively.

\begin{figure}[!ht]
\begin{centering}
\includegraphics[clip,width=10cm]{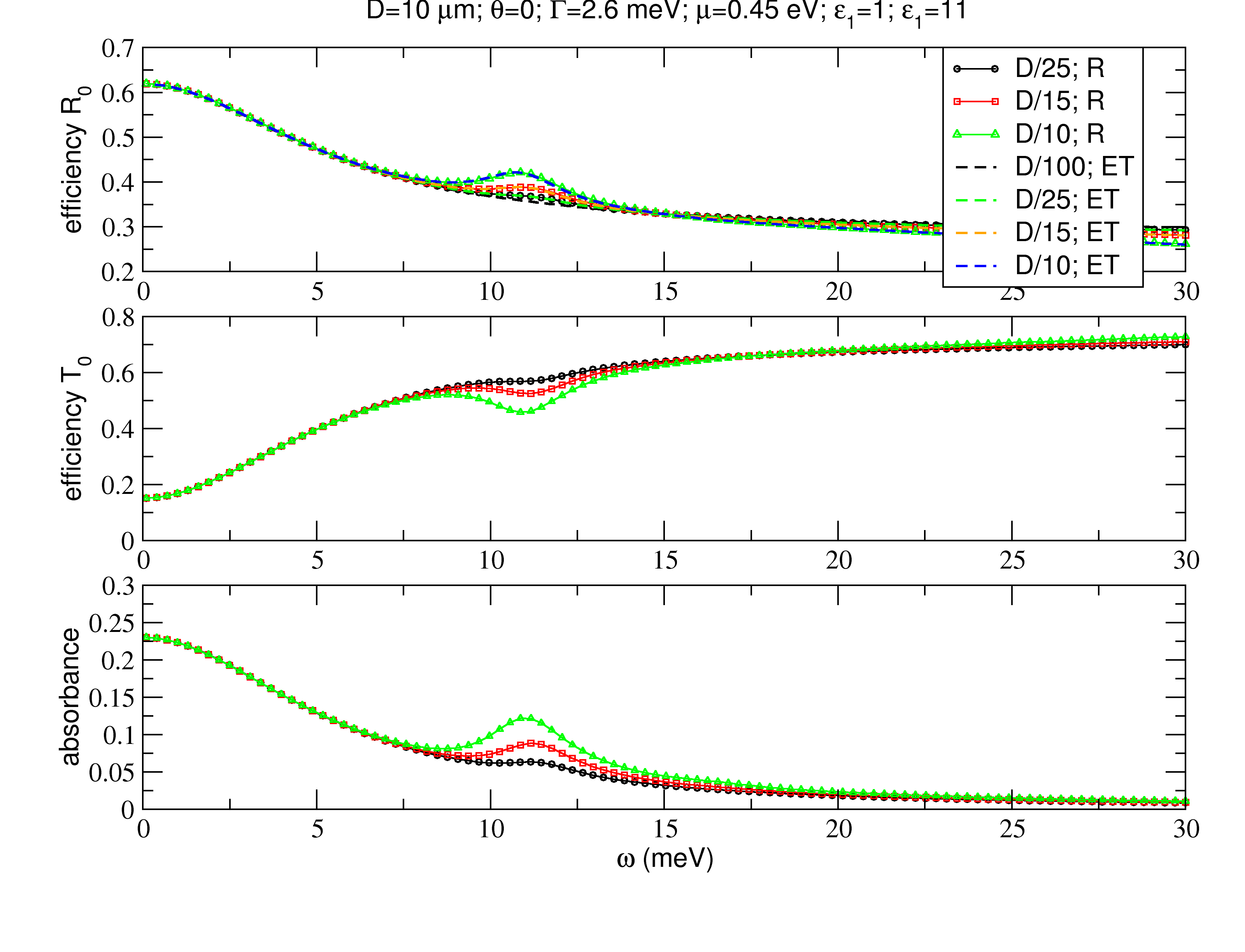}
\par\end{centering}

\caption{Dependence of the reflectance, transmittance and
absorbance on frequency for different depth of the grooves for sine profile gratings.
We compare the Rayleigh method with the results from the extinction
theorem. In the legend, the label $R$ stands for the Rayleigh approximation
whereas the label ET stands for the extinction theorem result. The parameters
are $E_{F}=0.45$~eV, $D=10$~$\mu$m, $\epsilon_{1}=1$, $\epsilon_{2}=11$
(silicon), $\Gamma=2.6$~meV, and $\theta=0$.}

\label{fig_depth}
\end{figure}

\begin{figure}[!ht]
\begin{centering}
\includegraphics[clip,width=10cm]{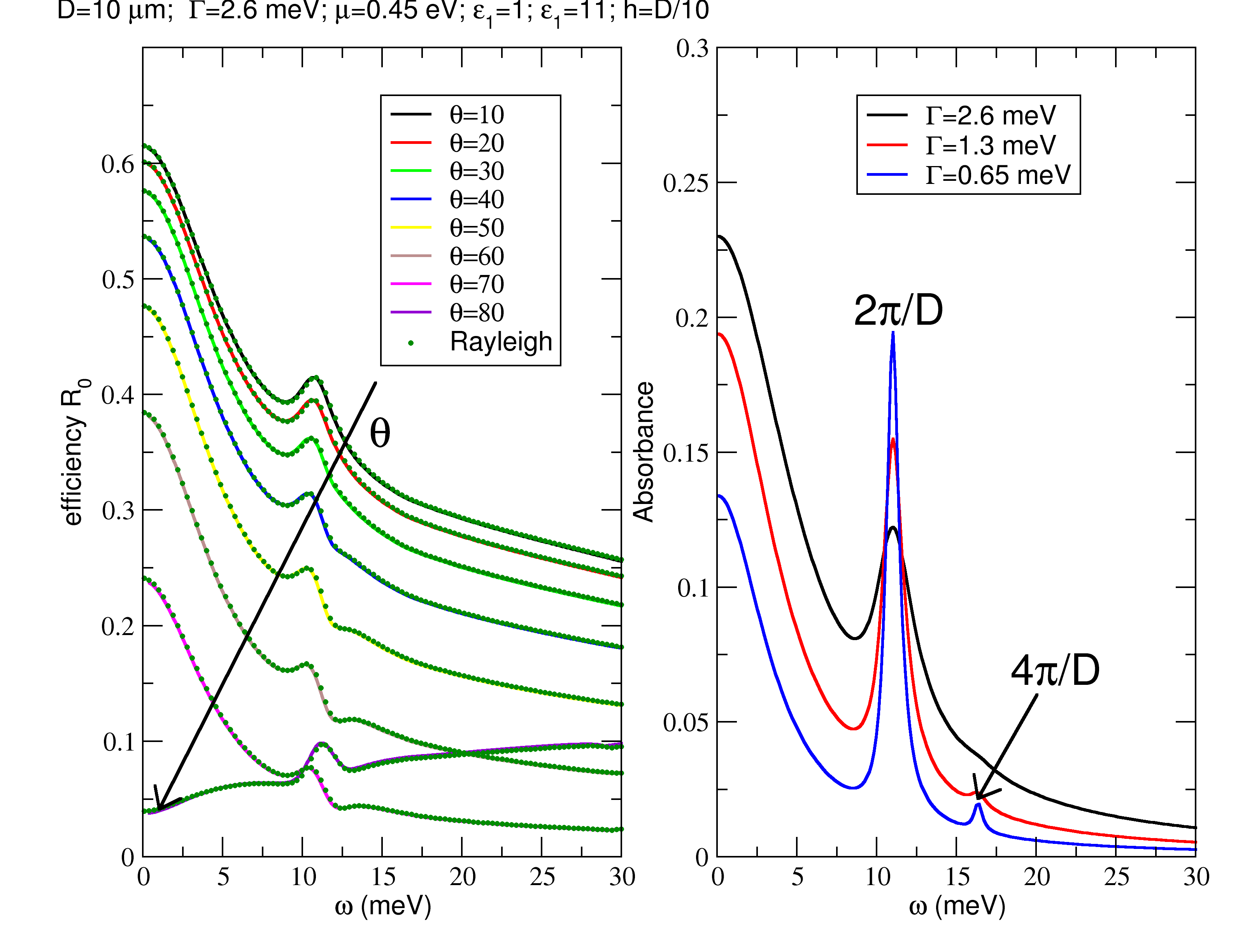}
\par\end{centering}

\caption{Left: Dependence of the reflectance on the angle of incidence for a sine profile grating. We
compare the Rayleigh method (solid circles) with the results from
the extinction theorem approach (solid lines). The arrow indicates the direction
of growth of the angle $\theta$. Right: absorbance for different
values of the broadening $\Gamma$, at normal incidence. Other parameters
are as in Fig.~\ref{fig_depth}.}

\label{fig_Sin_grating_angles}
\end{figure}

In Fig.~\ref{fig_depth} we represent the efficiencies ${\cal R}_{0}$
and ${\cal T}_{0}$ (top and central panels, respectively), and the
absorbance, ${\cal A}=1-{\cal R}_{0}-{\cal T}_{0}$, as functions of
the incoming photon energy, for different values of the ratio
$h/D$. When $h/D\ll1$, we recover the properties of a flat graphene
sheet. We compare results from the Rayleigh approximation described above with those produced by the extinction theorem method (dashed
curves), an exact integral-equation approach to the scattering problem \cite{Toigo,CorrugatedSLG}.
We see that the agreement is excellent. In Fig.~\ref{fig_Sin_grating_angles}
the quality of the agreement between the two methods is more evident.
In this figure we compare the dependence of the specular reflectance
${\cal R}_{0}$ on frequency, for several angles of incidence $\theta$
and $h/D=0.1$. The resonance seen in Figs.~\ref{fig_depth} and
\ref{fig_Sin_grating_angles} above 10 meV is due to the excitation
of a surface-plasmon-polariton of energy
\begin{equation}
\hbar\Omega_{p}=\sqrt{\frac{4\alpha}{\epsilon_{1}+\epsilon_{2}}E_{F}c\hbar\frac{2\pi}{D}}\approx11\mbox{ meV}\,.
\end{equation}

In the right panel of Fig.~\ref{fig_Sin_grating_angles} we present
the absorbance, ${\cal A}$, for different values of $\Gamma$. As
$\Gamma$ decreases, the resonance in the absorbance becomes more prominent.
Also, the coupling of the ER to the SPP of wave number $4\pi/D$ (extended band scheme)
becomes evident as a smaller resonance.


\subsection{A non-trivial example II: sawtooth profile}

\label{sec_Toigo_satooth}
\begin{figure}[!ht]
\begin{centering}
\includegraphics[clip,width=9cm]{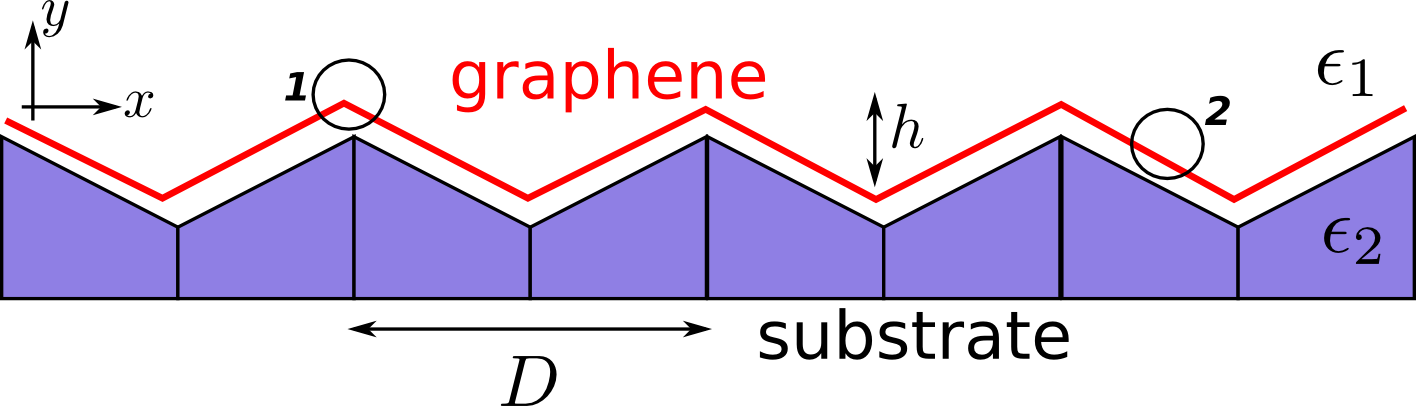}
\par\end{centering}

\caption{Periodic patterned substrate with a sawtooth profile. In regions with
the label 1 the strain is expected to be higher than in those with
label 2. Then, the problem of scattering by such grating also includes the previously
considered case of a periodically modulated conductivity.}

\label{fig_sawtooth}
\end{figure}

In the case of a sawtooth profile the function
$a(x)$ reads (see Fig.~\ref{fig_sawtooth}):
\begin{equation}
a(x)=\left\{ \begin{array}{c}
\frac{2h}{D}x+\frac{h}{2},\hspace{0.5cm}-D/2<x<0\,,\\
-\frac{2h}{D}x+\frac{h}{2},\hspace{0.45cm}0<x<D/2\,.
\end{array}\right.
\end{equation}

In this case the functions $M_{p}^{(i)}$ and $M_{m||p-n}$ can be
written as
\begin{eqnarray}
M_{p}^{(i)}=\frac{ik_{z}h}{\pi^{2}p^{2}-k_{z}^{2}h^{2}}\left[e^{ik_{z}h/2}-(-1)^{p}e^{-ik_{z}h/2}\right]\,,\\
M_{m||p-n}=\frac{(-1)^{m}\kappa_{m||n}h}{\pi^{2}(p-n)^{2}+\kappa_{m||n}^{2}h^{2}}\times\nonumber \\
\left[e^{(-1)^{m}\kappa_{m||n}h/2}-(-1)^{p-n}e^{(-1)^{m+1}\kappa_{m||n}h/2}\right]\,.
\end{eqnarray}
Contrary to the case of the sine profile, the Rayleigh method for
the sawtooth profile is neither convergent for some values of $h$, for a given $\omega $, nor
for a given $h$ for all values of $\omega$. We have checked that
partial convergence requires $h\lesssim D/10$. The convergence over
an energy range has to be checked case by case.
\begin{figure}[!ht]
\begin{centering}
\includegraphics[clip,width=10cm]{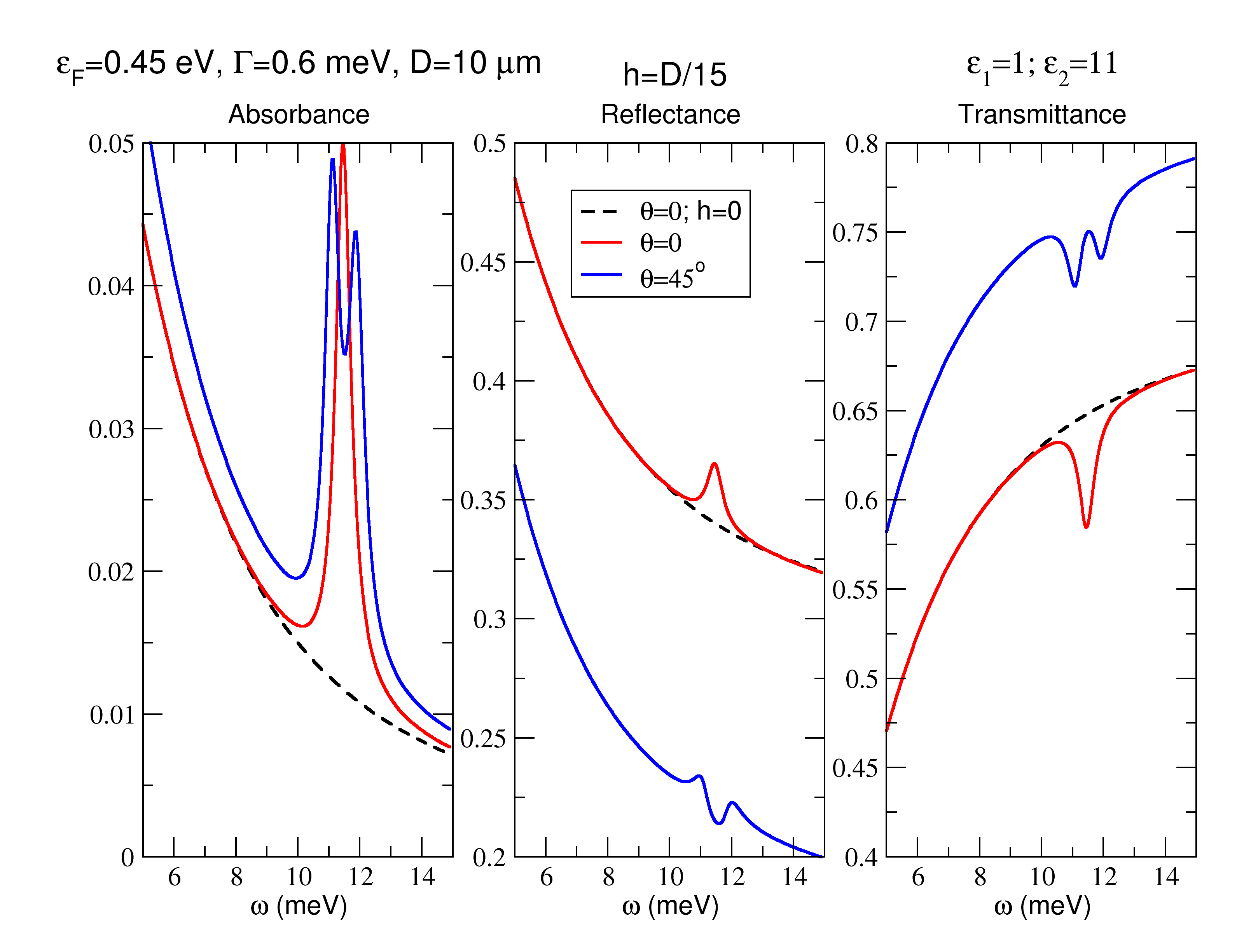}
\par\end{centering}

\caption{Absorbance, reflectance and transmittance for the sawtooth profile
at two angles of incidence, $\theta=0$ and $\theta=45^{o}$. The
dashed lines refer to the non-corrugated limit for $\theta=0$. Here
$h/D=1/15$ and $\Gamma=0.6$~meV; other parameters as in Fig.~\ref{fig_depth}.}

\label{fig_Toigo_sawtooth}
\end{figure}

In Fig.~\ref{fig_Toigo_sawtooth} we plot the absorvance, reflectance,
and transmittance for a grating with a sawtooth profile where the convergence could be achieved. As in the
case of the sinusoidal profile discussed in Sec.~\ref{subsec_sine_profile},
we see the presence of resonances associated with the coupling of
the impinging ER with the graphene SPPs. For $\theta=0$
the resonance is located at $\hbar\omega\approx11.6$~meV. As before,
the value can be predicted from:
\begin{eqnarray}
\hbar\Omega_{p}=\sqrt{\frac{4\alpha}{\epsilon_{1}+\epsilon_{2}}E_{F}\hbar cq_{p}}\\
=\sqrt{\frac{4}{137\times12}0.452\times0.2\frac{2\pi}{10}}\approx\mbox{11.8 meV}\,.
\end{eqnarray}
At finite angles of incidence, the single peak seen at $\theta=0$
splits into two peaks because the ATR scan line intercepts the $n=-1,1$
bands at different energies. The prediction for the positions of the
two peaks follows from the calculation of the band structure of the
SPPs for the sawtooth profile. It is interesting to note that the peak splitting
does not occur in the case of the sine profile.


\section{Graphene on a metallic grating}
\label{sec_metal_grating}
\subsection{Relation between the conductivity and the dielectric function of
a 3D metal}
\label{sec_epsilon_sigma_3D}

So far we have discussed the resonant coupling of ER to SPPs in graphene lying on a dielectric
grating. In this section we shall consider the case where the underlying
grating is made of a conductor, be it a metal, a doped semiconductor, or a mesostructure \cite{Pendry}.

Let us first derive a simple relation between
the conductivity and the dielectric function of a 3D metal, based
on Drude's model \cite{Fox}. The equation of motion of an electron
in a metal subject to an oscillating electric field is
\begin{equation}
m_{e}\frac{d^{2}x}{dt^{2}}+m_{e}\gamma\frac{dx}{dt}=-eE(\omega)^{-i\omega t}\,,
\label {eq:oscillatot}
\end{equation}
where $m_{e}$ denotes the electron mass and $-e$ is the electron
charge. Substituting $x=x_{0}e^{-i\omega t}$, into Eq. (\ref {eq:oscillatot})
we obtain
\begin{equation}
x_{0}=\frac{eE(\omega)}{m_{e}\omega(\omega+i\gamma)}\,.
\end{equation}
The polarization is defined as $P(\omega)=-en_{e}x_{0}$ (units of
C/m$^{2}$), where $n_{e}$ is the electronic density (per unit volume)
of the gas. The electric displacement field is defined as
\begin{eqnarray}
\nonumber
D(\omega)=\varepsilon_{0}\epsilon(\omega)E(\omega)=\varepsilon_{0}E(\omega)+P(\omega)\\
\qquad =\varepsilon_{0}E(\omega)-\frac{e^{2}n_{e}E(\omega)}{m_{e}\omega(\omega+i\gamma)}\,.
\label{eq_dielectric_function}
\end{eqnarray}
Therefore, we have
\begin{equation}
\epsilon(\omega)=1-\frac{e^{2}n_{e}/\varepsilon_{0}}{m_{e}\omega(\omega+i\gamma)}\,.\label{eq_dielectric_function_relative}
\end{equation}
On the other hand, Drude's model for the conductivity reads \cite{Fox}
\begin{equation}
\sigma_{\textrm{3D}}=\frac{e^{2}n_{e}}{m_{e}(\gamma-i\omega)}\,.
\label{eq_sigma_drude_3D}
\end{equation}
Comparing Eqs.~(\ref{eq_dielectric_function_relative}) and (\ref{eq_sigma_drude_3D})
we obtain
\begin{equation}
\epsilon(\omega)=1+i\frac{\sigma_{\textrm{3D}}}{\varepsilon_{0}\omega}\,.
\label{eq_epsilon_drude_1}
\end{equation}
If the dielectric screening by core electrons of the metal atoms is taken into account, the unity in Eq. (\ref{eq_epsilon_drude_1}) must be replaced by a background dielectric constant $\epsilon _\infty >1$ \cite {Born}. For instance, for gold $\epsilon _\infty =1.53$ \cite{Torrell}, while for most semiconductors it is of the order of 10.

Thus, when dealing with a 3D metal the conductivity enters the problem
through the dielectric function. Often it can be assumed that $\gamma\ll\omega$,
leading to a purely real dielectric function, that is, the metal acts
as a dispersive dielectric with a dielectric constant that can negative. In this limit, the dielectric function has the simple form
\begin{equation}
\epsilon(\omega)=\epsilon _\infty -\frac{\omega_{p}^{2}}{\omega^{2}}\,,\label{eq_epsilon_dispersice_cond}
\end{equation}
where $\omega_{p}^{2}=n_{e}e^{2}/(m_{e}\epsilon_{0})$ denotes the
plasma frequency of the bulk conductor. Although it is, of course, measured in $s ^{-1}$, for convenience we shall use values in meV, which correspond to plasma energy $\hbar\omega_{p}$.


\subsection{Surface plasmon--polaritons at a dielectric-metal interface}

\label{sec_metal_SPP} Before discussing the form of the spectrum
of SPPs when graphene is placed on a bulk conductor, we analyze the simpler
case of a dielectric-metal interface. We assume a system where a dielectric
of constant permittivity $\epsilon_{2}$ is in contact with a bulk
metal of dielectric function $\epsilon(\omega)$. The dispersion relation
can be obtained from Eq.~(\ref{eq_W_SPP_2D}) by putting $\sigma\equiv0$
and substituting $\epsilon_{1}$ by $\epsilon(\omega)$, namely
\begin{equation}
\frac{\epsilon(\omega)}{\kappa_{1}}+\frac{\epsilon_{2}}{\kappa_{2}}=0\,,\label{eq_eigen_3D_plasmon}
\end{equation}
where $\kappa_{1}^{2}=q^{2}-\epsilon(\omega)\omega^{2}/c^{2}$ and $\kappa_{2}^{2}=q^{2}-\epsilon _2\omega^{2}/c^{2}$. As
a result, for real values of $\kappa_{1}$ and $\kappa_{2}$, $\epsilon(\omega)$
has to be real and negative, implying $\omega<\omega_{p}$.
Sometimes in the literature the eigenvalue
equation (\ref{eq_eigen_3D_plasmon}) appears in the form
\begin{equation}
q=\frac{\omega}{c}\sqrt{\frac{\epsilon_{2}\epsilon(\omega)}{\epsilon_{2}+\epsilon(\omega)}}\,,
\end{equation}
which can be easily obtained from Eq.~(\ref{eq_eigen_3D_plasmon}).
Note that, since the argument in the square root above must be positive, it is necessary that $\epsilon(\omega)<-\epsilon_{2}$ and
$\omega<\omega_{p}/\sqrt{\epsilon _\infty +\epsilon_{2}}$,
which gives an upper bound for the SPP frequencies.

We can solve Eq.~(\ref{eq_eigen_3D_plasmon}) for $\omega$:
\begin{eqnarray}
\omega^{2} & = & \frac{\omega_{p}^{2}}{2\epsilon _\infty}+\frac{c^{2}q^{2}}{2}\frac{\epsilon _\infty +\epsilon_{2}}{\epsilon_{2}\epsilon _\infty}\nonumber \\
 & - & \frac{1}{2\epsilon_{2}\epsilon _\infty}\sqrt{[\omega_{p}^{2}\epsilon_{2}+c^{2}q^{2}(\epsilon _\infty+\epsilon_{2})]^{2}-4\epsilon_{2}\epsilon _\infty c^{2}q^{2}\omega_{p}^{2}}\,.
\end{eqnarray}
The minus sign in front of the square root is necessary to guarantee that $\omega<\omega_{p}/\sqrt{\epsilon _\infty +\epsilon_{2}}$. This relation takes simple forms in two limiting cases. When $q\rightarrow0$, we have
\begin{equation}
\omega^{2}=\frac{c^{2}q^{2}}{\epsilon_{2}}-\frac{c^{4}q^{4}(\epsilon_{2}+\epsilon _\infty)^{2}}{4\epsilon_{2}^{2}\epsilon _\infty \omega_{p}^{2}}\,.
\end{equation}
In this case, the dispersion relation of SPPs is always below the
spectrum of the ER in the dielectric, a situation that does not occur
for SPPs on graphene (recall Sec.~\ref{sec_TM_spectrum}).
\begin{figure}[!ht]
\begin{centering}
\includegraphics[clip,width=10cm]{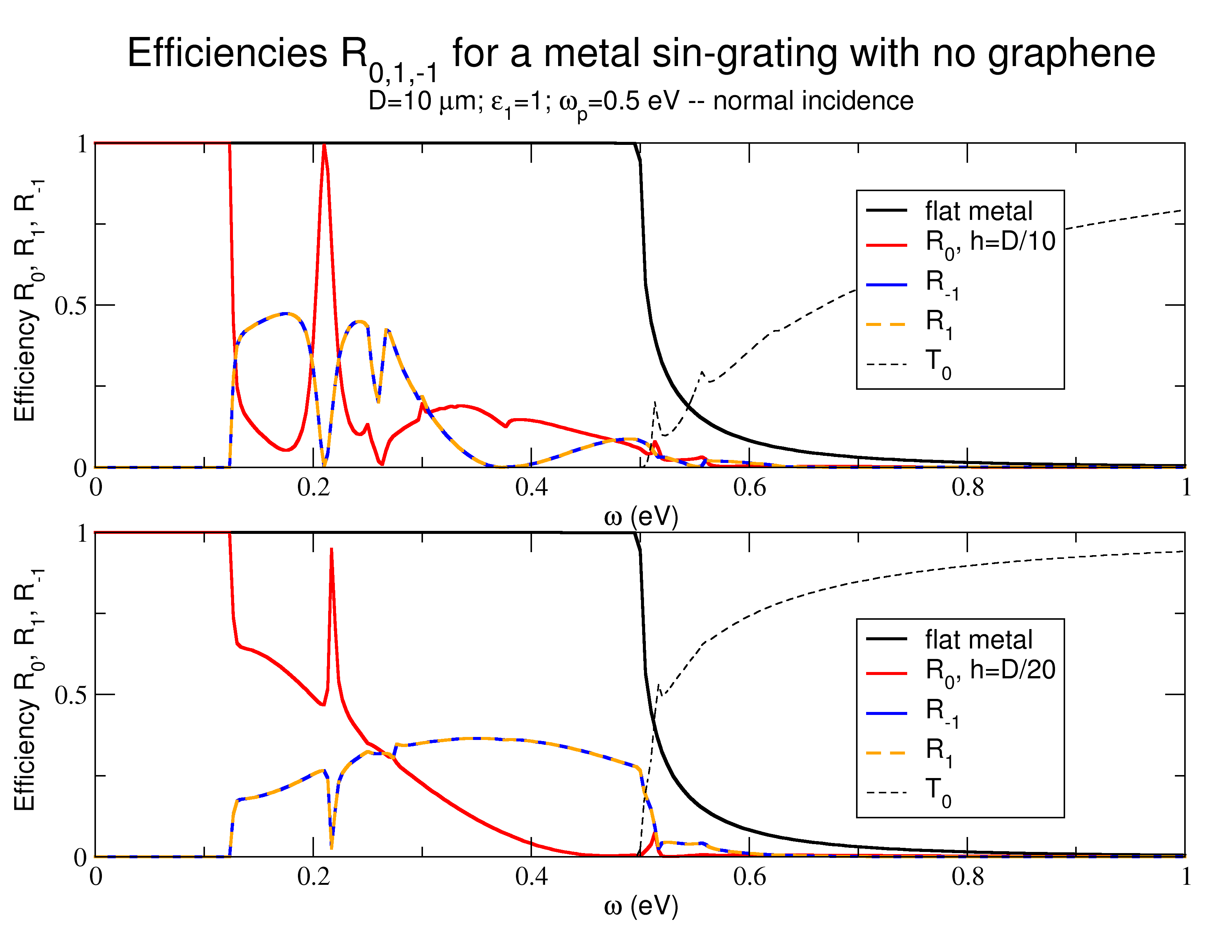}
\par\end{centering}

\caption{Reflectance and transmittance of a sinusoidal metallic grating (without
graphene). We depict three reflectance efficiencies, ${\cal R}_{0}$,
${\cal R}_{1}$, and ${\cal R}_{-1}$. The specular transmittance ${\cal T}_{0}$
is also depicted as a dashed line; ${\cal T}_{0}\ne0$ only for $\omega>\omega_{p}$.
The parameters are: $\epsilon_{1}=\epsilon _\infty =1$, $\hbar\omega_{p}=0.5$~eV,
$D=10$~$\mu$m, $h=D/10$ (top), and $h=D/20$ (bottom). }

\label{fig_metal_grid_no_graphene}
\end{figure}

For $q\rightarrow\infty$ we obtain
\begin{equation}
\omega^{2}=\frac{\omega_{p}^{2}}{\epsilon _\infty+\epsilon_{2}}-\frac{\epsilon_{2}\omega_{p}^{4}}{4\epsilon _\infty(\epsilon _\infty+\epsilon_{2})c^{2}q^{2}}\,,
\end{equation}
that is, the dispersion relation is always below $\omega_{p}/\sqrt{\epsilon _\infty+\epsilon_{2}}$.
Clearly, the dispersion curve of the surface plasmon--polaritons at the surface
of a 3D metal is quite different from that for a 2D conductor (see Sec.~\ref{sec_TM_spectrum}).


\subsection{Graphene on a metallic grating}
\label{sec_graphene_on_metal_grating}

As known, for frequencies below the plasma frequency a bulk metal reflects all light
impinging on it, a consequence of the negative value of its dielectric permittivity.\footnote {Strictly speaking, this is true for $\epsilon _\infty =1$, zero damping and neglecting interband or impurity-related optical transitions that may take place in the same spectral region.}
If the surface of the metal has a periodic corrugation, the energy of the
reflected wave is distributed among the specular and the diffracted
orders, as seen in Fig.~\ref{fig_metal_grid_no_graphene}. In this
figure we depict three reflectance efficiencies, ${\cal R}_{0}$,
${\cal R}_{1}$, and ${\cal R}_{-1}$, for normal incidence of the
impinging radiation. As expected for normal incidence, we have ${\cal R}_{1}={\cal R}_{-1}$.
Since we are considering the limit of a dispersive metal with no
absorption, there is no dissipation due to SPPs. The rich structure
seen in ${\cal R}_{0}$, ${\cal R}_{1}$, and ${\cal R}_{-1}$ spectra is
due to the distribution of the energy through the different diffraction
orders. The interpretation of the figure is rather difficult because
the problem is non-linear in the frequency $\omega$. For energies
below $0.12$ eV (smaller than the plasma frequency taken equal to 0.5 meV) only the specular
(${\cal R}_{0}$) order is propagating. Since the absorption of
the metal was neglected the reflectance is equal to unity; this corresponds
to frequencies $\omega/c<G$. Above $0.12$ eV the orders
$n=\pm1$ (${\cal R}_{\pm1}$) also become propagating and we have
Bragg diffraction (orders with $|n|>1$ are evanescent; this corresponds
to frequencies $G<\omega/c<2G$). In this case SPPs with the wavevectors
$\pm G$, excited by the incoming light thanks to the presence of the
grating, become radiative (non-evanescent) and the diffracted energy
is distributed among the three orders $n=0,\pm1$ in a non-trivial
way. In this case we have ${\cal R}_{0}+{\cal R}_{1}+{\cal R}_{-1}=1$,
since there is no dissipation in the metal. Above the plasma frequency
the grating becomes partially transparent. As the frequency of the
incoming ER increases the dielectric function of the metal tends to
$\epsilon _\infty$ and ${\cal T}_{0}\rightarrow1$, as seen in Fig.~\ref{fig_metal_grid_no_graphene}.

If we deposit graphene on top of a corrugated metal surface, the system
has two different regimes. For frequencies smaller than $\omega_{p}/\sqrt{\epsilon _\infty +\epsilon_{2}}$
the system behaves essentially as the surface of a bulk conductive
system. When $\omega>\omega_{p}/\sqrt{\epsilon _\infty +\epsilon_{2}}$ the system
behaves as a graphene sheet on a dispersive dielectric. The problem
of the plasmon spectrum of a graphene sheet in the vicinity of a thick
plasma-containing substrate was considered by Horing \cite{Norman}, who
derived the dispersion relation of the surface plasmons of the system.

\begin{figure}[!ht]
\begin{centering}
\includegraphics[clip,width=10cm]{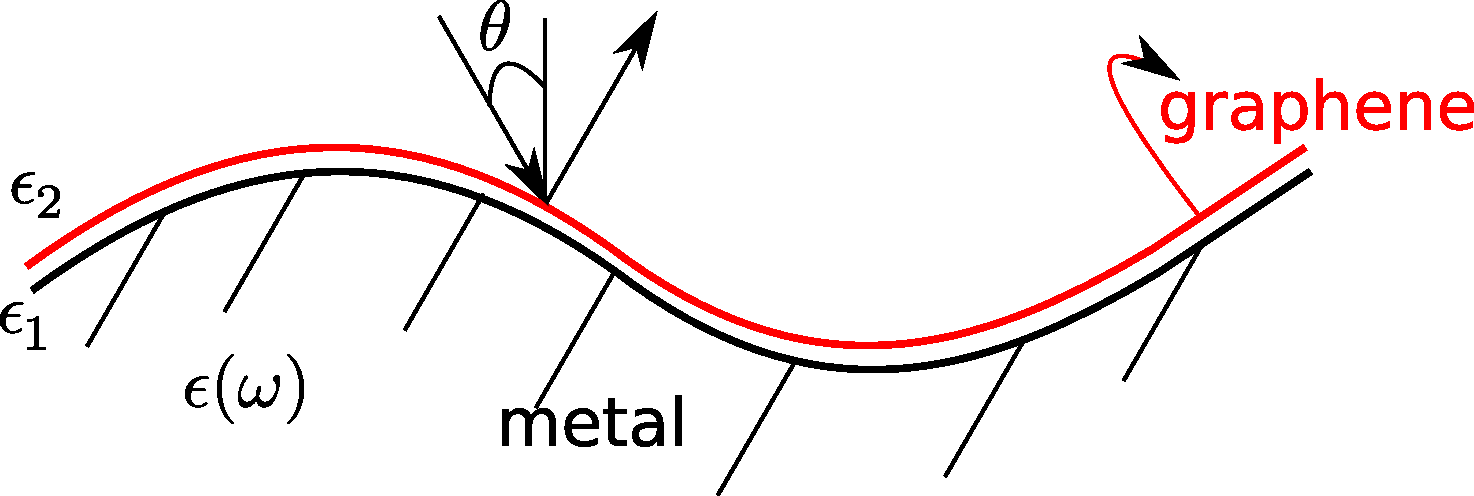}
\par\end{centering}

\caption{Graphene on a metal. The upper and lower media have dielectric permittivities $\epsilon_{2}$ and
$\epsilon_{1}\equiv \epsilon(\omega)$, respectively.
}
\label{fig_graphene_on_a_metal}
\end{figure}

In what follows, we assume that graphene is deposited on a metallic grating,
as illustrated in Fig.~\ref{fig_graphene_on_a_metal}.

For the metal we take a dielectric function corresponding to the limit of
a dispersive conductor, that is, given by Eq.~(\ref{eq_epsilon_dispersice_cond}). For simplicity, we shall assume $\epsilon _\infty =1$.
We want to study the form of the dispersion relation of the surface plasmon--polaritons.
In this case, the eigenvalue equation has the same form as Eq.~(\ref{eq_W_SPP_2D}),
\begin{equation}
1+\frac{\kappa_{2}\epsilon(\omega)}{\kappa_{1}\epsilon_{2}}+i\sigma_{g}\frac{\kappa_{2}}{\omega\epsilon_{0}\epsilon_{2}}=0\,.
\end{equation}
Let us approximate the conductivity of graphene by its imaginary part
only (the dispersive conductor limit). It allows for writing the eigenvalue equation as
\begin{equation}
1+\frac{\kappa_{2}\epsilon(\omega)}{\kappa_{1}\epsilon_{2}}-\frac{\alpha}{\epsilon_{2}}\frac{4E_{F}}{\hbar\omega}\frac{\hbar c\kappa_{2}}{\hbar\omega}=0\,.\label{eq_eigen_3D_2D_metal}
\end{equation}
\begin{figure}[!ht]
\begin{centering}
\includegraphics[clip,width=8cm]{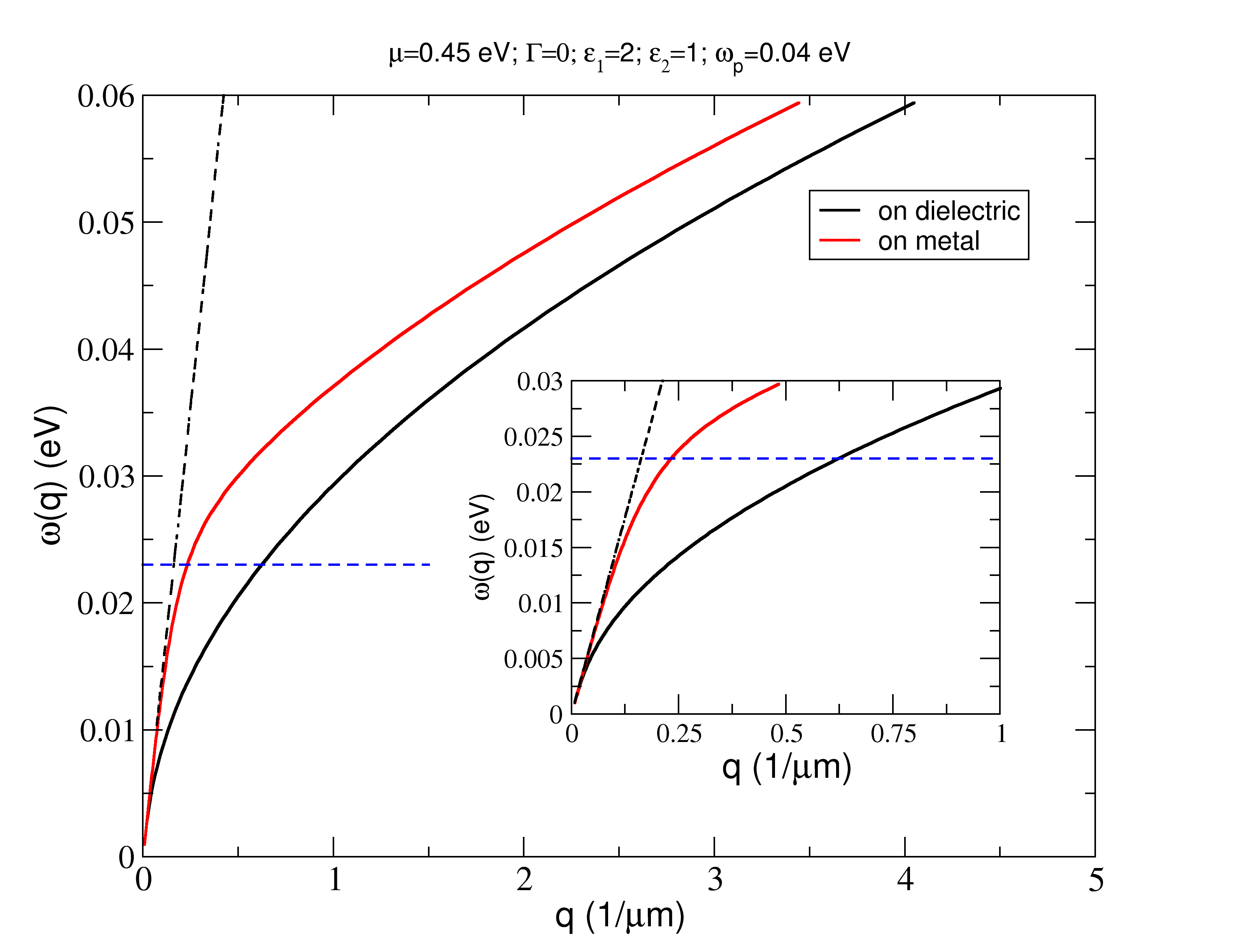}
\par\end{centering}

\caption{SPP dispersion curves for graphene on a conductor with $\omega_{p}=40$~meV and $\epsilon_\infty=1$. The dashed black line is
the light cone, $q/\sqrt{\epsilon_{2}}$, and the horizontal line marks
the value $\omega=\omega_{p}/\sqrt{1+\epsilon_{2}}$.}

\label{fig_graphene_on_a_metal_plasmon}
\end{figure}

Equation (\ref{eq_eigen_3D_2D_metal}) gives the SPP dispersion relation for graphene placed on a metal. When the conductivity of
graphene vanishes, we recover Eq.~(\ref{eq_eigen_3D_plasmon}). In
Fig.~\ref{fig_graphene_on_a_metal_plasmon} we plot the solution
of Eq.~(\ref{eq_eigen_3D_2D_metal}) for $\omega_{p}=0.04$ eV.
The general trend of the solution is the following: when $\omega<\omega_{p}/\sqrt{1+\epsilon_{2}}$,
the SPP dispersion relation is that of the conductor underneath graphene;
when $\omega>\omega_{p}/\sqrt{1+\epsilon_{2}}$ there is a change
of the regime and the dispersion curve follows that of SPPs in graphene cladded by two dielectrics.
It should be noted that, for a bulk
conductor alone, the SPP dispersion relation lies below the value $\omega_{p}/\sqrt{1+\epsilon_{2}}$
for all $q$ (see previous section). We also note that the spectrum for the combined system
\{ graphene+conductor \} lies above that for the graphene alone (with $\epsilon_{1}=1$)
because for $\omega>\omega_{p}$ we have $0<\epsilon(\omega)<1$.
Indeed, when $\omega>\omega_{p}$ and $q\gg\omega/c$, we can find
an analytical expression for $\omega(q)$. In this regime, Eq. (\ref{eq_eigen_3D_2D_metal})
reads:
\begin{equation}
1+\frac{1-\omega_{p}^{2}/\omega^{2}}{\epsilon_{2}}+i\frac{\sigma_{g}q}{\epsilon_{0}\omega\epsilon_{2}}=0\,.
\end{equation}
Solving it in order of $\omega$ gives
\begin{equation}
\omega=\sqrt{\frac{\omega_{p}^{2}}{1+\epsilon_{2}}+4\frac{\alpha E_{F}}{1+\epsilon_{2}}\frac{qc}{\hbar}}\,.
\end{equation}
If $\omega_{p}\rightarrow0$, we recover the dispersion relation
of SPPs on a graphene sheet cladded between two media of relative
permittivities $\epsilon_{1}=1$ and $\epsilon_{2}$.

\begin{figure}[!ht]
\begin{centering}
\includegraphics[clip,width=9cm]{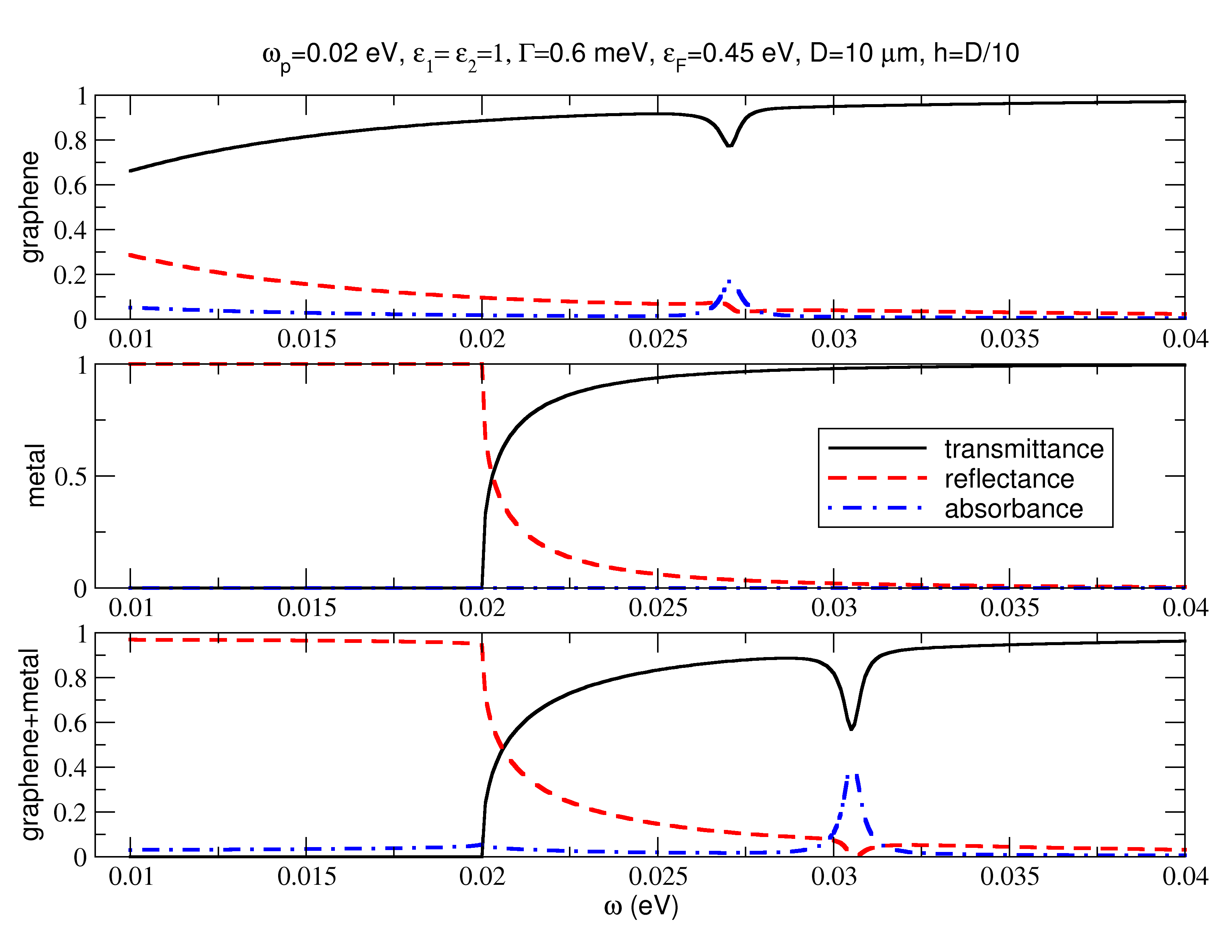}
\par\end{centering}

\caption{Reflectance, transmittance, and absorbance of graphene on a conductive
sinusoidal grating. Top panel: free standing graphene with a sinusoidal profile.
Central panel: sinusoidal metallic grating without graphene. Bottom panel: graphene
on a sinusoidal metallic grating. The parameters are: $\epsilon_{2}=\epsilon _\infty=1$,
$\Gamma=0.6$ meV, $\omega_{p}=0.02$ eV, $E_{F}=0.45$~eV, $D=10$~$\mu$m,
and $h=D/10$.}

\label{fig_metal_and_graphene}
\end{figure}

In Fig.~\ref{fig_metal_and_graphene} we plot the reflectance, transmittance,
and absorbance of graphene on a conductive sinusoidal grating. For
the plasma frequency we assume a value of 20~meV,
corresponding to a doped semiconductor (for example $n-$GaAs,
with $10^{18}$ electrons per cm$^{3}$); some types of mesostructures can also
have very low plasma frequencies \cite{Pendry}. The choice of such a
low plasma frequency simplifies the analysis, since, in this case,
the only propagating order is the specular one. If there is no graphene
on the grating (central panel of Fig.~\ref{fig_metal_and_graphene}
), we have total reflection for $\omega<\omega_{p}^\prime =\omega_{p}/\sqrt {\epsilon _\infty}$, as it should
be. When the graphene is present, the reflectance is smaller than 1
for $\omega<\omega_{p}^\prime$ because of the absorption in graphene. Note,
however, that if graphene were supported by a dielectric instead of the conductor, the reflectance would be
much lower (top panel of Fig.~\ref{fig_metal_and_graphene}). Above $\omega_{p}^\prime$, for this choice of parameters, the reflectance
and the transmittance show coupling of ER to the SPPs in graphene.
The presence of the conductor beneath the graphene sheet shifts the position
of the peak of the resonance toward higher energies because the its dielectric permittivity, although positive, is smaller than 1 (note that $\epsilon _\infty =1$ in this example).

As it can be seen in Fig.~\ref{fig_metal_and_graphene}, the presence of graphene introduces a dip in the
transmittance, at $\omega>\omega_{p}$. At the same time, an
enhancement of the absorption of graphene is observed. It is also
clear (bottom panel) that the reflectance drops to
zero at a given frequency above 0.03~eV, which does not occur when graphene stands
alone on a dielectric grating (top panel).
Note that the metal does not absorb electromagnetic radiation, since
in this model its dielectric constant is purely real. The absorption
occurs entirely within the graphene sheet. If the conductive substrate
also absorbes ER, the analysis becomes more complex \cite{CorrugatedSLG}.


\section{Summary}

\label{sec_summary} At present, graphene plasmonics is an active
field of research, both theoretical and experimental. The community
has been addressing a broad range of topics in this field, some of which are: (i) coupling of ER to SPPs in
graphene; (ii) use of nano-emitters for excitation of SPPs; (iii)
enhancement of light absorption in graphene owing to SPPs; (iv) filters
and polarizers exploring the propagation of SPPs along a graphene-covered surface; (v) launching and
detection of SPPs in graphene, aiming at nanoplasmonic-based circuitry;
(vi) exploring graphene double-layers (or multilayers) where two (or more) different
SPP branches emerge; (vii) ER switches and polarizers based on the ATR
configuration; and (viii) formation of polaritonic crystals with band
gaps controlled by gate voltage.

In this work we focused on the SPP spectra of single and double
layer graphene systems. We showed that the ATR configuration can work
as an optical switch and that the double-layer system allows for moving
the resonant energy towards higher energies. We showed further that, exploring
the different response of graphene to TM and TE waves, one can control
the state of polarization of the outgoing wave by changing the electron Fermi
energy in graphene. It was demonstrated that SPP excitation can be achieved directly by illuminating a thin metallic stripe deposited on top of graphene, which can be viewed as a topological defect in the otherwise uniform system, which enables to overcome the restriction imposed by the in-plane momentum conservation. We discussed the physics of graphene-based gratings where the underlying material is either a dielectric or a conductor.
We also showed that, even in the absence of a grating, an efficient
ER-SPP coupling exists in graphene systems with periodically modulated conductivity,
which can be induced by different mechanisms (e.g.,~strain, doping,
and gating). In particular, it was shown for an array of microribbons and for a continuous graphene sheet with a cosine-modulated
conductivity. Both systems have the properties of a polaritonic crystal, although there is some difference in details between them.
Another system from which one should expect qualitatively similar properties is graphene deposited on a periodically corrugated surface (a grating).
The presented computational description of structures of this type
in terms of Rayleigh-Fano expansions is demanding because the problem
is poorly convergent. Even though, the cases of a sinusoidal and a sawtooth
grating were considered and we showed that the
ER-SPP coupling manifests itself by resonances in the reflectance, transmittance,
and absorbance spectra. When graphene is placed on a metallic surface with smooth periodic profile, the
SPP dispersion is hybrid, showing the properties of a bulk metal at low
wavenumbers and those of an isolated graphene sheet as large wavenumbers.
Not surprisingly, the change of the regime takes place close to the plasma frequency of the metal underlying the graphene sheet.

\section*{Acknowledgements}

This work was partially supported by FEDER through the COMPTETE Program and by
the Portuguese Foundation for Science and Technology (FCT) through Strategic Project PEst-C/FIS/UI0607/2011.



\section*{References}

\begin{thebibliography}{100}
\expandafter\ifx\csname url\endcsname\relax
  \def\url#1{{\tt #1}}\fi
\expandafter\ifx\csname urlprefix\endcsname\relax\def\urlprefix{URL }\fi
\providecommand{\eprint}[2][]{\url{#2}}

\bibitem{Wood}
Wood R~W 1902 {\em Phil. Mag.\/} {\bf 4} 396

\bibitem{Lord}
Rayleigh L 1907 {\em Phil. Mag.\/} {\bf 14} 60

\bibitem{Fano1}
Fano U 1936 {\em Phys. Rev\/} {\bf 50} 573

\bibitem{Fano2}
Fano U 1937 {\em Phys. Rev\/} {\bf 51} 288

\bibitem{Kretschmann}
Kretschmann E and Reather H 1968 {\em Z. Naturf.\/} {\bf 23A} 2135

\bibitem{Otto}
Otto A 1968 {\em Z. Phys.\/} {\bf 216} 398

\bibitem{ExciationIntroduction}
Sambles J~R, Bradbary G~W and Yang F 1991 {\em Contemporary Phys.\/} {\bf 32}
  173

\bibitem{Toigo}
Toigo F, Marvin A, Celli V and Hill N~R 1977 {\em Phys. Rev. B\/} {\bf 15} 5618

\bibitem{Chandezon_1982}
Chandezon J, Dupuis M~T and Cornet G 1982 {\em J. Opt. Soc. Am.\/} {\bf 72} 839

\bibitem{Chandezon_1999}
Li L, Chandezon J, Granet G and Plumey J 1999 {\em Appl. Opt.\/} {\bf 38} 304

\bibitem{Ebbesen1998}
Ebbesen T~W, Lezec H~J, Ghaemi H~F, Thio T and Wolf P~A 1998 {\em Nature\/}
  {\bf 391} 667

\bibitem{Ebbesen2003}
Barnes W~L, Dereux A and Ebbesen T~W 2003 {\em Nature\/} {\bf 424} 824

\bibitem{Ebbesen2008}
Ebbesen T~W, Genet C and Bozhevolnyi S~I 2008 {\em Phys. Today\/} {\bf May} 44

\bibitem{theoryarray}
Mary A, Rodrigo S~G, {Martín-Moreno} L and {García-Vidal} F~J 2007 {\em Phys.
  Rev. B\/} {\bf 76} 195414

\bibitem{theoryarrayII}
Sturman B, Podivilov E and Gorkunov M 2008 {\em Phys. Rev. B\/} {\bf 77} 075106

\bibitem{Stockman2011}
Stockman M~I 2011 {\em Phys. Today\/} {\bf February} 39

\bibitem{PlasmonBook}
Maier S~A 2007 {\em Plasmonics: Fundamentals and Applications\/} (Springer)

\bibitem{SERS1}
Kneipp K 2007 {\em Phys. Today\/} {\bf Movember} 40

\bibitem{SERS2}
Lee J, Shim S, Kim B and Shin H S
 2011 {\em Chem. Eur. J. \/} {\bf 17} 2381

\bibitem{Zhang}
Zhang J, Zhang L and Xu W 2012 {\em J. Phys. D: Appl. Phys.\/} {\bf 45} 113001

\bibitem{Langmuir}
Jung L~S, Campbell C~T, Chinowsky T~M, Mar M~N and Yee S~S 1998 {\em
  Langmuir\/} {\bf 14} 5636

\bibitem{Amanda2005}
Haes A~J, Haynes C~L, DMcFarland A, Schatz G~C, {Van Duyne} R~P and Zou S 2005
  {\em MRS Bulletin\/} {\bf 30} 368

\bibitem{Willets2007}
Willets K~A and {Van Duyne} R~P 2007 {\em Annu. Rev. Phys. Chem.\/} {\bf 58}
  267

\bibitem{Shalabney2011}
Shalabney A and Abdulhalim I 2011 {\em Laser Photon. Rev.\/} {\bf 5} 571

\bibitem{Green2011}
Green M~A and Pillai S 2012 {\em Nature Photonics\/} {\bf 6} 130

\bibitem{Reece2008}
Reece P~J 2008 {\em Nature Photonics\/} {\bf 2} 333

\bibitem{Juan2011}
Juan M~L, Righini M and Quidant R 2011 {\em Nature Photonics\/} {\bf 5} 349

\bibitem{Ozbay2006}
Ozbay E 2006 {\em Science\/} {\bf 311} 189

\bibitem{Han-Bozhevolnyi2013}
Han Z and Bozhevolny S~I 2013 {\em Rep. Prog. Phys. \/} {\bf 76} 016402

\bibitem{Ashkan}
Vakil A and Engheta N 2011 {\em Science\/} {\bf 332} 1291

\bibitem{Ashkan2}
Vakil A and Engheta N 2012 {\em Phys. Rev. B\/} {\bf 85} 075434

\bibitem{Xu}
Xu H~J, Lu W~B, Zhu W, Dong Z~G and Cui T~J 2012 {\em Appl. Phys. Lett.\/} {\bf
  100} 243110

\bibitem{Shung}
Shung K~W~K 1986 {\em Phys. Rev. B\/} {\bf 34} 979

\bibitem{SarmaPlasmon}
Hwang E~H and {Das Sarma} S 2007 {\em Phys. Rev. B\/} {\bf 75} 205418

\bibitem{Wunsch}
Wunsch B, Stauber T, Sols F and Guinea F 2006 {\em New J. Phys.\/} {\bf 8} 318

\bibitem{StauberFull}
Stauber T, Schliemann J and Peres N~M~R 2010 {\em Phys. Rev. B\/} {\bf 81}
  085409

\bibitem{Jablan}
Jablan M, Buljan H and {Solja\v{c}i\'{c}} M 2009 {\em Phys. Rev. B\/} {\bf 80}
  245435

\bibitem{StauberRS}
Stauber T, Peres N~M~R and Guinea F 2007 {\em Phys. Rev. B\/} {\bf 76} 205423

\bibitem{Mishchenko}
Mishchenko E~G, Shytov A~V and Silvestrov P~G 2010 {\em Phys. Rev. Lett.\/}
  {\bf 104} 156806

\bibitem{Popov}
Popov V~V, Bagaeva T~Y, Otsuji T and Ryzhii V 2010 {\em Phys. Rev. B\/} {\bf
  81} 073404

\bibitem{Kinaret}
Wang W, Apell P and Kinaret J 2011 {\em Phys. Rev. B\/} {\bf 84} 085423

\bibitem{Pedersen}
Schultz M~H, Jauho A~P and Pedersen T~G 2011 {\em Phys. Rev. B\/} {\bf 84}
  045428

\bibitem{Thongrattanasiri}
Thongrattanasiri S, Manjavacas A and {de Abajo} F~J~G 2012 {\em ACS Nano\/}
  {\bf 6} 1766

\bibitem{Nikitin1}
Nikitin A~Y, Guinea F, Garcia-Vidal F~J and Martin-Moreno L 2011 {\em Phys.
  Rev. B\/} {\bf 85} 081405

\bibitem{Nikitin2}
Nikitin A~Y, Guinea F, Garcia-Vidal F~J and Martin-Moreno L 2011 {\em Phys.
  Rev. B\/} {\bf 84} 161407

\bibitem{AbajoPlasmons}
Thongrattanasiri S, Silveiro I and {de Abajo} F~J~G 2012 {\em Appl. Phys.
  Lett.\/} {\bf 100} 201105

\bibitem{Raza}
Andersen D~R and Raza H 2012 {\em Phys. Rev. B\/} {\bf 85} 075425

\bibitem{Christensen}
Christensen J, Manjavacas A, Thongrattanasiri S, Koppens F~H~L and {de Abajo}
  F~J~G 2012 {\em ACS Nano\/} {\bf 6} 431

\bibitem{Pellegrino}
Pellegrino F~M~D, Angilella G~G~N and Pucci R 2010 {\em Phys. Rev. B\/} {\bf
  82} 115434

\bibitem{Pellegrino2}
Pellegrino F~M~D, Angilella G~G~N and Pucci R 2011 {\em Phys. Rev. B\/} {\bf
  84} 195407

\bibitem{Koppens2011}
Koppens F~H~L, Chang D~E and {de Abajo} F~J~G 2011 {\em Nano Lett.\/} {\bf 11}
  3370

\bibitem{Nikitin3}
Nikitin A~Y, Guinea F, Garcia-Vidal F~J and Martin-Moreno L 2011 {\em Phys.
  Rev. B\/} {\bf 84} 195446

\bibitem{Santos}
{G\'omez-Santos} G and Stauber T 2011 {\em Phys. Rev. B\/} {\bf 84} 165438

\bibitem{Huidobro}
Huidobro P~A, Nikitin A~Y, {González-Ballestero} C, {Martín-Moreno} L and
  {García-Vidal} F~J 2012 {\em Phys. Rev. B\/} {\bf 85} 155438

\bibitem{Manjavacas}
Manjavacas A, Nordlander P and {de Abajo} F~J~G 2012 {\em ACS Nano\/} {\bf 6}
  1724

\bibitem{Sarma}
Hwang E~H, Sensarma R and {Das Sarma} S 2010 {\em Phys. Rev. B\/} {\bf 82}
  195406

\bibitem{PlasmonsBL}
Jablan M, Buljan H,  and {Solja\v{c}i\'{c}} M 2011 {\em OPTICS EXPRESS\/} {\bf
  19} 11236

\bibitem{StauberDL}
Stauber T and {G\'omez-Santos} G 2012 {\em Phys. Rev. B\/} {\bf 85} 075410

\bibitem{ProfumoDL}
Profumo R~E~V, Asgari R, Polini M and MacDonald A~H 2012 {\em Phys. Rev. B\/}
  {\bf 85} 085443

\bibitem{GanDL}
Gan C~H, Chu H~S and Li E~P 2012 {\em Phys. Rev. B\/} {\bf 85} 125431

\bibitem{OgnjenDL}
Ilic O, Jablan M, Joannopoulos J~D, Celanovic I, Buljan H and
  {Solja\v{c}i\'{c}} M 2012 {\em Phys. Rev. B\/} {\bf 85} 155422

\bibitem{WangDL}
Wang B, Zhang X, Yuan X and Teng J 2012 {\em Appl. Phys. Lett.\/} {\bf 100}
  131111

\bibitem{Schedin2010}
Schedin F, Lidorikis E, Lombardo A, Kravets V~G, Geim A~K, Grigorenko A~N,
  Novoselov K~S and Ferrari A~C 2010 {\em ACSNano\/} {\bf 4} 5617

\bibitem{LongJuPlasmonics}
Ju L, Geng B, Horng J, Girit C, Martin M~C, Hao Z, Bechtel H~A, Liang X, Zettl
  A, Shen Y~R and Wang F 2011 {\em Nature Nanotechnology\/} {\bf 6} 630

\bibitem{EchtermeyerPhoto}
Echtermeyer T~J, Britnell L, Jasnos P~K, Lombardo A, Gorbachev R~V, Grigorenko
  A~N, Geim A~K, Ferrari A and Novoselov K~S 2011 {\em Nature Communications\/}
  {\bf 2} 458

\bibitem{BasovPlasmons}
Fei Z, Andreev G~O, Bao W, Zhang L~M, McLeod A~S, Wang C, Stewart M~K, Zhao Z,
  Dominguez G, Thiemens M, Fogler M~M, Tauber M~J, {Castro-Neto} A~H, Lau C~N,
  Keilmann F and Basov D~N 2011 {\em Nano Lett.\/} {\bf 11} 4701

\bibitem{rmp}
{Castro Neto} A~H, Guinea F, Peres N~M~R, Novoselov K~S and Geim A~K 2009 {\em
  Rev. Mod. Phys.\/} {\bf 81} 109

\bibitem{rmpPeres}
Peres N~M~R 2010 {\em Rev. Mod. Phys.\/} {\bf 82} 2673

\bibitem{Terahertz}
Zhang X~C and Xu J 2010 {\em Introduction to THz Wave Photonics\/} (Springer)

\bibitem{BasovSPP}
Fei Z, Rodin A~S, Andreev G~O, Bao W, McLeod A~S, Wagner M, Zhang L~M, Zhao Z,
  Dominguez G, Thiemens M, Fogler M~M, {Castro-Neto} A~H, Lau C~N, Keilmann F
  and Basov D~N 2012 {\em arXiv\/}  1202.4993

\bibitem{KoppensSPP}
Chen J, Badioli M, Alonso-González P, Thongrattanasiri S, Huth F, Osmond J,
  Spasenovic M, Centeno A, Pesquera A, Godignon P, Zurutuza A, Camara N, {de
  Abajo} J~G, Hillenbrand R and Koppens F 2012 {\em arXiv\/}  1202.4996

\bibitem{Avouris}
Yan H, Li X, Chandra B, Tulevski G, Wu Y, Freitag M, Zhu W, Avouris P and Xia F
  2012 {\em Nature Nano.\/} {\bf 7} 330

\bibitem{Chen}
Chen P~Y and Al\`u A 2011 {\em ACS Nano\/} {\bf 5} 5855

\bibitem{YuliyEPL}
Bludov Y~V, Vasilevskiy M~I and Peres N~M~R 2010 {\em EPL\/} {\bf 92} 68001

\bibitem{Kuzmenko}
Crassee, Orlita M, Potemski M, Walter A~L, Ostler M, Seyller T, Gaponenko I,
  Chen J and Kuzmenko A~B 2012 {\em Nano Lett.\/} {\bf 12} 2470

\bibitem{Ting-Yu_2012}
Sreekanth K~V, Zen S, Shang J, Yong K-T, and Ting Yu 2012 {\em Sci. Rep.\/} {\bf 2} 737

\bibitem{CorrugatedSLG}
Ferreira A and Peres N~M~R 2012 {\em Phys. Rev. B\/} {\bf 86} 205401

\bibitem{nunoSPP}
Peres N~M~R, Ferreira A, Bludov Y~V and Vasilevskiy M~I 2012 {\em J. Phys.:
  Condens. Matter\/} {\bf 24} 245303

\bibitem{Davoyan2012}
Davoyan A~R, Popov V~V and Nikitov S~A 2012 {\em Phys. Rev. Lett.\/} {\bf 108}
  127401

\bibitem{YuliyPRB}
Bludov Y~V, Peres N~M~R and Vasilevskiy M~I 2012 {\em Phys. Rev. B\/} {\bf 85}
  245409

\bibitem{Shen}
{De Martini} F and Shen Y~R 1976 {\em Phys. Rev. Lett.\/} {\bf 36} 216

\bibitem{Georges}
Georges A~T and Karatzas N~E 2012 {\em Phys. Rev.\/} {\bf 85} 155442

\bibitem{Renger}
Renger J, Quidant R, van Hulst N, Palomba S and Novotny L 2009 {\em Phys. Rev.
  Lett.\/} {\bf 103} 266802

\bibitem{semiconductorfilms}
Aharonian K~H and Tilley D~R 1989 {\em J. Phys.: Condens. Matter\/} {\bf 1}
  5391

\bibitem{OpticsBook}
Fowles G~R 1989 {\em Introduction to Modern Optics\/} (Dover)

\bibitem{nmrPRB06}
Peres N~M~R, Guinea F and {Castro Neto} A~H 2006 {\em Phys. Rev. B\/} {\bf 73}
  125411

\bibitem{falkovsky}
Falkovsky L~A and Pershoguba S~S 2007 {\em Phys. Rev. B\/} {\bf 76} 153410

\bibitem{StauberGeim}
Stauber T, Peres N~M~R and Geim A~K 2008 {\em Phys. Rev. B\/} {\bf 78} 085432

\bibitem{Polini}
Abedinpour S~H, Vignale G, Principi A, Polini M, Tse W~K and MacDonald A~H 2011
  {\em Phys. Rev. B\/} {\bf 84} 045429

\bibitem{EOM}
Ferreira A, Viana-Gomes J, Bludov Y~V, Pereira V, Peres N~M~R and {Castro Neto}
  A~H 2011 {\em Phys. Rev. B\/} {\bf 84} 235410

\bibitem{Ziman}
Ziman J~M 2001 {\em Electrons and Phonons\/} (Oxford University Press)

\bibitem{Dubinov}
Dubinov A~A, Aleshkin V~Y, Mitin V, Otsuji T and Ryzhii V 2011 {\em J. Phys.:
  Condens. Matter\/} {\bf 23} 145302

\bibitem{KianPing}
Bao Q, Zhang H, Wang B, Ni Z, Lim C~H~Y~X, Wang Y, Tang D~Y and Loh K~P 2011
  {\em Nature Photonics\/} {\bf 5} 411

\bibitem{Born} Born M and Wolf E 1989 {\it Principles of Optics} (Pergamon)

\bibitem{Falko}
Falko V~I and Khmelnitskii D~E 1989 {\em Sov. Phys. JETP\/} {\bf 68} 1150

\bibitem{Hecht}
Hecht E 2003 {\em Optics\/} 4th ed (Pearson)

\bibitem{ConfinedMPP}
Ferreira A, Peres N~M~R and {Castro Neto} A~H 2012 {\em Phys. Rev. B\/} {\bf
  85} 205426

\bibitem{Satou_Mikhailov}
Satou A and Mikhailov S~A 2007 {\em Phys. Rev. B\/} {\bf 75} 045328

\bibitem{Morse} P. M. Morse and H. Feshbach 1953 {\it  Methods of theoretical physics} (McGraw-Hill, New York)

\bibitem{Roldan2009}
Roldan R, Fuchs J~N and Goerbig M~O 2009 {\em Phys. Rev. B\/} {\bf 80} 085408

\bibitem{Chui1974}
Chiu K~W and Quinn J~J 1974 {\em Phys. Rev. B\/} {\bf 9} 4724

\bibitem{Kukushkin2006}
Kukushkin I~V, Muravev V~M, Smet J~H, Hauser M, Dietsche W and von Klitzing K
  2006 {\em Phys. Rev. B\/} {\bf 73} 113310

\bibitem{ResonantCoupling}
Park S, Lee G, Song S~H, Oh C~H and Kim P~S 2003 {\em Optics Lett.\/} {\bf 28}
  41870

\bibitem{Chen-Friedman} Chen X and Friedman A 1991 {\em Trans. Amer. Math. Soc. \/}
  {\bf 323} 465

\bibitem{Smirnov}
Smirnov G 1997 {\em J. Math. Anal. Appl. \/}
  {\bf 214} 395

\bibitem{Pendry}
Pendry J~B, Holden A~J, Stewart W~J and Youngs I 1996 {\em Phys. Rev. Lett.\/}
  {\bf 76} 4773

\bibitem{Fox}
Fox M 2007 {\em Optical Properties of Solids\/} (Oxford)

\bibitem{Torrell} Torrell M, Kabir R, Cunha L, Vasilevskiy M~I, Vaz F, Cavaleiro A, Alves E and
Barradas N P 2011 {\em J. Appl. Phys.\/}
  {\bf 109} 074310

\bibitem{Norman}
Horing N~J~M 2009 {\em Phys. Rev. B\/} {\bf 80} 193401

\end{thebibliography}

\providecommand{\newblock}{}


\end{document}